\documentclass{article}

\usepackage{PRIMEarxiv}

\usepackage[utf8]{inputenc} 
\usepackage[T1]{fontenc}    
\usepackage{hyperref}       
\usepackage{url}            
\usepackage{booktabs}       
\usepackage{amsfonts}       
\usepackage{nicefrac}       
\usepackage{microtype}      
\usepackage{lipsum}
\usepackage{fancyhdr}       
\usepackage{appendix}
\usepackage{graphicx}       
\usepackage[justification=centerlast, indention=.5cm]{caption}
\usepackage[position=b]{subcaption}
\usepackage{amsmath}
\usepackage{comment}
\usepackage{tabularx}
\usepackage{dsfont}
\graphicspath{{media/}}     

\usepackage{footnote}
\usepackage{bbold}
\usepackage{ragged2e}
\usepackage{xcolor}

\title{A comprehensive guideline for regularization-path variable selection in high-dimensional Gaussian linear regression
}

\author{
  Perrine Lacroix \\
  Université Paris-Saclay, CNRS, INRAE, Laboratoire de Mathématiques d’Orsay (LMO) -  \\
  91405, Orsay, France, \\
   Université Paris-Saclay, CNRS, INRAE, Institute of Plant Sciences Paris-Saclay (IPS2) - \\
     91190, Gif sur Yvette, France, \\
  Nantes Université, CNRS, Laboratoire de Mathématiques Jean Leray (LMJL) UMR 6629, \\
   F-44000 Nantes, France \\
    \texttt{perrine.lacroix@univ-nantes.fr} \\
   \And
  Mélina Gallopin \\
  Université Paris-Saclay, CNRS, CEA, Institut de Biologie Intégrative de la Cellule (UMR 9198), \\
  91405, Orsay, France \\
  \texttt{melina.gallopin@universite-paris-saclay.fr} \\
  \And
  Marie-Laure Martin \\
   Université Paris-Saclay, CNRS, INRAE, Université Evry, Institute of Plant Sciences Paris-Saclay (IPS2), \\ 91190, Gif sur Yvette, France.\\
    Université Paris Cité, Institute of Plant Sciences Paris-Saclay (IPS2), \\
    91190, Gif sur Yvette, France. \\
    Université Paris-Saclay, AgroParisTech, INRAE, UMR MIA Paris-Saclay, \\
     91120, Palaiseau, France \\
  \texttt{marie-laure.martin@inrae.fr} \\
}

\begin{document}
\maketitle

\begin{abstract}
This paper provides a comprehensive comparison of complete regularization-path-based variable selection procedures in high-dimensional Gaussian linear regression. Our simulation study stands out from previous ones for several reasons. First, it offers comprehensive guidelines for complete variable selection, from regularization path construction to final variable subset selection. In particular, we compare jointly two convex regularization functions (Lasso and Elastic-Net) with two optimization algorithms (LARS and cyclic coordinate descent) and a range of model selection and variable identification techniques. Second, we also incorporate methods with non-asymptotic theoretical guarantees, which are typically not included in existing reviews.
It is unrealistic to expect a single method that works best. Still, we provide detailed guidance on which methods perform best under different evaluation criteria, including pROC-AUC, MSE, recall, specificity, and FDP, and that are robust to data scenarios, including varying variable dependency structures.
Overall, Elastic-Net combined with the LARS algorithm provides the most reliable regularization path, while the preferred final selection procedure depends on whether prediction performances, support recovery or false discovery control is prioritized. We show that some methods are empirically better for a given criterion than others, even though the latter were designed to control for it theoretically.
Regarding new developments in a non-asymptotic setting, we highlight the quality of LinSelect and the need, as future work, to fine-tune the unknown constants of data-dependent penalties in high-dimensional Gaussian linear regression models.
We further assess the robustness of these methods by relaxing key model assumptions, such as the Gaussian distribution and the high-dimensional setting, and show that relaxing the Gaussian assumption or reducing the number of observations consistently degrades the performance of all procedures. 
\end{abstract}


\section{Introduction}
Recent scientific advances have fundamentally transformed the nature of available datasets and the methods used to analyze them. The dataset size explodes, as does the complexity of each dataset (including heterogeneous data types, complex variable structures, time-series components...). 
For example, in genomics, microarray and RNA-sequencing technologies that allow simultaneous quantification of gene expression facilitate the study of their interactions (across genomic, proteomic and metabolomic data) and thereby improve the understanding of molecular activities.
In medicine, developments in biomedical imaging, treatment monitoring, and market strategies have generated massive and complex datasets, as well as GPS video-surveillance data in the digital domain.
From a statistical perspective, the number of parameters to estimate explodes, hindering the use of traditional estimation methods. Dimension reduction techniques are essential to identify influential variables while removing redundant information from a given model. 

In this paper, we consider frameworks where the objective is to investigate the relationship between a set of variables and a response variable. Depending on the context, the goal may be either to predict the response variable for new observations accurately or to better understand the relationships among the variables and the response variable. 
Assuming that this relation is linear, high-dimensional Gaussian linear regression provides a suitable statistical framework for identifying either the variables related to the response (called active variables) or the predictors of the response. 
Many variable selection procedures have been proposed in the literature. 
In this paper, we focus on regularization-path-based variable selection procedures, which represent one of the most widely used classes of variable selection procedures in high-dimensional linear regression models. 

\subsection{State of the art}
The literature reviewing variable selection in high-dimensional Gaussian linear regression has predominantly focus only on regularization path construction methods. 
These approaches minimize a cost criterion penalized by a regularization function and yield an ordering of variables. 
The authors of \cite{Wainwright2009} provided a meticulous theoretical analysis of the $\ell_1$ regularization function (Lasso) and notably discussed the required number of observations to ensure asymptotic properties of recovering the active variables. 
\cite{buhlmann2014high} compared several regularization functions through a simulation study based on semi-real datasets. They considered several numbers of observations, variables and active variables and varied the signal-to-noise ratio and the correlation. structure among variables. The results of the different regularization functions are inspected with ROC curves and partial ROC curves when 0.5$n$ and 0.9$n$ variables are selected. 
\cite{wang2020high} extended comparisons of \cite{buhlmann2014high} to a large set of regularization functions under similar design settings and evaluated both prediction and variable identification, but they considered only one model selection applied from the regularization path (cross-validation approach). 
More recently, \cite{hanke2024variable} showed, through a neutral empirical comparison across various settings, parameter choices and evaluation metrics, that the $\ell_0$ regularization function, often considered as a benchmark for optimal subset selection in terms of prediction, can be outperformed by the Elastic-Net (weighted combination of Lasso and $\ell_2$ penalties) regularization function, especially when correlation between variables exists.
Complementarily, \cite{zhao2023survey} focused on algorithms that can solve the Lasso optimization problem. They provide a theoretical comparison of how quickly the sequences of estimators converge to the optimal one, of the algorithmic complexities, and of the robustness of the solution. However, they didn't rely on the calibration of the unknown regularization hyperparameter, which can be addressed using model selection or variable identification approaches. 
At the opposite end of the complete regularization-path-based variable selection procedures, \cite{wu2020survey} reviewed some approaches to tune the unknown Lasso (and variants) regularization hyperparameter, including model selection criteria (e.g., BIC and variants) and variable identification methods (e.g., Bolasso). Their theoretical comparisons emphasized that the choice of the best method should be based on the target performance and data structure. \\

Overall, existing simulation studies focus on comparing individual part of the complete regularization-path-based variable selection procedures: either regularization paths, iterative algorithms for only one regularization function, model selection procedures used to extract a final model from a fixed regularization path, or regularization paths combined with a single variable selection strategy. 
To the best of our knowledge, no comparison study exists for complete regularization-path-based variable selection procedures, which combine both path construction and final subset selection. 
Moreover, despite their optimal theoretical guarantees, non-asymptotic model selection criteria are generally not included in existing reviews, which have concentrated only on classical asymptotic model selection criteria. 
We propose filling these two gaps in this simulation study.

\subsection{Scope of the simulation study}
The scope of this simulation study is restricted to complete regularization-path-based variable selection procedures, with an emphasis on methods based on non-asymptotic theoretical performances. We do not consider Adaptive Best Subset Selection (ABESS) \cite{zhu2020polynomial} method that are not based on regularization paths. To construct the regularization path, we focus on convex penalized least-squares problems (allowing the use of computationally efficient optimization algorithms) which excludes the Dantzig Selector \cite{candes2009dantzig} and the Smoothy Clipped Absolute Deviation (SCAD) \cite{fan2001variable} from the paper. We do not consider multiple testing procedures based on p-values, such as the Benjamini-Hochberg (BH) \cite{benjamini1995controlling} and Benjamini-Yekutieli (BY) \cite{benjamini2001control} procedures.
Finally, post-selection approaches are outside the scope of this paper discarded methods like the Gaussian mirror \cite{xing2023controlling} or data splitting \cite{dai2023false}, recently proposed to control the False Discovery Rate in variable selection. \\

We note that some simulation studies have explored different contexts not considered in this paper. \cite{wu2015selective} investigated robust variable selection strategies in the presence of heavy-tailed errors and outliers in response variables. They detailed the different steps from modified least squares criteria to parameter selection, tuning the model selection through a presentation of algorithms that account for outliers. 
\cite{vinga2021structured} considered a range of models from survival ones to generalized linear ones, commonly used in biomedical research. \cite{desboulets2018review} considered a wide range of model structures (including linear, grouped, additive, partially linear and non-parametric models) and discussed three main categories of variable selection algorithms. Methods based on multiple testing procedures \cite{kos2020asymptotic} and Bayesian approaches are beyond the scope of this paper and we refer the readers to \cite{celeux2012regularization} for an empirical comparison between frequentist and Bayesian approaches. We also assume no prior known information like interactions between variables, spatial localizations or chronological ordering and refer to \cite{bondell2012consistent, razaghi2020supervised} for such approaches. 
Finally, for a general overview of the main dimension reduction methods, including the inverse regression problem, the moments method,  the sequential test method, the bootstrap method, the penalty criterion and the spectral eigen-decomposition method, we refer the reader to \cite{ma2013review}. 
 
\subsection{Main contributions} 
\label{Contributions}
The main contributions of this simulation study are multiple. 

\subsubsection{A comprehensive benchmark of complete regularization-path-based variable selection procedures} 
\label{contrib_complete}
Previous work mainly focuses on comparing either regularization paths or model selection procedures separately. Consequently, practitioners still lack practical guidance on how to combine the different steps of a complete regularization-path-based variable selection procedure. 
Therefore, we propose, in this simulation study, the joint evaluation of both the construction of regularization paths and the selection of the final subset of variables along them. 
More precisely, we assess complete selection strategies obtained by combining regularization functions, optimization algorithms (both for the regularization path) and model selection or variable identification approaches (for the selection of the final variable subset).
Specifically, a total of $33$ combinations are compared in this article. \\
More precisely : 
\begin{itemize}
\item To construct the regularization paths, we consider two penalty functions: Lasso \cite{tibshirani1996regression} and Elastic-Net \cite{zou2005regularization}, combined with two algorithms: LARS \cite{efron2004least} and the cyclic coordinate descent algorithm \cite{friedman2010regularization}. 
\end{itemize}
Each regularization path provides a collection of variable subsets. To select the final subset, we compare both model selection and variable identification approaches. 
\begin{itemize}
\item Model selection methods are based on penalized least squares criteria. We include the eBIC penalization \cite{chen2008extended}, data-driven calibration strategies \cite{birge2007minimal, lebarbier2005detecting, baudry2012slope, arlot2019minimal} and LinSelect \cite{baraud2009gaussian,giraud2012high}.
\item On the other hand, variable identification methods rely on sampling strategies to stabilize the final subset selection process while limiting the selection of non-active variables. We consider ESCV \cite{lim2016estimation}, Bolasso \cite{bach2008bolasso}, Stability Selection \cite{meinshausen2010stability}, Tigress \cite{haury2012tigress} and the knockoffs method \cite{barber2015controlling}. 
\end{itemize}

After describing the methods and outlining their theoretical properties, we evaluate and compare their performance through a simulation study based on Gaussian linear models. 
We vary the correlation structure among variables to evaluate how dependencies between variables influence the behavior of the methods. We also consider the \textit{independent} setting, which corresponds to the theoretical framework in which methods are typically developed, and treat it as a benchmark. 
To evaluate performances, several metrics are computed. To test the ability to discriminate between active and non-active variables, we use the pROC-AUC metric. Predictive performance is evaluated with the mean squared error (MSE). The qualities of the selected variables in terms of recovering the active ones are evaluated using recall, specificity, and false discovery proportion (FDP). 
In line with previous simulation studies \cite{buhlmann2014high, wang2020high, wu2020survey}, no single method consistently outperforms all others across all metrics we evaluate.

\subsubsection{Recent advances in non-asymptotic model selection procedures}
\label{contrib_non_asymptotic}
Non-asymptotic model selection criteria are attractive methods because of their finite-sample optimal theoretical guarantees. 
However, they are rarely considered in the literature and practitioners generally rely on asymptotic approaches, even when the number of observations is relatively small. 
Therefore, we propose one of the first comparisons between classical asymptotic model selection criteria and recent non-asymptotic model selection ones within the complete regularization-path-based variable selection procedures. More precisely, we consider the non-asymptotic criteria LinSelect and data-driven calibration methods, whereas eBIC penalization serves as our asymptotic model selection criterion benchmark.
Furthermore, these model selection criteria are based on theoretical predictive performance and, as a result, are rarely compared with variable identification approaches that recover the true set of active variables. We include these two approaches in a unified comparison study. 

\subsubsection{Practical guidelines for practitioners derived from the benchmark} 
\label{contrib_guidelines}
As there is no an unambiguous winner, we provide practical recommendations, based on the simulation results and the metric of interest, for selecting appropriate methods that are robust to the (usually unknown) dependency structure of the variables. 
In particular, our simulation study suggests that Elastic-Net provides a more robust choice than Lasso across the considered scenarios, and the LARS algorithm is more effective than cyclic coordinate descent. 
To ensure predictive ability, eBIC, ESCV and the knockoffs are the most judicious choices. 
If the goal is to limit the selection of non-active variables while maximizing the selection of active variables,  ESCV and the knockoffs method are preferable. 
LinSelect, Tigress, Bolasso and the knockoffs method should be preferred to limit the selection of non-active variables. 
Notably, Bolasso achieves intermediate performance across metrics, with high performance and very rarely non-active selected variables. 

\subsubsection{Robustness from the model assumptions} 
\label{contrib_robust}
Finally, we investigate the robustness of the considered methods by relaxing key model assumptions. In particular, we explore high dimensionality by increasing the number of observations and we consider datasets generated by the \textit{FRANK} algorithm \cite{carre2017reverse} which mimic the biological complexity of transcription factor regulation and thus deviate from the Gaussian model. 

\subsection{Outline}
The rest of the paper is organized as follows. Section \ref{methods_review} describes the statistical framework and all the methods we compare in this simulation study. 
Section \ref{parameters_methods} presents the simulation settings, technical aspects about the implementation of methods and the
evaluation metrics we use. 
Section \ref{results} is devoted to all the results and their interpretations. 
Lastly, a conclusion is provided in Section \ref{conclusions}.

\section{Methods}
\label{methods_review}

\subsection{\textbf{Statistical framework}}
For the sequel, the norms $|.|_0$, $|.|_1$ and $||.||$  are defined for a vector $\beta$$ \in \mathbb{R}^q$ ($q \in \mathbb{N}^{*}$) by :
\begin{equation*}
    |\beta|_0 = \sum_{j=1}^q \mathbb{1}_{\{\beta_j \neq 0 \}}; \;\;
    |\beta|_1 = \sum_{j=1}^q |\beta_j|; \;\;
     ||\beta|| = \sqrt{\sum_{j=1}^q \beta_j^2}. \\
\end{equation*}

We consider the Gaussian linear regression model where the response variable $Y$ is explained by a linear combination of $p$ quantitative variables $X=(X_1,\ldots,X_p)$ : 
$$ Y = X  \beta^* + \varepsilon, $$
where $\beta^* \in \mathbb{R}^p$ is the unknown parameter vector of interest and $\varepsilon$ is a centered Gaussian noise vector with an unknown variance denoted $\sigma^2$. To estimate $\beta^*$ and $\sigma^2$, independent observations $y_i \in \mathbb{R}$ and $(x_{i1},..., x_{ip}) \in \mathbb{R}^p$ are available for $i \in \{1, ..., n\}$. 

We consider the high-dimensional framework where $p \sim n$ or $p > n$, preventing the traditional least squares estimation. In this context, we assume that only a small number of the $p$ variables explains the response variable. 
These variables are named active variables and are associated to a non-zero coefficient in $\beta^*$. Under this sparsity assumption, the equation to solve becomes for $t >0$ :
 $$\underset{\beta \in \mathbb{R}^{p} : |\beta|_0 \leq t}{\text{min}}  ||Y-X \beta||^2. $$
An equivalent penalized formulation is obtained through the Lagrangian problem for $\lambda>0$ :
\begin{eqnarray}
\underset{\beta \in \mathbb{R}^{p}}{\text{min}} \left\{ ||Y-X \beta||^2 + \lambda |\beta|_0 \right\}. 
   \label{L0_penalty}
\end{eqnarray}
The proof of the equivalence and the link between $t$ and $\lambda$ are provided in \cite{tibshirani1996regression}. 
A large value of $\lambda$ provides a small subset of variables involved in the regression, but it may  correspond to a fit far from the response variable. A small value of $\lambda$ corresponds to a fit close to the response variable, but it may  provide a large subset of variables. Determining the hyperparameter $\lambda$ is therefore one of the major issues and the challenge lies in its calibration to adjust a trade-off between sparsity and good adjustment. 
Moreover, the criterion being non-convex, the existence and the uniqueness of the solution are not guaranteed. So, as presented in \cite{vinga2021structured}, Equation~(\ref{L0_penalty}) is often replaced with the optimization problem :
\begin{eqnarray}
\underset{\beta \in \mathbb{R}^{p}}{\min} \left\{ ||Y-X \beta||^2 + \lambda F(\beta) \right\},
   \label{penalty}
\end{eqnarray}
where $F$ is a continuous and convex  function satisfying the existence of a minimum for any $\lambda$. 

\subsection{\textbf{Regularization functions}}
The choice of the regularization functions $F$ in (\ref{penalty}) is based on a trade-off between sparsity, computational tractability, and the ability to account for the variable structure dependencies. In this paper, we focus on the two commonly used functions : Lasso and Elastic-Net. 

The first one is the $\ell_1$ regularization, named Lasso \cite{tibshirani1996regression} with 
$$F(\beta) = |\beta|_1.$$ 
The Lasso procedure achieves the best trade-off between regularity (convexity, reasonable computational solution) and sparsity for independent variables. Moreover, if $\lambda$ is well chosen and under suitable assumptions, it provides a consistent estimator of $\beta^*$. 
However, when some variables are correlated, Lasso tends to select randomly only one of them rather than selecting none or all of them (see \cite{freijeiro2022critical} for a complete overview of the Lasso regression, an explanation of Lasso limitations in the presence of variable dependency structures and an empirical comparison of some Lasso procedures). 

Another well-known regularization function is the Ridge regularization \cite{hoerl1988ridge} where $F(\beta) = ||\beta||^2$. Ridge is efficient at accounting for variable dependencies and provides a strictly convex and differentiable optimization problem with an explicit estimator of $\beta^*$. However, this estimator is not sparse. 

The Elastic-Net regularization \cite{zou2005regularization}, defined as 
  $$ F(\beta) = (1-\alpha)|\beta|_1 + \alpha ||\beta||^2, $$
where $\alpha \in ]0,1[$, combines the sparsity property of the Lasso with the dependency grouping effects of the Ridge, making it a particularly popular smoothing function. Parameter $\alpha$ controls the trade-off between sparsity and the ability to take correlation between variables into account. 

When prior knowledge on variable dependencies is available, there exist other regularization functions, not considered here : the Adaptive Lasso \cite{zou2006adaptive}, the Group Lasso \cite{yuan2006model}, the Overlap Group Lasso \cite{jacob2009group}, the Hierarchical Group Lasso \cite{zhao2009composite}, the double sparse Lasso \cite{li2023sharp} and the fused Lasso \cite{tibshirani2005sparsity}.

\subsection{\textbf{Optimization algorithm for the regularization path construction}}
The optimization problem~(\ref{penalty}) has generally no explicit solution and requires a computational approach. As in the previous section, we consider the two most common algorithms: LARS and cyclic coordinate descent. 

A first algorithm is LARS \cite{efron2004least} providing a collection of subsets of variables, called a regularization path. Briefly speaking, the first subset contains the variable $X_j$ which has the largest absolute correlation with $Y$. The second subset contains exactly two variables : $X_j$ and the variable which is the most correlated with the residuals of the regression of $Y$ on $X_j$. Hence, each step of the LARS algorithm corresponds to a selection of one supplementary variable and is associated to a specific value of $\lambda$. 
LARS provides an exact solution of the optimization problem with nested subsets, an important property for theoretical considerations.  

A second algorithm is based on a cyclic coordinate descent procedure \cite{friedman2010regularization}, a variant of the gradient descent method. This algorithm constructs a regular grid $\Lambda$ of a given size by starting with the largest $\lambda$ corresponding to the first nonempty subset of variables. Then, a subset of variables is obtained for each $\lambda$ of this grid by solving~(\ref{penalty}) with the cyclic coordinate descent method. In contrast to LARS, the cyclic coordinate descent method provides a proxy of the optimization problem with independent solutions along the grid and possibly several subsets of variables of the same dimension. \\

From the choice of a regularization function and an algorithm, 
a collection of subsets $\left(m_\lambda\right)_{\lambda \in \Lambda}$ is obtained along the regularization path. 
Each $m_\lambda$ is associated with an estimator of $\beta^*$, solution of (\ref{penalty}). This estimator being biased \cite{meinshausen2007relaxed}, it is commonly replaced with the least-squares estimator $\Hat{\beta}_\lambda$ calculated on the subset $m_\lambda$ \cite{connault2011calibration} :
$$ \Hat{\beta}_\lambda = \underset{\{\beta, \ X \beta \in m_\lambda\}}{\arg \min} \ ||Y-X\beta||^2.$$ 
The number of non-zero coefficients of $\Hat{\beta}_\lambda$ is denoted $D_\lambda$ and corresponds to the number of variables in $m_\lambda$. \\

Once a regularization path has been computed, providing a collection $\left(m_\lambda, \Hat{\beta}_\lambda\right)_{\lambda \in \Lambda}$, the remaining question is how to obtain the final set of selected variables. In this paper, we consider two families of approaches: model selection (section \ref{model_selection}) and variable identification procedures (section \ref{variable_identification}). 

\subsection{\textbf{Model selection criteria}}
\label{model_selection}
Model selection approaches consist of minimizing on the collection $\left(m_\lambda, \Hat{\beta}_\lambda\right)_{\lambda \in \Lambda}$ a goodness-of-fit criterion penalized by the model complexity  in $\lambda \in \Lambda$ :
\begin{eqnarray}
   \gamma(m_\lambda) + \text{pen}(n,p,D_\lambda).
\label{model_select}
\end{eqnarray}
Here, a subset of variables is called a model. The loss function $\gamma(m_\lambda)$, quantifying the quality of the model fit, is either the least-squares function $||Y - X \Hat{\beta}_\lambda||^2$ or the deviance $-2 \log(L(Y,X ; \Hat{\beta}_\lambda, \Hat{\sigma}_\lambda^2))$, where $L$ is the likelihood function calculated with the empirical estimators $\Hat{\beta}_\lambda$ and $\Hat{\sigma}^2_\lambda$ on  $m_\lambda$. The penalty function pen$(n,p,D_\lambda)$ accounts for the model complexity, especially the size of the model, and the characteristics of the sample. It aims to prevent overfitting and promotes sparsity. Choosing a relevant form for pen$(n,p,D_\lambda)$ is a real challenge, still widely studied. 
 In this simulation study, we propose comparing families of model selection methods based on asymptotic and non-asymptotic performance. One of our objectives is to highlight the practical performance of non-asymptotic criteria in high-dimension, which practitioners rarely consider. \\
 
\noindent \textit{\textbf{Asymptotic criteria}.}
The model selection criterion the most widely used in practice is based on asymptotic theoretical guarantees, meaning that theoretical performance holds as the sample size $n$ tends to infinity. In this simulation study, we focus on the more recent asymptotic criterion called eBIC \cite{chen2008extended} where the loss function is the deviance and the penalty function is: 
\begin{equation*}
   \text{pen}_{\text{eBIC}}(n,p,D_\lambda) = D_\lambda \log(n) + 2\delta \log(\binom{p}{D_\lambda}),
\end{equation*}
where $\delta$ is a value in $[0,1]$. The eBIC estimator is consistent. \\
While this criterion is simple and easy to implement, it is not adapted when the number of observations is relatively small. \\

\noindent \textit{\textbf{Non-asymptotic criteria}.}
In practice, having theoretical guarantees for $n$ going to infinity makes no sense and applying criteria with properties confirmed for any fixed sample size $n$ is more relevant \cite{giraud2012high}. Introduced by \cite{birge2001gaussian}, the objective of non-asymptotic criteria is to achieve the risk oracle: $$\underset{\lambda \in \Lambda}{\inf} \mathbb{E} [\;||X \beta^*  - X \Hat{\beta}_\lambda ||^2], $$
and instead of getting asymptotic equality of the kind :
$$ \mathbb{P} \left( \underset{n \rightarrow + \infty}{\lim}
\frac{\mathbb{E} [ \; ||X \beta^* - X \Hat{\beta}_{\Hat{\lambda}}||^2] }
{\underset{\lambda \in \Lambda}{\inf} \mathbb{E} [\;||X \beta^*  - X \Hat{\beta}_\lambda ||^2]} = 1 \right) = 1, $$
they get an inequality holding for any value of $n$ :
\begin{equation*}
 \mathbb{E} [ \; ||X \beta^*  - X \Hat{\beta}_{\Hat{\lambda}}||^2 ] \leq C_n \underset{\lambda \in \Lambda}{\inf} \{ \mathbb{E} [\; ||X \beta^*  - X \Hat{\beta}_\lambda ||^2 ]\; \} + R_n,
\end{equation*}
where $C_n \approx 1$ at least for $n$ large and $R_n$ is small comparable to the risk oracle.
The selected model is the minimizer of Equation~(\ref{model_select}) where $\gamma(m_\lambda)$ is the least-squares function. In other words, the model selection procedure selects the best model from the collection in terms of prediction performance.
In this simulation study, we consider the two types of penalty functions available in the literature that do not require knowledge of the variance. 
The first type consists of data-driven penalties \cite{birge2007minimal}: 
\begin{equation}
    \text{pen}_{\mbox{Data-driven}}(n, p, D_\lambda) = 2 \kappa D_\lambda \left( 2.5 + \log\left(\frac{p}{D_\lambda}\right) \right),
    \label{penalty_EL_Review}
\end{equation}
where the constant $2.5$ has been fixed in a context of changepoint detection in a signal \cite{lebarbier2005detecting}. The constant $\kappa$ is calibrated entirely from the data. For that, two strategies exist. The first one is the slope heuristics: assuming that the least-squares function is linear in $D_\lambda \left( 2.5 + \log(\frac{p}{D_\lambda}) \right)$ as soon as $D_\lambda$ is large enough (see Figure 2 of \cite{baudry2012slope}),  the constant $\kappa$ is equal to the estimated slope. The second strategy is the dimension jump: the constant $\kappa$ is such that for all the values smaller than $\kappa$, the associated model has a very high dimension, whereas for all the values greater than $\kappa$, the associated model has a reasonable dimension (see Figure 1 of \cite{baudry2012slope}). For more practical and theoretical details, we refer the reader to \cite{baudry2012slope, arlot2019minimal}. 

The second type of penalty function is LinSelect \cite{baraud2009gaussian,giraud2012high} where variance is estimated simultaneously with model selection, meaning that the empirical estimator of the variance on each $m_\lambda$ is computed, leading to the penalty form :
\begin{equation*}
\text{pen}_{\mbox{LinSelect}}(n, p, D_\lambda) = 1.1 \times \frac{n-D_\lambda}{n-D_\lambda-1} \Psi \left( D_\lambda + 1, n-D_\lambda-1, e^{-L_\lambda} \right),
\end{equation*}
where the $L_\lambda$ is weights satisfying some properties and the function $\Psi[D,N,q]$ is the unique solution of the equation :
$$ \phi\left[D,N,\Psi(D,N,q)\right] = q, $$
where $\phi[D,N,x]$ is defined for $x \geq 0$ :
$$ \phi[D,N,x] = \frac{1}{D} \mathbb{E}\left[ \max \left(0, \chi^2_D -x \frac{\chi^2_N}{N}\right) \right], $$
for $\chi^2_D$ and $\chi^2_N$ two independent $\chi^2$ random variables with degrees of freedom $D$ and $N$ respectively. 

\subsection{\textbf{Variable identification}}
\label{variable_identification}
Model selection procedures lead to select from the model collection a final model with the best (non-)asymptotic prediction performance. As several models may have similar prediction performances in the high-dimensional framework, these approaches may provide unstable results : addition, suppression, or modification of some observations could radically change the selected subset of variables. \\
Family of variable identification approaches is built to circumvent this problem. Methods are built with perturbed datasets generated from the original sample. 
In this simulation study, we compare three types of variable identification approaches: stability-based methods, resampling-based methods and false-discovery-control methods. \\

\noindent \textit{\textbf{Stability-based methods}.}
Cross-validation \cite{allen1974relationship,stone1974cross} is the most common one. It consists of splitting the original sample $ K$ times into a training set and a test set. For each split, the training set is used to compute an estimator $\Hat{\beta}_\lambda$ and the test set is used to evaluate the mean squared error. The final subset of variables minimizes the $K$ mean squared errors in $\lambda$. But applying cross-validation in a high-dimensional context is computationally expensive and known to be unstable. An alternative is ESCV \cite{lim2016estimation} which estimates the instability along the regularization path with the $K$ perturbed datasets and selects the subset of variables by minimizing in $\lambda$  :
$$ \frac{\frac{1}{K}\underset{k=1}{\overset{K}{\sum}} ||X \Hat{\beta}_{\lambda}^k - \frac{1}{K} \underset{\ell=1}{\overset{K}{\sum}} X \Hat{\beta}_{\lambda}^\ell||^2}{||\frac{1}{K} \underset{\ell=1}{\overset{K}{\sum}} X \Hat{\beta}_{\lambda}^\ell||^2}. $$
In other words, ESCV directly selects the most stable model from the collection using $K$ perturbed datasets, making it computationally faster than cross-validation. \\

\noindent \textit{\textbf{Resampling-based methods}.}
An alternative approach is the sampling strategy. The two most widely used approaches are Bolasso \cite{bach2008bolasso} and Stability Selection \cite{ meinshausen2010stability}. They mainly differ in the dataset perturbation strategy: Bolasso generates datasets of $n$ data uniformly chosen with replacement among the original sample, whereas Stability Selection generates datasets of $\left\lfloor \frac{n}{2} \right\rfloor$ distinct data randomly chosen and considers both the obtained datasets and their complement to limit the subsampling effects. 
For both sampling strategies, a collection of variable subsets is obtained, providing an occurrence frequency for each variable. Variables with the highest occurrence frequency are retained to form the final subset. 
Lastly, we consider the Tigress method \cite{haury2012tigress}, which extends Stability Selection by considering the entire regularization path rather than only the final selected variable subset. More precisely, the occurrence frequency is a weighted average along the regularization path, with weights defined by the step number of the LARS algorithm at which the variable is selected. \\

\noindent \textit{\textbf{False-discovery-control methods}.}
The last type of variable identification methods we consider in this article is the false-discovery-control methods. These approaches differ fundamentally from the previous ones. Instead of relying on stability, they are built to limit the selection of non-active variables. 
In this article, we consider the knockoffs method \cite{barber2015controlling}. It is constructed to control the False Discovery Rate (FDR), which is the expectation of the False Discovery Proportion (FDP).  
This method starts by constructing artificial variables in a matrix $\Tilde{X}$ with the same covariance structure as $X$ and such that each copy $\Tilde{X_j}$ of $X_j$ is uncorrelated with $Y$ \cite{barber2015controlling, candes2016panning}. In other words, artificial variables mimic the dependence structure of the original variables but are independent of the response variable. 
Then, a regularization path is constructed on the augmented matrix $X \Tilde{X}$ of size $n \times 2p$ where the active variables are expected to be selected much earlier than their copies. Let denote :
\begin{equation*}
W_j = \max \left(Z_j, \Tilde{Z_j} \right) \times \text{sign} \left(Z_j - \Tilde{Z_j}\right),
\end{equation*}
where $Z_j$ and $\Tilde{Z_j}$ correspond to the largest $\lambda$ for which $X_j$ and $\Tilde{X_j}$ are selected respectively. A positive value of $W_j$ states that $X_j$ is selected before its copy $\Tilde{X}_j$ and a large positive value indicates that $X_j$ is selected rapidly. Let $q$ be the target FDR, the final subset of variables is composed of the $X_j$ such that $W_j \geq T$ with :
$$ T = \min  \left\{ t \in  \{ |W_j|, j=1,...,p\} \setminus \{0\}, \ \frac{1 + \# \{ j : W_j \leq -t\}}{\min \left(1, \# \{ j : W_j \geq t \}\right)} \leq q\right\}. $$ 

\medskip

Model selection and variable identification procedures yield a final subset of variables called the selected variables. They are built for complementary objectives, which makes comparing them interesting.  

\section{Comparison study}
\label{parameters_methods}

This section presents the simulation protocols we used to process our analyses. 
We recall that our contributions are multiple (see section \ref{Contributions}). In particular, we compare complete regularization-path-based variable selection procedures under the high-dimensional Gaussian linear model. And we propose testing robustness to model assumptions by increasing the problem dimensionality and considering non-Gaussian datasets generated with the FRANK algorithm. Moreover, there is no single metric that characterizes the performance of a variable selection procedure, so we rely on many criteria. All the details of the generated synthetic data, the method parameters and the metrics considered in this article are presented below.

\subsection{Three simulation settings}
\label{simulation_settings}
\noindent \textit{\textbf{Simulation under a Gaussian graphical model}.} The main model of the article is the Gaussian linear regression. To obtain data with Gaussian distributions and interactions described by linear links, we use the equivalence between network inference via Gaussian graphical models and support estimation in Gaussian linear regression \cite{meinshausen2006high}. An edge between nodes $i$ and $j$ in the network means either that $X_j$ is an active variable when $X_i$ is the response variable, or that $X_i$ is an active variable when $X_j$ is the response variable. 
We generate a dataset of size $n$ from a $(p+1)$ multivariate centered Gaussian distribution with covariance matrix $\Sigma$ where the dependency structure is encoded in the precision matrix $\Sigma^{-1}$ \cite{lauritzen1996, cordoba2019generating}. 
The response variable $Y$ is chosen as a column of the $(p+1)$ multivariate centered Gaussian and the remaining columns constitute the matrix $X$ of size $n \times p$. It differs from \cite{zou2008regularized,wang2012quantile,peng2015iterative,wang2020high}, where the response variable is simulated once the matrix $X$ is fixed. 
We consider two graph patterns :
\begin{itemize}
    \item \textit{cluster}: the precision matrix is simulated as a block diagonal matrix with $B$ blocks of equal size. The response variable $Y$ is defined as the first variable.  
    \item \textit{scale-free} : 
    a few variables have many neighbors, while the others have few. We consider two response variables corresponding to the variables having the highest and the smallest number of neighbors. These simulation designs are called \textit{scale-free-max} and \textit{scale-free-min} respectively. \\
\end{itemize}

\noindent \textit{\textbf{Simulation under \textit{independent} design}.}
The independent design serves as a benchmark since most theoretical guarantees are established under weak dependency assumptions and it is the simplest setting in which the high-dimensional framework is the single handicap \cite{fan2014adaptive,wang2020high}.
The matrix $X$ is simulated by the concatenation of $p$ independent standard Gaussian vectors of size $n$.
The number of non-zero coefficients of the vector $\beta^*$ is generated from a uniform variable on integers between $10$ and $15$. Their values are generated from $\mathcal{U}([0.5 , 2])$ and the response variable $Y$ is defined by $Y = X  \beta^* + \varepsilon, \ \text{where} \ \varepsilon \sim \mathcal{N}(0, I_n)$. \\

\noindent \textit{\textbf{Simulation under a dynamical process}.}
To test the behavior of methods in a non-Gaussian data distribution, 
we use the algorithm \textit{FRANK} \cite{carre2017reverse} which simulates large networks with characteristics of gene regulatory networks. This type of data is very attractive in biology. Indeed, understanding gene regulation is a real challenge in molecular biology, and the emergence of high-throughput technologies, such as microarrays and RNA sequencing, has enabled measuring the activities of thousands of genes simultaneously and performing genome-scale inference of transcriptional gene regulation. Several studies have already shown that 
inferring gene networks is a difficult task and no single inference method performs optimally \cite{Marbach2012}. Regression methods are usually included in such studies, but only with Lasso. In this algorithm, variables are categorized into transcription factors that activate or inhibit target genes and the FRANK data are generated by a dynamic process that deviates from the statistical model assumptions, especially the Gaussian distribution. We use FRANK with only transcription factor variables to compare results with those from the other settings. We consider the variables with the highest and lowest numbers of neighbors as response variables. These simulation designs are called \textit{FRANK-max} and \textit{FRANK-min} respectively. \\

The design of the simulation study is intended to examine the behavior of the methods with respect to the dependency structures among the variables. Compared to the \textit{independent} setting, the \textit{cluster} setting evaluates the impact of the dependency structure (with the number of active variables similar; see later). Using the \textit{scale-free} design, we investigate the method's behavior as a function of the number of active variables. The data distribution in FRANK is far from Gaussian, allowing testing the behavior of methods on non-Gaussian data.  \\

 For all settings, we set $n=150$ and $p=199$. We generate $100$ samples of size $2n$ to create a training set of size $n$ for the estimation and a test set of size $n$ to evaluate the methods. Before use, each sample's variables are centered and scaled. To generate data from a Gaussian graphical model, we use the function \textit{huge.generator} from the R package \textit{huge} (version 1.3.4.1). For the \textit{cluster} design, the block number $B$ is equal to $5$ and the probability of connection within a component is set to the default value $0.3$. For the FRANK algorithm, we use the online version available on the website \url{https://m2sb.org/?page=FRANK} with 
 $p$ transcription factors and $2n$ observations. The number of eigenvalues of the matrix on the unit circle is fixed to $2$ and the minimum and maximum of sparsity are set to $1$ and $50$. Other parameters are set to default values and  $40$ samples are generated. 
 To test the impact of high dimensionality on the performance of the methods, we increase $n$ to $ 300$, $ 600$ and $1200$ for Gaussian and independent data. The observations generated for $n=150$ are included in the datasets of size $n=\{300, 600, 1200\}$. 
 
\subsection{Investigated methods and their parameters}
A total of $16$ combinations using model selection methods are defined by the choice of a regularization function (Lasso or Elastic-net), an algorithm (LARS or the gradient descent algorithm) and a penalty function (eBIC, LinSelect or the $2$ data-driven penalties). 
A total of $17$ combinations using variable identification methods are defined : 
when the sampling strategy is performed before the definition of the grids, $8$ combinations are defined by the choice of a regularization function (Lasso or Elastic-net), an algorithm (LARS or the gradient descent) and a sampling strategy (Bolasso or Stability Selection). When the grid $\Lambda$ is fixed, the sampling strategy is performed for each $\lambda$ of the grid, which implies using the gradient descent algorithm only; hence, $4$ combinations are defined by the choice of a regularization function (Lasso or Elastic-net) and a sampling strategy (Bolasso or Stability Selection). Furthermore, we include Tigress using LARS, the knockoffs method and ESCV with a gradient descent algorithm and either Lasso or Elastic-net.

For the LARS algorithm, we use the function \textit{enet} of the R package \textit{elasticnet} (version 1.1.1) and the maximal number of steps to define the grid size is the default value $50 \times \min(p, n-1)$. For the gradient descent method, we use the function \textit{glmnet} of the R package \textit{glmnet} (version 3.0) and set the grid size at $1000$. The functions \textit{enet} and \textit{glmnet} implement Lasso and elastic-net regularization, respectively. We set $\alpha = 0.5$ for Elastic-Net. 

To perform model selection, eBIC is implemented with $\delta=1$. LinSelect is implemented in the function \textit{tuneLasso} of the R package \textit{LinSelect} (version 1.1.3). The data-driven penalties are calculated using the \textit{capushe} function in the \textit{capushe} R package (version 1.1.1). All the parameters are set to their default values, except for the minimum percentage of points for plateau selection, which we set to $0.1$. 

To perform variable identification, the function \textit{escv.glmnet} of the R package \textit{HDCI} (version 1.0.2) is used for the ESCV strategy, with a number of groups $K$ fixed to $10$. Bolasso and Stability Selection are implemented with $100$ samples. 
A variable is selected when its occurrence frequency is higher than $0.8$. For Tigress, we use the \textit{tigress} function from the \textit{tigress} R package (version 0.1.0), with the number of steps in LARS fixed at $50$. For the knockoffs method, we use the function \textit{knockoff.filter} with option $\textit{\text{create.second\_order}}$ of the R package \textit{knockoff} (version 0.3.2), we calculate the $W_j$'s with the function $\textit{\text{stat.lasso\_lambdasmax}}$ and we set the FDR to $0.1$.

\subsection{Evaluation metrics}
To evaluate the quality of a regularization path, we use the partial area under the receiver operating characteristic curve (pROC-AUC). The x-axis of the curve is the proportion of non-active variables in each subset of the collection among the non-active variables and the y-axis is the proportion of active variables in each subset of the collection among the active variables. Since the length of the regularization paths differs according to the choice of regularization function and algorithm, all the curves are truncated at the x-axis corresponding to the largest common x-value to all the regularization paths, so they can be fairly compared. The pROC-AUC ranges from $0$ to $1$, with $1$ indicating that the active variables are distinguished from the others. 

To evaluate the quality of a subset of variables, we consider the mean squared error (MSE), recall, specificity, and false discovery proportion (FDP). The MSE evaluates the predictive performance of the methods using a test set. First, the methods are applied on a training set $\left((Y_i,X_i) \in \mathbb{R} \times \mathbb{R}^p\right)_{\{1 \leq i \leq n\}}$, then the MSE is calculated on a test set $\left((\Tilde{Y}_i,\Tilde{X}_i) \in \mathbb{R} \times \mathbb{R}^p\right)_{\{1 \leq i \leq n\}}$ independent of  $\left((Y_i,X_i) \in \mathbb{R} \times \mathbb{R}^p\right)_{\{1 \leq i \leq n\}}$, and is equal to :
\begin{equation*}
\frac{1}{n} \underset{i=1}{\overset{n}{\sum}} \left( \Tilde{Y}_i - (\Tilde{X} \Hat{\beta}_{\Hat{\lambda}})_i \right)^2,
\end{equation*}
where $\Hat{\beta}_{\Hat{\lambda}}$ is the estimator of $\beta^*$ calculated on  $\left((Y_i,X_i) \in \mathbb{R} \times \mathbb{R}^p\right)_{\{1 \leq i \leq n\}}$.  
As data are centered and scaled, a MSE value ($\in \mathbb{R}_+$) lower than $1$ means that the selected variables predict $Y$ better than the empty set. 
The recall is the proportion of selected active variables among the active variables, the specificity is the proportion of non-active variables not selected among the non-active variables and the FDP is the proportion of selected non-active variables among the selected variables across the samples for which at least one variable is selected. By construction, recall and specificity cannot be separated in the analyses and they jointly quantify the ability to limit the selection of non-active variables while selecting as many active variables as possible. 
In many application domains, it is very important to limit the selection of non-active variables and controlling the FDP is preferred. This metric evaluates the quality of the selected subset of variables by ensuring control of the percentage of selected non-active variables among the selected variables. Recall, specificity and FDP values belong to $[0,1]$. Recall and specificity are expected to be close to $1$ while the FDP is expected to be low. 

\section{Results}
\label{results}
Section \ref{classic} is dedicated to the results obtained in the high-dimensional Gaussian linear regression. As a reminder, $n$ and $p$ are fixed at $n=150$ and $p=199$, respectively, and the \textit{independent} setting is used as a benchmark in the analyses. Each setting has been run $100$ times. Over these runs, the mean of the sparsity parameter is $12.59$, $11.63$, $31.41$, and $1$ for respectively the \textit{independent}, \textit{cluster}, \textit{scale-free-max} and \textit{scale-free-min} design with standard deviations to $1.76$, $2.75$, $9.70$ and $0$ respectively. So, the sparsity assumption is verified. 
In Section \ref{impact_high_dim}, we discuss how the sample size $n$ impacts the method's performance using datasets of sizes $n=300, 600$ and $1200$. In Section \ref{frank_section}, we investigate the impact of the non-Gaussian assumption through the FRANK datasets. \\
For the sequel, the notations \textit{GD} and \textit{E-Net} denote respectively the cyclic coordinate descent algorithm and the Elastic-Net regularization function and  \textit{grid} and \textit{sub} denote the strategies for generating the grids or the samples first. We discuss the median of the evaluation metrics obtained with the $100$ simulated samples of each setting.

\subsection{Method performances for Gaussian data in a high dimensional context}
\label{classic}

\subsubsection{Discrimination of the active variables from the others with pROC-AUC (regularization paths)}
\label{AUC_section}
\begin{figure}[!htb]
    \centering
    \includegraphics[width=1\linewidth,scale=0.75]{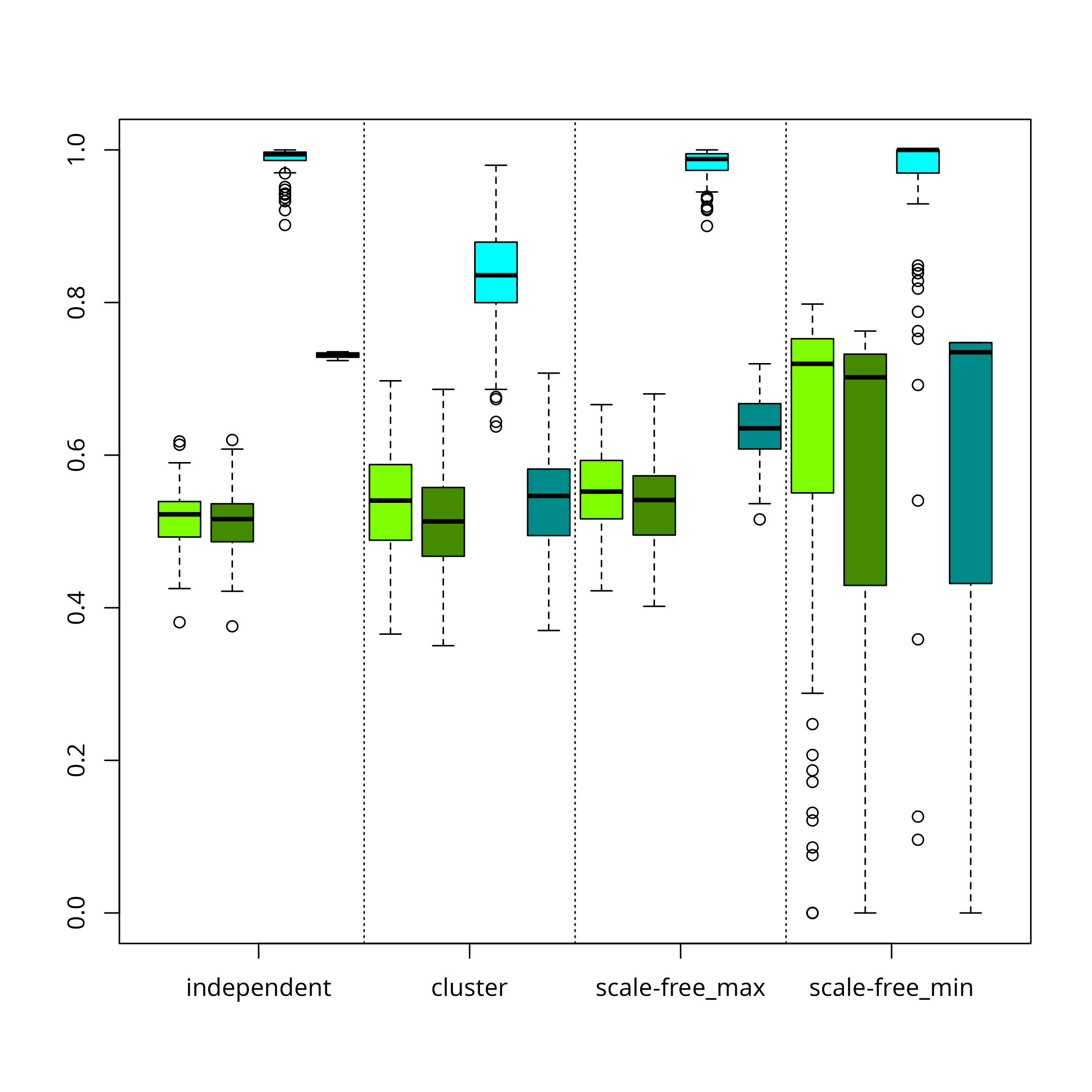}
\caption{Boxplots of the pROC-AUC values calculated on 100 samples of size $n=150$.
The  cyclic coordinate descent algorithm combined with \textit{E-Net} is colored light green, the  cyclic coordinate descent algorithm combined with $\ell_1$ regularization is colored dark green, LARS combined with \textit{E-Net} is colored cyan and LARS combined with $\ell_1$ regularization function is colored dark cyan.}
\label{auc}
\end{figure}

Figure~\ref{auc} summarizes the pROC-AUC values of the different regularization paths for the \textit{independent}, \textit{cluster}, \textit{scale-free-max} and \textit{scale-free-min} designs. 

For the \textit{independent} design, values obtained from LARS  combined with \textit{E-Net} are around $0.99$, whereas the use of Lasso with LARS provides a median value around $0.73$. In general, the GD algorithm discriminates less well between active and non-active variables. \\
For the other three settings, the highest median values are always obtained with LARS combined with the \textit{E-Net} regularization:  $0.84$ and $0.98$ for \textit{cluster} and \textit{scale-free-max} respectively and $1$ for \textit{scale-free-min}. LARS combined with Lasso does not differ from the GD algorithm and provides values not exceeding $0.73$.  \\
In conclusion, LARS combined with \textit{E-Net} is always the best combination for discriminating active variables from the others. This result is consistent with the grouping effect of the Elastic-Net penalty.  \\

The number of methods compared in this study is large. Therefore, for the sequel, we represent, only results obtained with the best combination (LARS with \textit{E-Net} regularization), for the model selection methods, Bolasso and Stability Selection when samples are first generated ($sub$). For ESCV and the knockoffs method, which are based on the cyclic coordinate descent algorithm, we show the MSE obtained with \textit{E-Net}, as this regularization method appears to be slightly better than Lasso. For Tigress, no choice is required, since the method is implemented only with LARS and Lasso. 

\subsubsection{Prediction performances with MSE (subsets of selected variables)}
\label{MSE_section}

\begin{figure}[!htb]
    \centering
    \includegraphics[width=1\linewidth,scale=0.5]{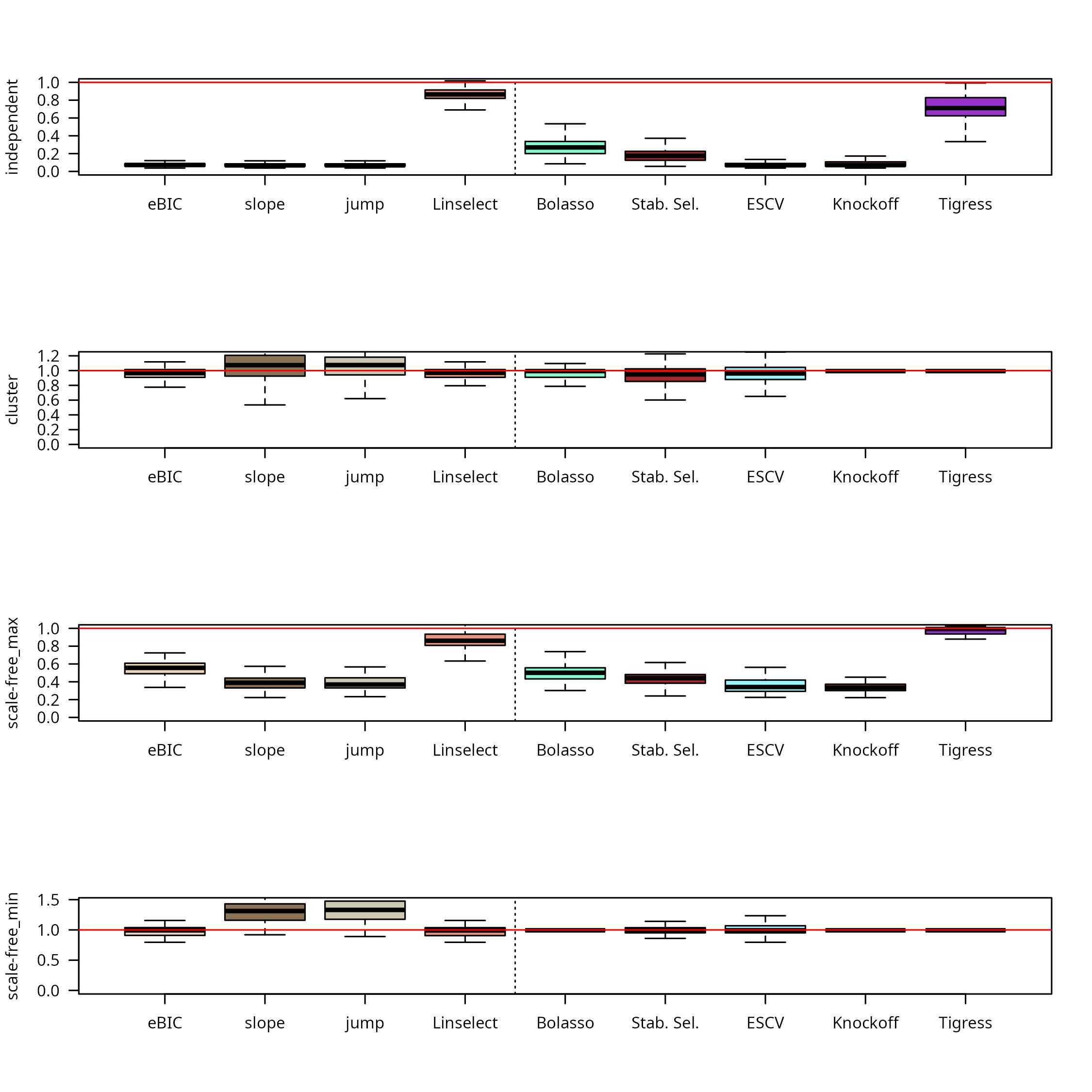}
\caption{Boxplots of the MSE calculated on 100 samples of size $n=150$ in the four settings. The red line 1, the value below which the methods have a prediction ability.
Results are presented with LARS with \textit{E-Net} regularization for the model selection methods, Bolasso and Stability Selection. For ESCV and the knockoffs method which are based on the cyclic coordinate descent algorithm, the MSE is shown with \textit{E-Net}. Tigress is implemented with LARS and Lasso. }
\label{mse_150}
\end{figure}

We measure the performance of prediction by the MSE shown in Figures \ref{mse_150}, \ref{MSE1} and \ref{MSE2}. 

For the \textit{independent} setting, all procedures remain predictive, a MSE lower than $1$, although substantial differences appear between model selection and variable identification approaches. For the model selection strategies, the median value is around $0.07$ for eBIC and the data-driven penalties but very close to $1$ ($0.96$) for LinSelect. For the variable identification methods, the smallest median values are around $0.07$ for ESCV and the knockoffs method but Tigress has a median value ten times higher. Regarding the Bolasso and Stability Selection, the median MSE values are $0.27$ and $0.17$, respectively. \\
When a dependency structure exists, the MSE of all methods increases significantly. For model selection, the median values of eBIC and LinSelect are greater than $0.92$ for the \textit{cluster}. The data-driven penalties, the penalties eBIC and LinSelect achieve $0.37$, $0.56$ and $0.87$ respectively for the \textit{scale-free-max}. The MSE values for the data-driven penalties are larger than $1$ on the \textit{cluster} and \textit{scale-free-min} meaning that they are not predictive. 
The variable identification methods are not better. The median values are very close to $1$ for the \textit{cluster} and \textit{scale-free-min} designs. For the \textit{scale-free-max} design, the median value is around $0.35$ for ESCV and the knockoffs method. The values for Bolasso and Stability Selection are $0.5$ and $0.44$ respectively. Tigress values are very close to $1$. \\
In summary, when the variables are independent, the best methods are eBIC, the data-driven penalties, ESCV and the knockoffs method with a small MSE. When a dependency structure exists, predictive performance is significantly deteriorated. The data-driven penalties provide some values strictly larger than $1$. All the values are very close to $1$ for the \textit{cluster} and \textit{scale-free-min} designs. For \textit{scale-free-max} design, the best methods are eBIC, ESCV and the knockoffs method.

\subsubsection{Discrimination of the active variables from the others with recall and specificity (subsets of selected variables)}
\label{Recall_section}

\begin{figure}[!htb]
    \centering
\includegraphics[width=1\linewidth,scale=0.5]{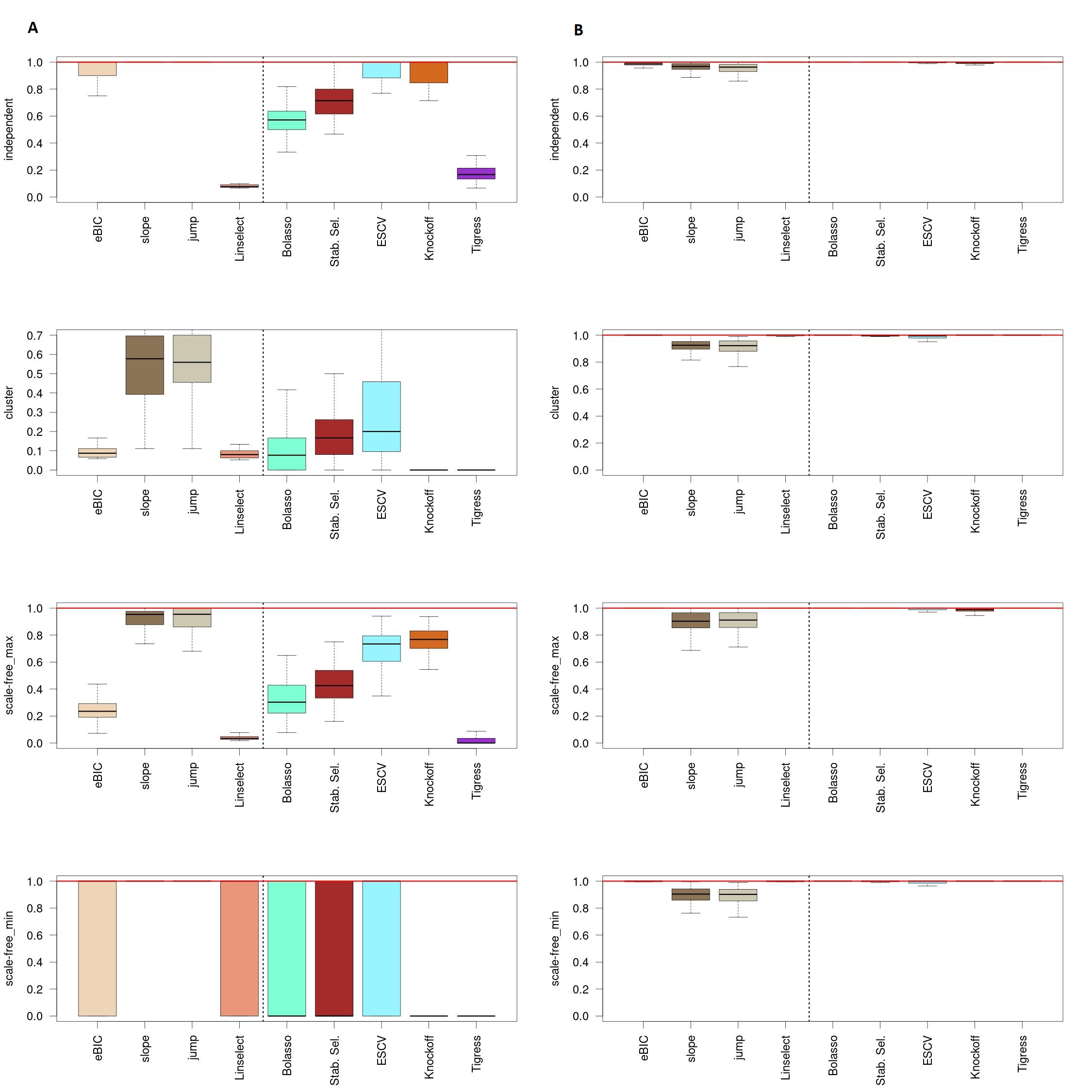}
\caption{Column A : Boxplots of the recall calculated on 100 samples of size $n=150$ in the four settings. The red line at 1 indicates the ability to recover all the active variables. Column B : Boxplots of the specificity  calculated on the same 100 samples of size $n=150$ in the four settings. The red line at 1 indicates the ability to not select all the non-active variables.
 Results are presented with LARS with \textit{E-Net} regularization for the model selection methods, Bolasso and Stability Selection. For these latter, the sampling strategy is $sub$. For ESCV and the knockoffs method which are based on the cyclic coordinate descent algorithm, the recall and the specificity are showed with \textit{E-Net}. Tigress is implemented with LARS and Lasso.  }
\label{recall-150}
\end{figure}

We measure the ability to limit the selection of non-active variables while selecting as many active variables as possible through the simultaneous study of the recall and the specificity. Results are shown in Figures \ref{recall-150} (column A for recall, column B for specificity), \ref{recall1}, \ref{recall2}, \ref{specificity1} and \ref{specificity2}. 

For the \textit{independent} setting, all the model selection methods except LinSelect (with a recall very close to $0$) select all the active variables. Regarding specificity, only the data-driven penalties show a specificity different from $1$, indicating that some non-active variables are selected for these penalties, which might be explained by the size of the estimated support being larger than the number of active variables (see Figure \ref{support1}). 
Among the variable identification methods, ESCV and the knockoffs method also select all the active variables. Bolasso and Stability Selection have median values of $0.57$ and $0.71$ respectively. Finally, Tigress has a significantly smaller median value (around $0.18$). Based on Figures~\ref{recall1} and~\ref{recall2}, we observe that the choice of the regularization function for Bolasso and Stability Selection is very important since with Lasso, the recall equals $0$, meaning that no active variables are selected. 
Regarding specificity, all the variable identification methods achieve a value of $1$, probably because they tend to be conservative (see Figure \ref{support2}). \\
When a dependency structure exists, conclusions about specificity are exactly the same, but differences are observed in recall. Among the model selection methods, data-driven penalties achieve the best recall across the three settings. However, in Figures~\ref{recall1}, we observe that their results vary depending on the choice of the regularization function and the algorithm. Among the variable identification methods, the results are highly dependent on the setting. For the \textit{cluster} design, the best median value is obtained by ESCV and Stability Selection. However, the median value remains low (around $0.2$) with a large variability. For the \textit{scale-free-max} design, the knockoffs method and ESCV provide the best results ($0.75$). The other methods are clearly worse. For the \textit{scale-free-min} design, only ESCV selects the active variables. \\
In summary, to control a relevant trade-off between recall and specificity, the data-driven penalties are discarded due to their specificity. 
When variables are independent, eBIC, ESCV and the knockoffs method achieve the best trade-off with a specificity value close to $1$. When a dependency structure exists, the recall values decrease significantly, and the best methods are ESCV and the knockoff methods. 

\subsubsection{Quality of the selected variables with FDP (subsets of selected variables)}
\label{FDR}
\begin{figure}[!htb]
    \centering
    \includegraphics[width=1\linewidth,scale=0.5]{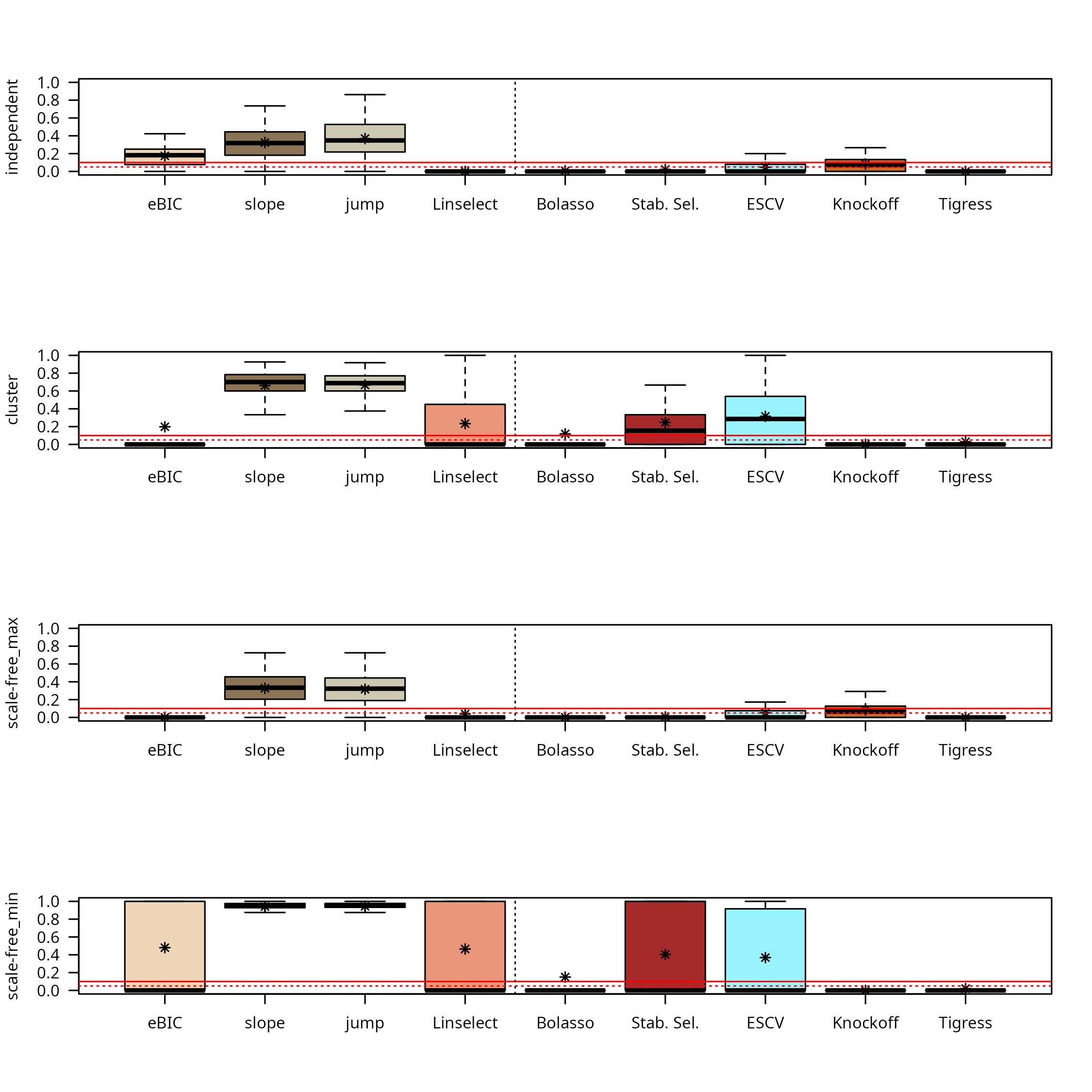}
\caption{Boxplots of the FDP calculated on 100 samples of size $n=150$ in the four settings. The star indicates the expected threshold of the FDR (expectation of the FDP). The red line indicates $0.1$, the threshold for the knockoffs method. The dashed red line indicates $0.05$. Results are presented with LARS with \textit{E-Net} regularization for the model selection methods, Bolasso and Stability Selection. For these latter, the sampling strategy is $sub$. For ESCV and the knockoffs method which are based on the cyclic coordinate descent algorithm, the FDP is showed with \textit{E-Net}. Tigress is implemented with LARS  and Lasso.}
\label{fdp-150}
\end{figure}

To limit the selection of non active variables, we calculate the FDP. Figures \ref{fdp-150}, \ref{fdp1} and \ref{fdp2} show the results for the four settings. 

In the \textit{independent} setting, among the model selection methods, eBIC and data-driven penalties have median FDP values above $0.1$, indicating that they select too many variables (see Figure \ref{support1}). Conversely, LinSelect has a median FDP value equal to $0$, and as it selects variables (a few but not none (see Figure \ref{support1})), it means that it selects variables that are always active. For the variable identification methods, which tend to select a few variables (see Figure \ref{support2}), the FDP is very often lower than $0.05$, except for the knockoffs method, which has an FDP of $0.08$, close to the threshold given as input to the method. \\
Unsurprisingly, with a dependency structure, the FDP values increase. For model selection methods, eBIC and LinSelect have median FDP values below $0.05$ in the \textit{scale-free-max} setting and about $0.2$ in the \textit{cluster} setting. In contrast, they are higher than $0.4$ for the \textit{scale-free-min} setting. For eBIC, we observe in Figure~\ref{fdp1} that LARS combined with Lasso yields an FDP below $0.1$. The FDP of the data-driven penalties is always too large. Among the variable identification methods, for the \textit{scale-free-max} setting, the FDP value is always lower than $0.1$. For the other two settings, only Bolasso, the knockoffs method and Tigress still have a value of FDP below $0.1$ with values very close to $0$ for the knockoffs method and Tigress meaning that these two methods are very conservative. \\
In summary, when variables are independent, only LinSelect and all variable identification methods yield an FDP below $0.1$. When variables are dependent, LinSelect, Bolasso, Tigress and the knockoffs method should be preferred. The data-driven penalties, Stability Selection, and ESCV yield values larger than $0.1$. \\

\subsubsection{Practical guidelines}
To facilitate the practical use of our benchmark, we propose Table \ref{Table_resume} as a take home message for complete regularization-path-based variable selection procedures under the high-dimensional Gaussian linear model. 
It summarizes the main recommendations which are valid whatever the dependency structure between variables. This stands for a global conclusion of Section \ref{classic}. 

\begin{table}[!htb]
\tabcolsep=9pt
\centering
\begin{tabular}{||>{\RaggedRight}p{1.7cm}||>{\RaggedRight}p{2.4cm}||>{\RaggedRight}p{2.5cm}||>{\RaggedRight}p{2.5cm}||>{\RaggedRight}p{2.5cm}||}
  \hline
   Metric : & \textbf{pROC-AUC} & \textbf{MSE} & \textbf{Recall and Specificity} & \textbf{FDP} \\
   \hline 
   \hline
    Best \newline methods : & E-net + LARS & eBIC, ESCV, \newline the knockoffs method & ESCV, \newline the knockoffs methods & LinSelect, \newline Tigress, \newline Bolasso, \newline the knockoffs methods \\
    \hline
    Methods \newline to avoid : & & data-driven penalties, \newline LinSelect, \newline Tigress  & LinSelect, data-driven penalties, eBIC, Tigress & data-driven penalties, ESCV, \newline Stability \newline Selection \\
  \hline
\end{tabular}
\caption{Recommendations for the high-dimensional linear Gaussian regression}
\label{Table_resume}
\end{table}

Throughout the study, we observe that data-driven penalties perform poorly across all evaluated metrics. It may be due to both the shape of the penalty and the fixed constant value $2.5$ in Equation~(\ref{penalty_EL_Review}), which are previously fixed in the detection of changepoints in a signal \cite{lebarbier2005detecting}. These quantities may not be well-suited for a high-dimensional Gaussian linear regression. Alternative methodologies will be investigated in future work, particularly to make the procedure more conservative (Figure \ref{support1}).

\subsection{Impact of the sample size on the method performances}
\label{impact_high_dim}
We evaluate how the number of observations $n$ 
affects the performance of the methods considered in this article, keeping in mind that if the metric improves as the sample size increases, then the behaviors observed in the previous section are actually due to the high dimension. 

\subsubsection{Discrimination of the active variables from the others with pROC-AUC (regularization paths)}

\begin{figure}[!htb]
    \centering
    \includegraphics[width=1\linewidth,scale=0.75]{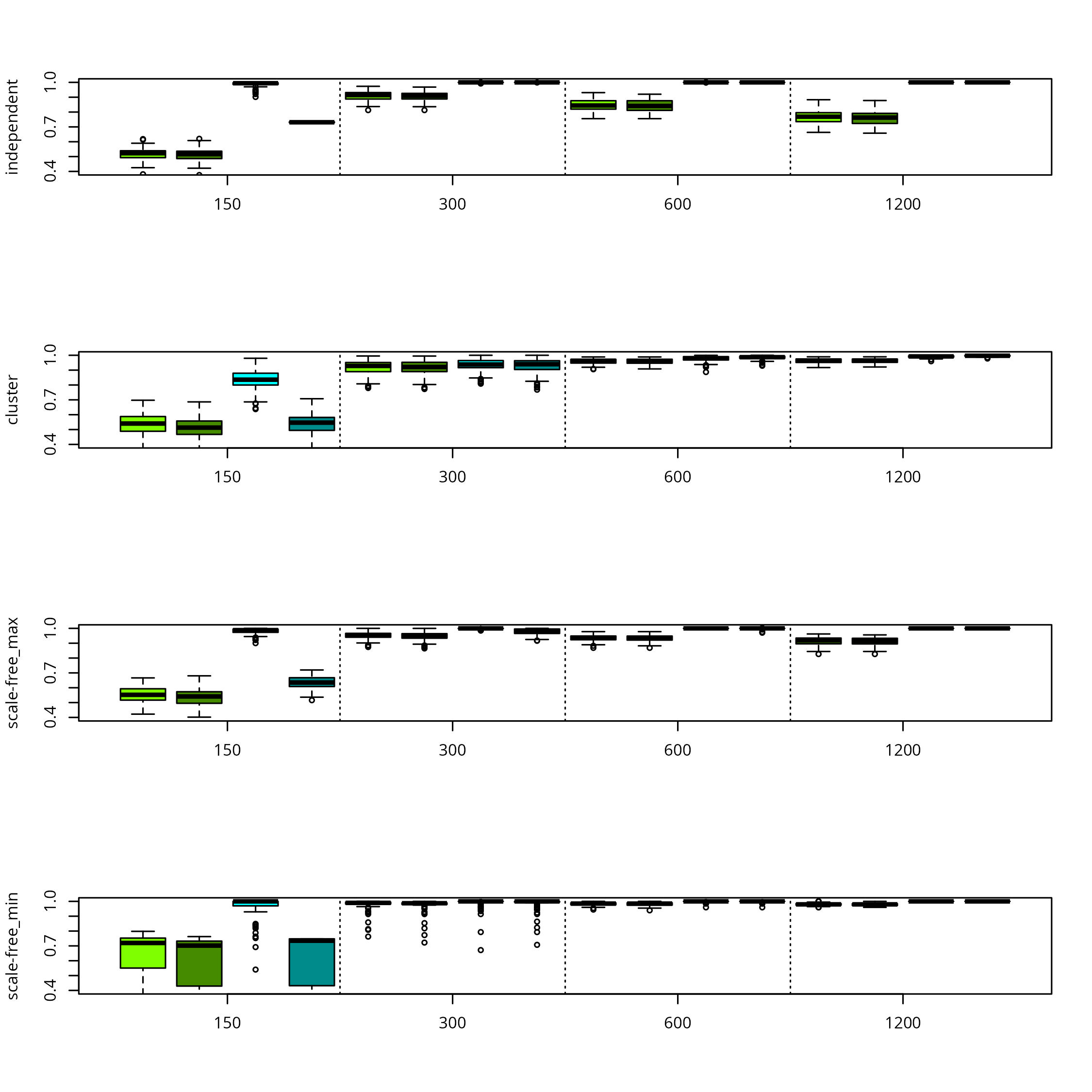}   
\caption{Boxplots of the pROC-AUC values calculated on 100 samples of size  $n=\{150, 300, 600, 1200\}$. The  cyclic coordinate descent algorithm combined with \textit{E-Net} is colored light green, the  cyclic coordinate descent algorithm combined with Lasso is colored dark green, LARS combined with \textit{E-Net} is colored cyan and LARS combined with Lasso is colored dark cyan.}
\label{auc-evolution}
\end{figure}

The evolution of the pROC-AUC is shown in Figure \ref{auc-evolution}. 

The behavior of each combination with respect to $n$ cannot be compared, since the length of the regularization path can differ across sample sizes. However, with a fixed sample size, we can compare the four combinations. On Figure \ref{auc-evolution}, we observe that LARS combined with \textit{E-Net} remains the best in all the settings, although the differences between the four combinations decrease as $n$ increases. 

\subsubsection{Prediction performances with MSE (subsets of selected variables)}

The evolution of the MSE is shown in Figure \ref{mse_evol} and in Figures Suppl 9 to Suppl 16 in the supplementary material \cite{lacroix:hal-04998945}.

\begin{figure}[!htb]
    \centering
    \includegraphics[width=1\linewidth,scale=0.75]{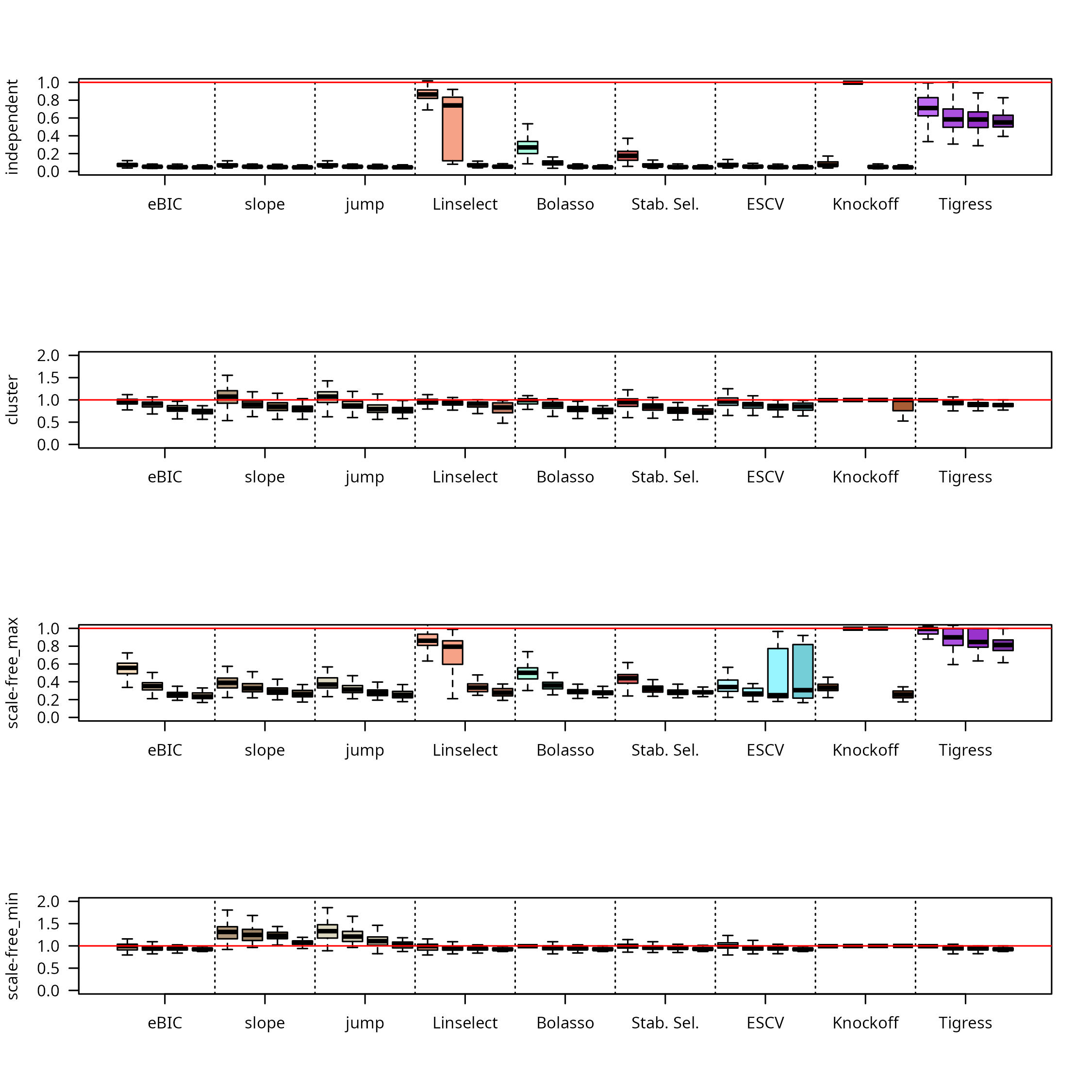}
\caption{Boxplots of the MSE calculated on 100 samples in the four settings. Within a method, the colour gradates from lightest to darkest depending on the size of the dataset to represent $n=150$ (the lightest, left-part), $ 300, 600$ and $n=1200$ (the darkest, right-part). The red line 1, the value below which the methods have a prediction ability.
Results are presented with LARS with \textit{E-Net} regularization for the model selection methods, Bolasso and Stability Selection. For ESCV and the knockoffs method which are based on the cyclic coordinate descent algorithm, the results are showed with \textit{E-Net}. Tigress is implemented with LARS  and the  $\ell_1$ regularization. }
\label{mse_evol}
\end{figure}

For the \textit{independent} design, the MSEs of eBIC and the data-driven penalties are very low, regardless of $n$, and decrease with $n$ until $0.05$. The MSE of LinSelect decreases significantly with $n$ and reaches $0.07$ with LARS as soon as $n=600$. For the variable identification methods, the MSE of Bolasso and Stability Selection decreases with $n$, except when the  $grid$ sampling strategy is used. For ESCV and Tigress, the MSE decreases gradually with $n$, but the results differ: ESCV always has a median MSE below $0.2$, while Tigress always has a median MSE above $0.6$. the knockoffs method keeps a MSE close to $1$ for $n=300$ but value drops to $0.1$ as $n=600$. \\
When there is a dependency, we observe that all the model selection methods have a MSE decreasing with $n$ but the median values remain higher than $0.5$ except for \textit{scale-free-max} or close to $1$ (often). Among the variable identification methods, only Bolasso and Stability Selection combined with LARS and \textit{E-Net} yield a decreasing MSE with $n$ up to $0.2$. 
Tigress gets MSE between $1$ and $0.8$. ESCV has an averaged MSE that increases slightly from $0.36$ to $0.52$ between $n=300$ and $n=1200$, with a significant increase in inter-sample variability, leading to an observable difference in the median value between the two regularization functions. The knockoffs method has an MSE comparable to that when $n=150$ when $ n=1200$, but between these two sample sizes, the MSE surprisingly increases to $0.85$ or more. \\
To summarize, the MSE of most methods decreases as the number of observations increases, but surprisingly increases with $n$ for some variable identification methods, except for Bolasso and Stability Selection, when a dependency structure exists. Moreover, the dependency structure again complicates the prediction task, as MSE values remain high across most methods, suggesting that this is not an effect of the high dimension.

\subsubsection{Discrimination of the active variables from the others with recall and specificity (subsets of selected variables)}

The evolution of the ability to limit the selection of non-active variables while selecting as many active variables as possible is shown in Figure \ref{recall-evol} (column A for recall, column B for
specificity) and in Figures Suppl 17 to Suppl 32 in the supplementary material \cite{lacroix:hal-04998945}.

\begin{figure}[!htb]
    \centering
    \includegraphics[width=1\linewidth,scale=0.75]{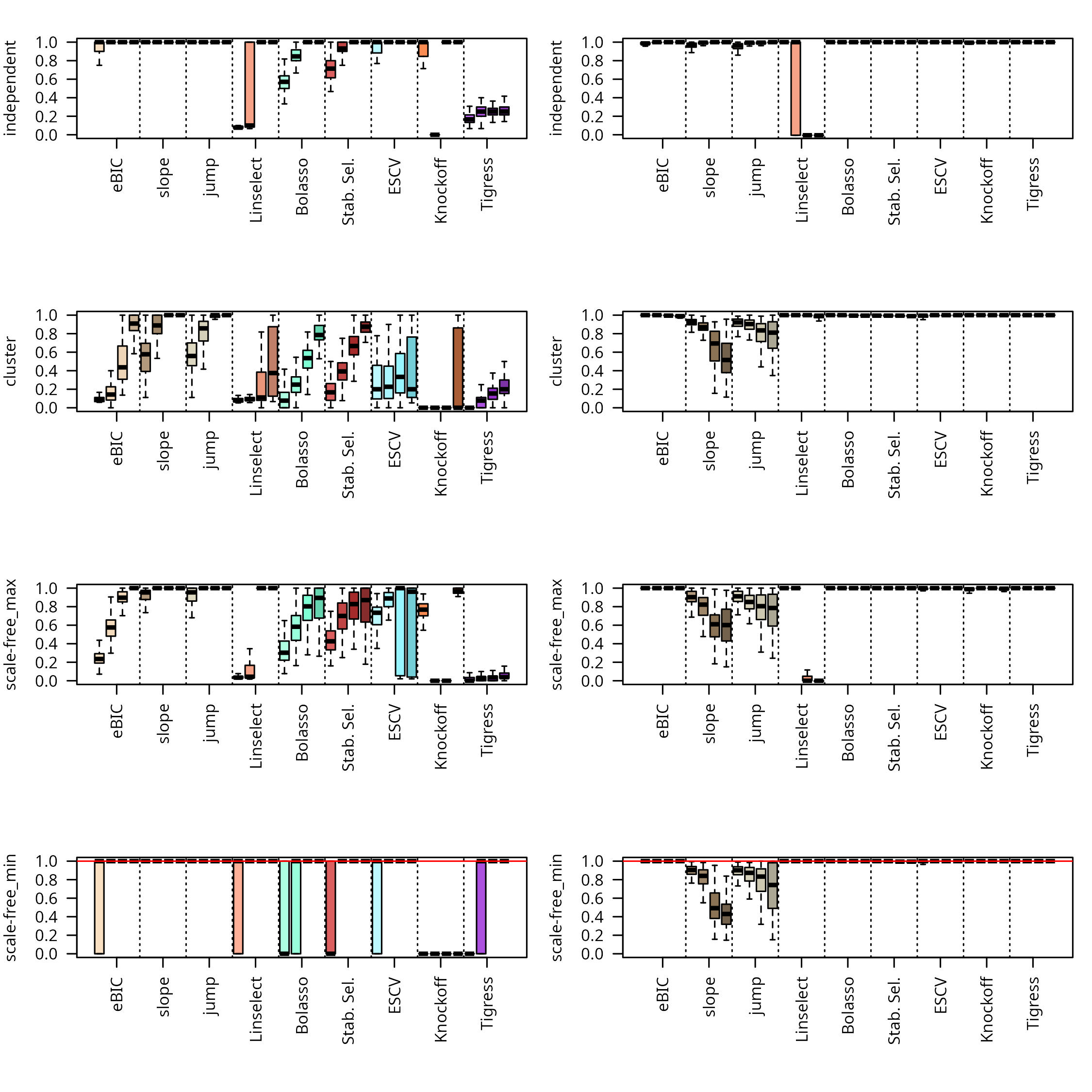}
\caption{Column A : Boxplots of the recall calculated on 100 samples in the four settings. The red line at 1 indicates the ability to recover all the active variables. Column B : Boxplots of the specificity calculated on 100 samples in the four settings. The red line at 1 indicates the ability to not select all the non-active variables. 
Within a method, the colour gradates from lightest to darkest depending on the size of the dataset to represent $n=150$ (the lightest, left-part), $ 300, 600$ and $n=1200$ (the darkest, right-part). Results are presented with LARS with \textit{E-Net} regularization for the model selection methods, Bolasso and Stability Selection. For these latter, the sampling strategy is $sub$. For ESCV and the knockoffs method which are based on the cyclic coordinate descent algorithm, the recall and the specificity are showed with \textit{E-Net}. Tigress is implemented with LARS and Lasso.  }
\label{recall-evol}
\end{figure}

For the \textit{independent} design, among the model selection methods, only LinSelect does not have a recall of $1$ for $n=150$, but as soon as $n=600$, it does. For the variable identification methods, the recall of Bolasso and Stability Selection increases with $n$, except when the $grid$ strategy is used. The recall of ESCV is $1$ when $n=150$ or more. The recall of Tigress increases gradually with $n$, but it remains low with a median value of $0.25$. This observation is consistent with 
Figures Suppl 1 to Suppl 8 in the supplementary material \cite{lacroix:hal-04998945} where we observe that the size of the selected subset of variables increases with $n$ but remains small with large $n$ for Tigress. 
The knockoffs method again shows unexpected behavior: the recall drops to $0.2$ when $n=300$, but equals $1$ when $n=600$. 
Regarding specificity, only the data-driven penalties do not have specificity equal to $1$ for $n=150$, but as soon as $n=300$, they do. As the sample size increases, the specificity of LinSelect decreases, likely probably because smaller subsets of selected variables are used. For the variable identification methods, the specificity equals $1$ for all the sample sizes, consistent with the size of the selected variables subset remaining smaller than the size of the active variables (see Figures Suppl 1 to Suppl 8 in the supplementary material \cite{lacroix:hal-04998945}). \\
For the three dependency structures, all the model selection methods have a recall increasing with $n$. The recall reaches $1$ for all the methods as soon as $n=300$ for the \textit{scale-free-min} setting. For the \textit{scale-free-max} and \textit{cluster} settings, the recall equals $1$ as soon as $n=600$, except for LinSelect, which seems sensitive to the dependency structure, since its recall is below $0.5$ in the \textit{cluster} setting. In contrast, it reaches $1$ in the \textit{scale-free-max} setting. 
For the variable identification methods, the results also vary between the three dependency structures. In the \textit{scale-free-min} setting, all the methods retrieve the active variable as soon as $n=300$, except the knockoffs method, which selects the empty set whatever the value of $n$. When many variables are active, Bolasso and Stability Selection have a recall that increases with $n$, and Stability Selection seems slightly better. Tigress also shows increasing recall, but it remains conservative, so the values remain low and differ between the \textit{cluster} and \textit{scale-free-max} settings. ESCV for \textit{cluster} shows a slight increase in recall up to $0.5$, but the inter-sample variability is large, leading to a median value lower for $n=1200$ than for $n=600$. For \textit{scale-free-max}, its recall increases with $n$ and reaches $1$ for $n=1200$ when \textit{E-Net} is used. 
Concerning the knockoffs method, it usually selects no variables in the \textit{cluster} setting, whereas it reaches $0.86$ when $n=1200$ in the \textit{scale-free-max} setting, which is better than $0.75$ obtained when $n=150$. However, between these two sample sizes, the recall decreases to $0$ and $0.2$ for $n=300$ and  $n=600$, respectively. Concerning the specificity, values decrease for the data-driven penalties. 
As soon as $n=600$, eBIC has decreasing specificity for the \textit{cluster} design, probably because the median size of the estimated support is smaller than the true support. LinSelect in \textit{scale-free-max}, Bolasso and Stability Selection in \textit{cluster} and \textit{scale-free-min} and the knockoffs method in \textit{scale-free-max} get specificity decreasing with $n$ (see Figures Suppl 1 to Suppl 8 in the supplementary material \cite{lacroix:hal-04998945}). \\
To summarize, as $n$ increases, we observe that the dependency structure significantly affects the values of recall and specificity. All methods except the knockoffs method have increasing recall and the model selection methods require fewer observations than the variable identification methods to achieve a recall equal to $ 1 $. Concerning specificity, values decrease with $n$ when a dependency structure exists. 
Hence, the selection tends to be more focused on the active variables and the discrimination between active and non-active variables is easier when $n$ is large enough.

\subsubsection{Quality of the selected variables with FDP (subsets of selected variables)}
To limit the selection of non active variables, we calculate the FDP. Figure \ref{fdp-evol} and Figures Suppl 33 to Suppl 40 in the supplementary material \cite{lacroix:hal-04998945} show the results for the four settings. 

\begin{figure}[!htb]
    \centering
    \includegraphics[width=1\linewidth,scale=0.75]{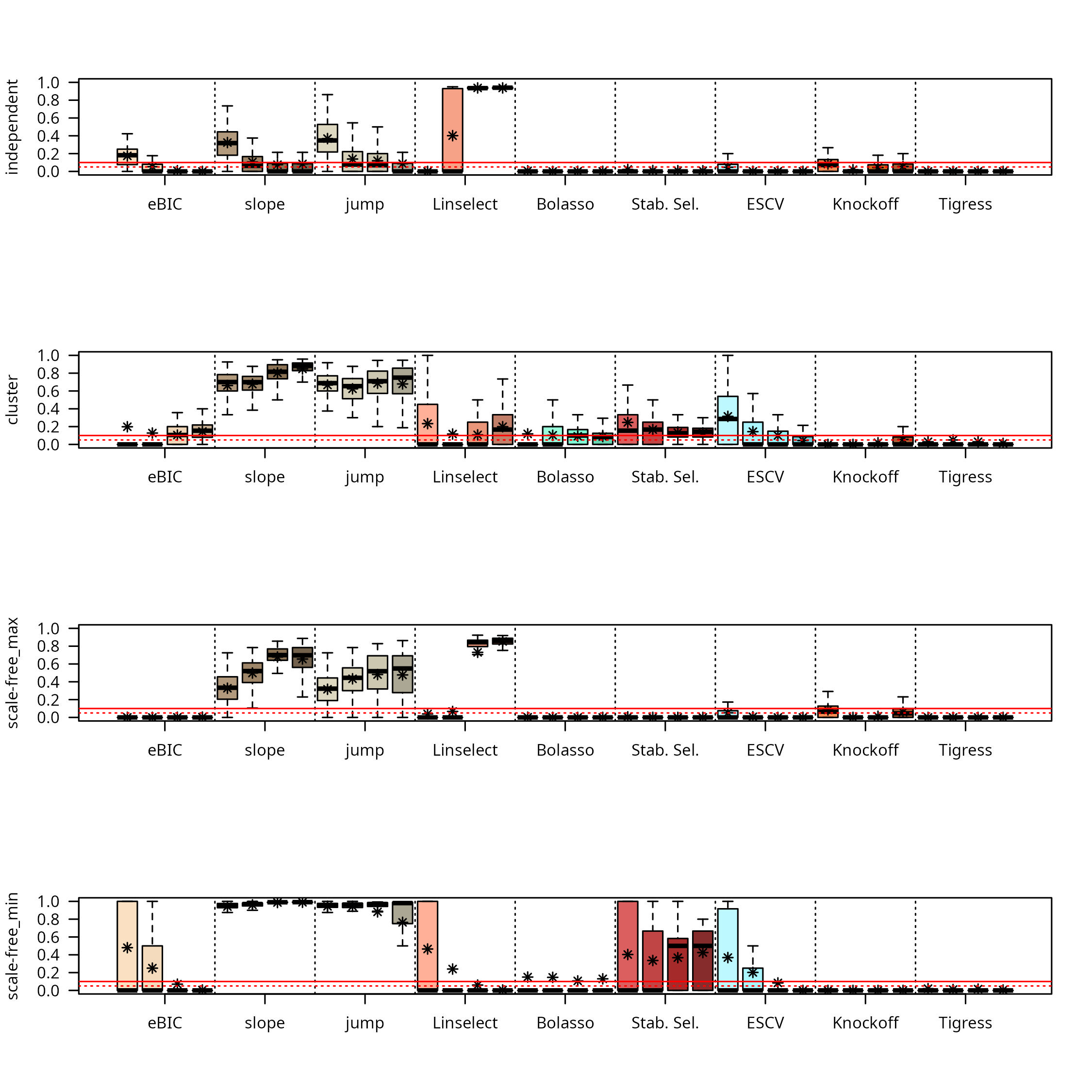}
\caption{Boxplots of the FDP calculated on 100 samples in the four settings. Within a method, the colour gradates from lightest to darkest depending on the size of the dataset to represent $n=150$ (the lightest, left-part), $ 300, 600$ and $n=1200$ (the darkest, right-part).
The star indicates the expected threshold of the FDR (expectation of the FDP). The red line indicates $0.1$, the threshold for the knockoffs method. The dashed red line indicates $0.05$.  Results are presented with LARS algorithm with \textit{E-Net} regularization for the model selection methods, Bolasso and Stability Selection. For these latter, the sampling strategy is $sub$. For ESCV and the knockoffs method which are based on the cyclic coordinate descent algorithm, the FDP is showed with \textit{E-Net}. Tigress is implemented with LARS  and the  $\ell_1$ regularization.}
\label{fdp-evol}
\end{figure}

In the \textit{independent} setting, we observe that eBIC and the data-driven penalties achieve a median FDP below $0.1$ as soon as $n=300$. The FDP of LinSelect increases with $n$ but is always larger than $0.1$. For all the variable identification methods, FDP is lower than $0.1$. \\
A dependency structure significantly increases the FDP values of the data-driven penalties across all sample sizes. The FDP of eBIC varies across settings: for \textit{scale-free-max}, it is always below $0.01$; for \textit{cluster} and \textit{scale-free-min}, it decreases with $n$ and reaches $0.1$ at $n=600$. For LinSelect, the behavior of the FDP also varies with the dependency structure: in the \textit{cluster} and  \textit{scale-free-min} settings, the value of FDP decreases with $n$ and is lower than $0.1$ as $n=300$ and $n=600$, respectively; for the \textit{scale-free-max}, it is lower than $0.1$ when $n=300$ and then increases strongly to $0.7$. 
For all variable identification methods, the FDP in the \textit{scale-free-max} setting is below $0.1$. For the \textit{cluster} setting, the FDP of Bolasso and Stability Selection decreases with $n$ and is lower than a$0.1$ for almost all the combinations of algorithm, regularization function, and sampling strategy. ESCV has a decreasing FDP as $n$ increases. With Lasso, the FDP median value of ESCV is lower than $0.1$. Tigress has a very low FDP because the number of selected variables is very small. The knockoffs method has an estimated value always lower than $0.1$. For the \textit{scale-free-min} setting, the FDP of Bolasso and Stability Selection decrease with $n$ and is lower than $0.1$ for some combinations of algorithm, regularization function and sampling strategy, but it is difficult to identify the best combination. ESCV maintains a FDP close to $0.1$ with Lasso. Tigress has a very low FDP because the number of selected variables is very small. The knockoffs method has an estimated value always lower than $0.1$. \\
To summarize, for most cases, the FDP values decrease when $n$ increases. This observation is consistent with the recall and specificity results. All the methods have a FDP value lower than $0.1$ when $n \geq 300$ in the \textit{independent} design. Again, the dependency structure still deteriorates the FDP with values that can exceed a reasonable threshold, even when $n$ is large and especially for model selection methods. 

\subsection{Behavior of the methods in a non-Gaussian framework}
\label{frank_section}

We evaluate how the non-Gaussian distribution affects the performance of the methods considered in this article, keeping in mind that if non-Gaussian data do not degrade the metric, then the data distribution does not actually explain the behaviors observed in the previous sections. To do that, we use the FRANK datasets described in subsection \ref{simulation_settings}.
We also study the non-paranormal transformation $shrinkage$ from the R package $huge$, known to help relax the normality assumption.

\subsubsection{Discrimination of the active variables from the others with pROC-AUC (regularization paths)}
The pROC-AUC values are shown in Figure \ref{franck_auc}. 

Similarly with scenarios from independent and Gaussian models, the combination of the \textit{E-Net} regularization with the LARS achieves the highest value of pROC-AUC. However, these values do not exceed $0.5$ for FRANK-max and $0.57$ for FRANK-min. All other combinations have values smaller than $0.3$. We do not observe improvement by using the non-paranormal transformation of the data. Hence, the quality of the regularization paths has clearly deteriorated on FRANK data. No combination of an algorithm and a regularization function clearly distinguishes active from non-active variables.   

\subsubsection{Prediction performances with MSE (subsets of selected variables)}
The MSE values are shown in Figure \ref{franck_mse_150} and in Figures Suppl 43 to Suppl 44 in the supplementary material \cite{lacroix:hal-04998945}.

For FRANK-max and FRANK-min, all methods yield an MSE value either very close to $1$ or strictly larger than $1$, indicating they are not predictive. Hence, the Gaussian assumption is important to achieve good predictive performance. The best methods remain eBIC, LinSelect and ESCV.  

\subsubsection{Discrimination of the active variables from the others with recall and specificity (subsets of selected variables)}
The recall and specificity values are shown in Figure \ref{franck_recall-150} and in Figures Suppl 45 to Suppl 48 in the supplementary material \cite{lacroix:hal-04998945}.

The recall is significantly deteriorated on the FRANK data. Its median value is often $0$, except for ESCV and the data-driven penalties for the FRANK-max design, which remain low around $0.05$. Regarding specificity, the median values are larger than $0.98$ for all methods except ESCV, which has a value of $0.96$, and the data-driven penalties, which decrease to $0.7$. We do not observe any improvement by using the non-paranormal transformation of the data. 
Hence, the recall is significantly lower with FRANK data than with Gaussian data, while specificity remains high. This observation suggests that the selection of any variable is difficult and methods tend to provide a very small set of selected variables (see Figure \ref{support_FRANK} and Figures Suppl 41 and Suppl 42 in the supplementary material \cite{lacroix:hal-04998945}).

\subsubsection{Quality of the selected variables with FDP (subsets of selected variables)}
The FDP values are shown in Figure \ref{franck_fdp-150} and in Figures Suppl 49 to Suppl 50 in the supplementary material \cite{lacroix:hal-04998945}.

For the FRANK-max design, the FDP median values are always greater than $0.88$ across all methods. For the FRANK-min design, all model selection methods achieve the value $1$. In contrast, all the variable identification methods achieve the value $0$ except for ESCV, Bolasso and Stability Selection with LARS, \textit{E-Net} and when samples are first generated where values are larger than $0.7$. 
Hence, the FDP values have drastically deteriorated in the FRANK data. When the methods select variables, these variables are mostly non-active variables meaning the methods fail to select the active variables. The exception is the Tigress and the knockoffs method, which are very conservative but seem to be able to select an active variable when the subset is not empty.

\section{Conclusions}
\label{conclusions}
This paper provides a comprehensive comparison of complete regularization-path-based variable selection procedures in high-dimensional Gaussian linear regression. 
By combining different regularization functions, optimization algorithms, and final variable selection strategies (across model selection or variable identification families), we provide detailed guidance for practitioners on selecting appropriate methods based on metrics that control for and are robust to the (usually unknown) dependency structure of the variables.
We also highlight methods with non-asymptotic theoretical guaranties, which practitioners rarely use despite their appealing finite-sample properties. 
We further explore the robustness of these methods by relaxing key model assumptions, such as the Gaussian distribution and high-dimensional setting.

Based on our large-scale comparison study, we recommend the LARS algorithm combined with the Elastic-Net regularization function (which benefits from its ability to handle grouping variable effects) for constructing the regularization path. 
No procedure dominates all performance criteria. We recommend eBIC, ESCV, and the knockoffs method for prediction performance; ESCV and the knockoffs method to control a relevant trade-off between recall and specificity; and LinSelect, Tigress, Bolasso, and the knockoffs method for the lowest FDP values. 
Bolasso exhibits intermediate behavior across metrics, achieves high performance with very few non-active selected variables, and selects variable subset sizes close to the active variable numbers.  
Tigress and LinSelect appear as conservative methods, often returning empty sets of variables. 
As already noted by \cite{Lacroix-Martin-2024}, all performance metrics deteriorate in the presence of correlation between variables. 
Finally, we highlight the importance of modeling choice, especially when dealing with real data that often violates the statistical assumptions required by many models. In particular, we show that high dimensionality and a non-Gaussian framework negatively affect the performance of any procedure, in agreement with previous results by \cite{clarte2024analysis} and \cite{sanchez2018comparison}. 
Even when data transformations are incorporated into the procedure, no clear improvement is observed in the metrics, underscoring the importance of pre-processing steps when the Gaussian assumption is not satisfied (see \cite{liu2012high}). These results corroborate previous studies on the difficulties of using high-dimensional Gaussian regression on transcriptomic data \cite{carre2017reverse, Marbach2012}, and we show that these difficulties arise primarily from the construction of the regularization path. It may be worth investigating this in the future. 

This paper contributes to the literature on variable selection in high-dimensional contexts. First, it shows that non-asymptotic methods outperform asymptotic methods. As future work, a meticulous investigation into calibrating shape and constants in data-driven penalties is required. Second, it provides a comprehensive comparison of procedures that jointly combine regularization path constructions and final variable subset selections. Third, it includes both model selection and variable identification methods and reveals that none of the model selection procedures appears to be the best for prediction performance.
Future work could extend this simulation study by incorporating additional evaluation criteria, such as the stability and sensitivity of the variable selection procedures, as in \cite{heinze2018variable}, or more sophisticated metrics, as in the recent work \cite{tang2023nonparametric}. 

\section{Additional information}
The R scripts used to reproduce all simulations are available on \url{https://forgemia.inra.fr/GNet/high-dimensional_regression_comparison}, as well as the supplementary material \cite{lacroix:hal-04998945}. 

\section*{Funding}
IPS2 benefits from the support of the LabEx Saclay Plant Sciences-SPS  (ANR-17-EUR-0007).  This work was supported by a public grant as part of the Investissement d'avenir project, reference ANR-11-LABX-0056-LMH, LabEx LMH.
We are grateful to the INRAE MIGALE bioinformatics facility (MIGALE, INRAE, 2020. Migale Bioinformatics Facility, doi: 10.15454/1.5572390655343293E12) for providing help and/or computing and/or storage resources.

\bibliographystyle{unsrt}  
\bibliography{reference}  

\newpage
\appendix
\renewcommand\thefigure{\thesection A\arabic{figure}}    
\setcounter{figure}{0}    
\section*{Appendixes}
\label{appendix_Review}

For the sequel, the data-driven penalty procedures are named \textit{slope} and \textit{jump} with the slope heuristics and the dimension jump strategies respectively.
Bolasso and Stability Selection are named \textit{bol} and \textit{ss} respectively.

\begin{figure}[h!]
\begin{subfigure}{0.49\linewidth}
    \centering
    \includegraphics[width=1\linewidth]{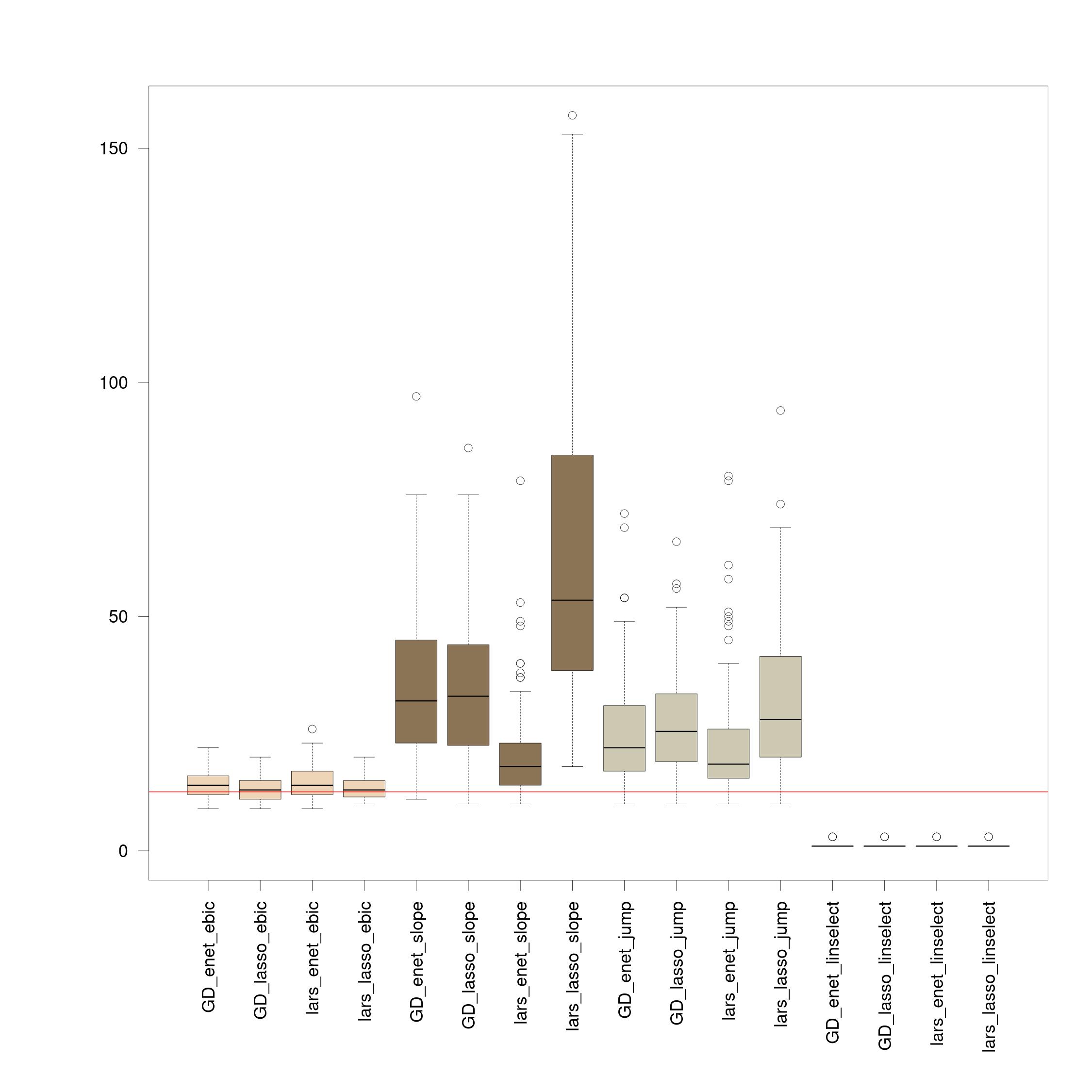}
    \caption*{\textit{independent}}
\end{subfigure}
\begin{subfigure}{0.49\linewidth}
    \centering
    \includegraphics[width=1\linewidth]{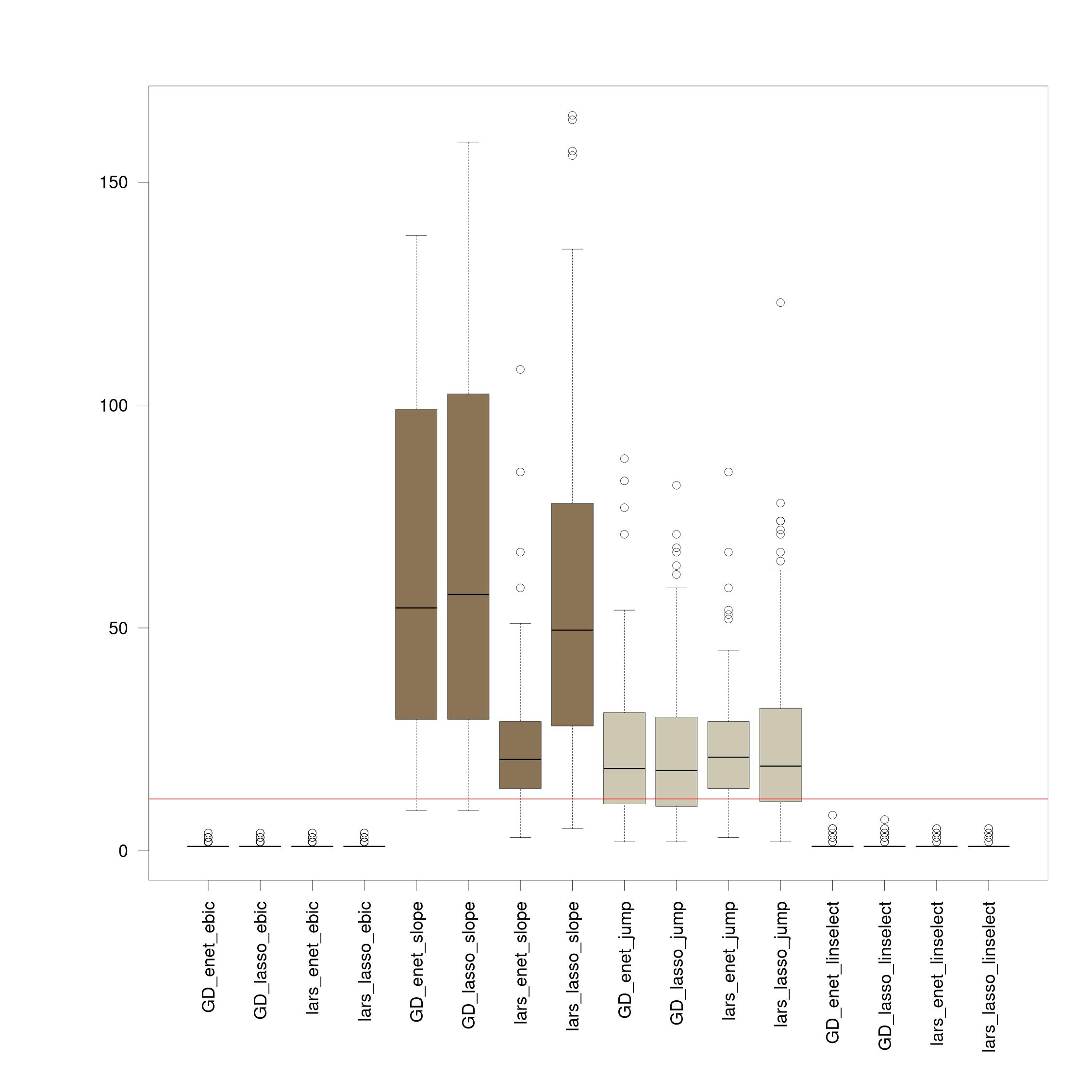}
    \caption*{\textit{cluster}}
\end{subfigure}
\begin{subfigure}{0.49\linewidth}
    \centering
    \includegraphics[width=1\linewidth]{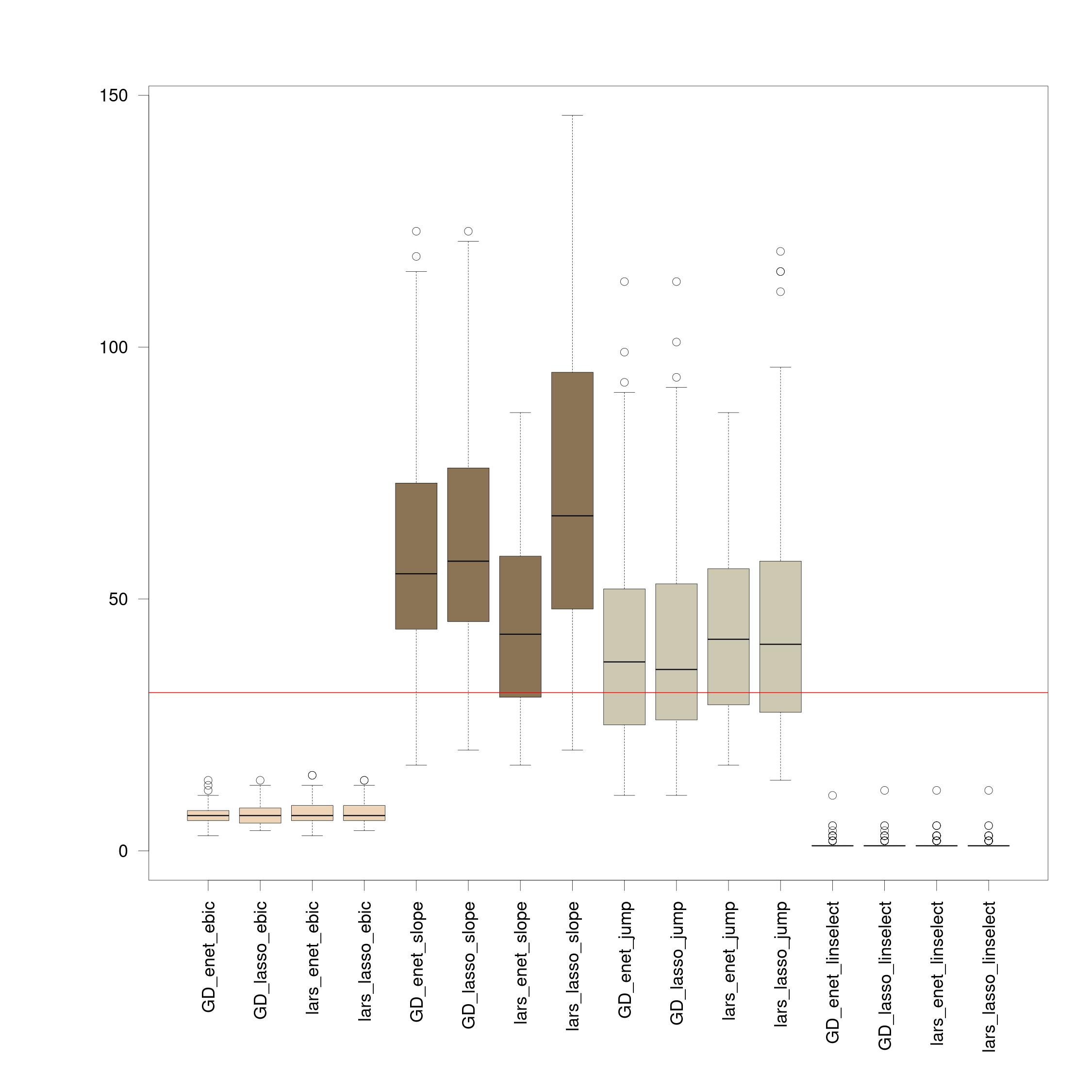}
    \caption*{\textit{scale-free-max}}
\end{subfigure}
\begin{subfigure}{0.49\linewidth}
    \centering
    \includegraphics[width=1\linewidth]{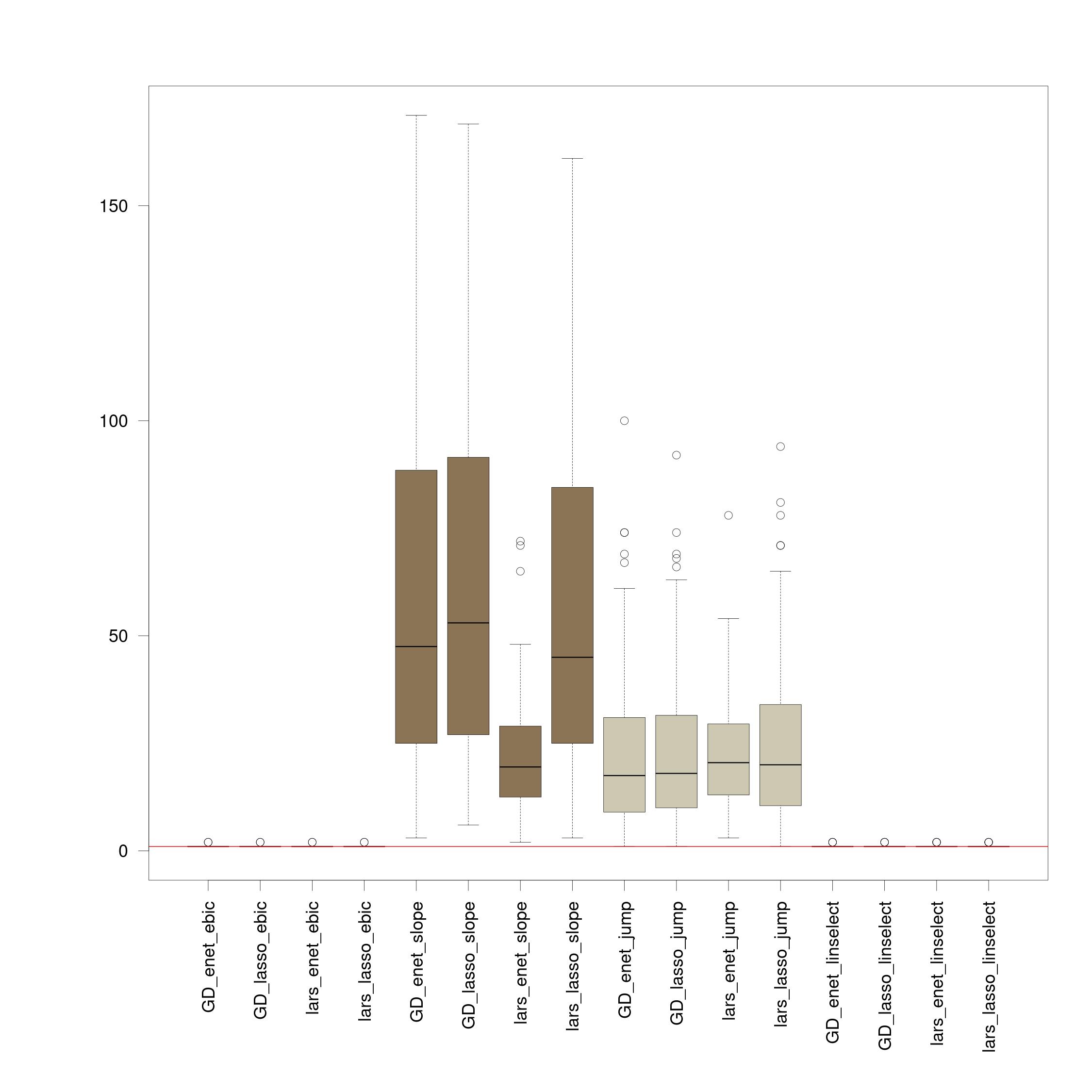}
    \caption*{\textit{scale-free-min}}
\end{subfigure}
\caption{Boxplots of the size of the selected variables subset obtained by the model selection procedures from dataset of size $n=150$ in the four different settings. The red line indicates the true number of  active variables. }
\label{support1}
\end{figure}

\begin{figure}[h!]
\begin{subfigure}{0.49\linewidth}
    \centering
    \includegraphics[width=1\linewidth]{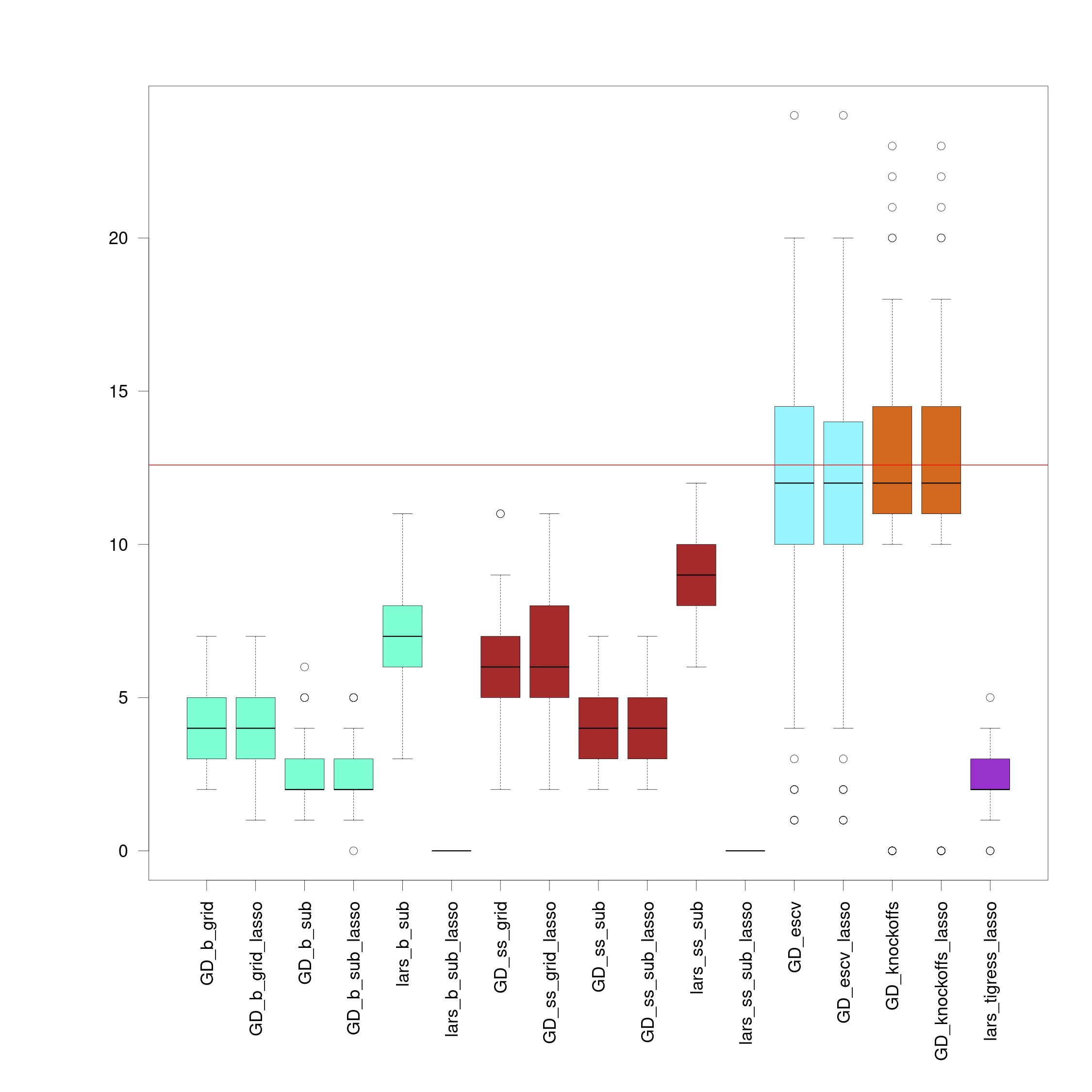}
    \caption*{\textit{independent}}
\end{subfigure}
\begin{subfigure}{0.49\linewidth}
    \centering
    \includegraphics[width=1\linewidth]{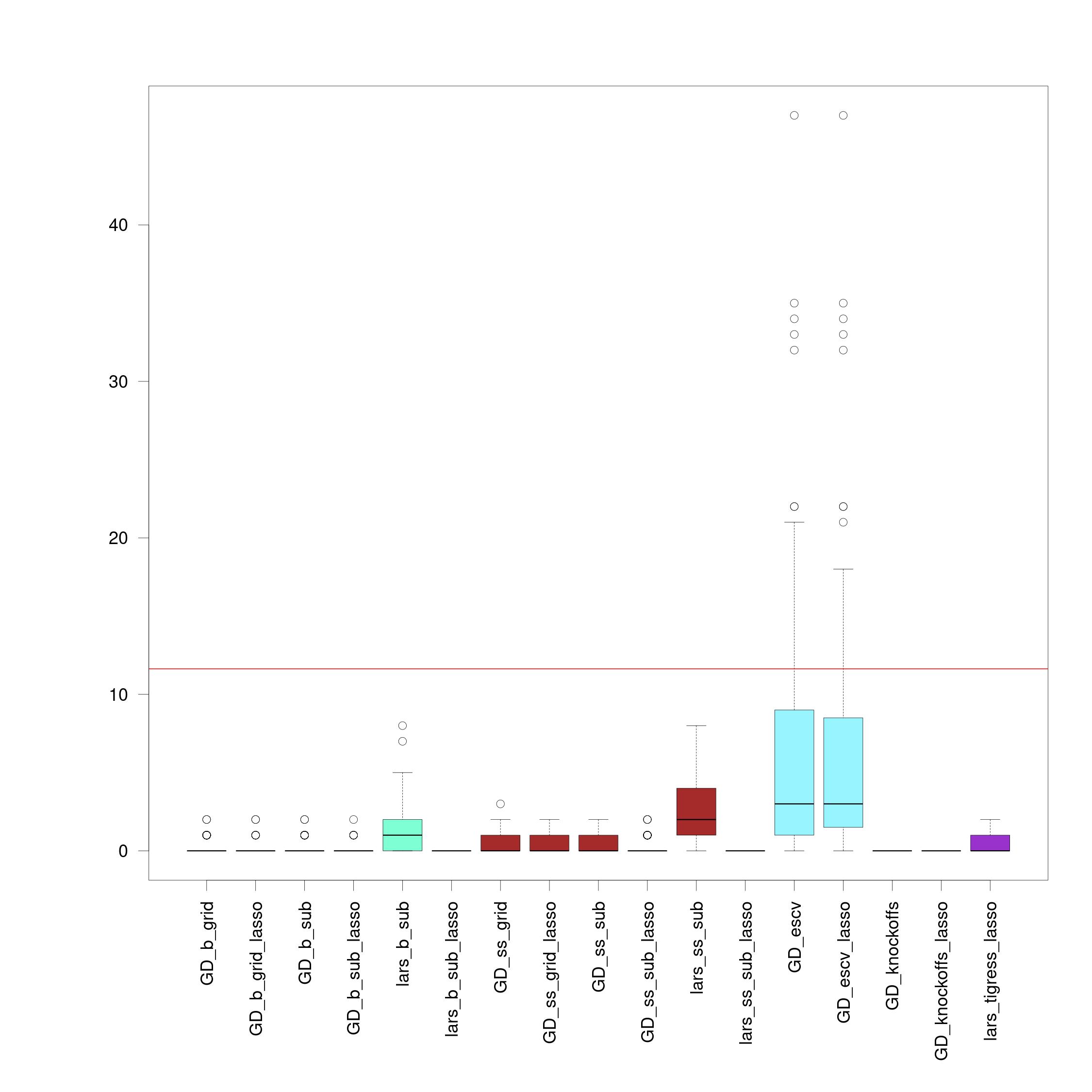}
    \caption*{\textit{cluster}}
\end{subfigure}
\begin{subfigure}{0.49\linewidth}
    \centering
    \includegraphics[width=1\linewidth]{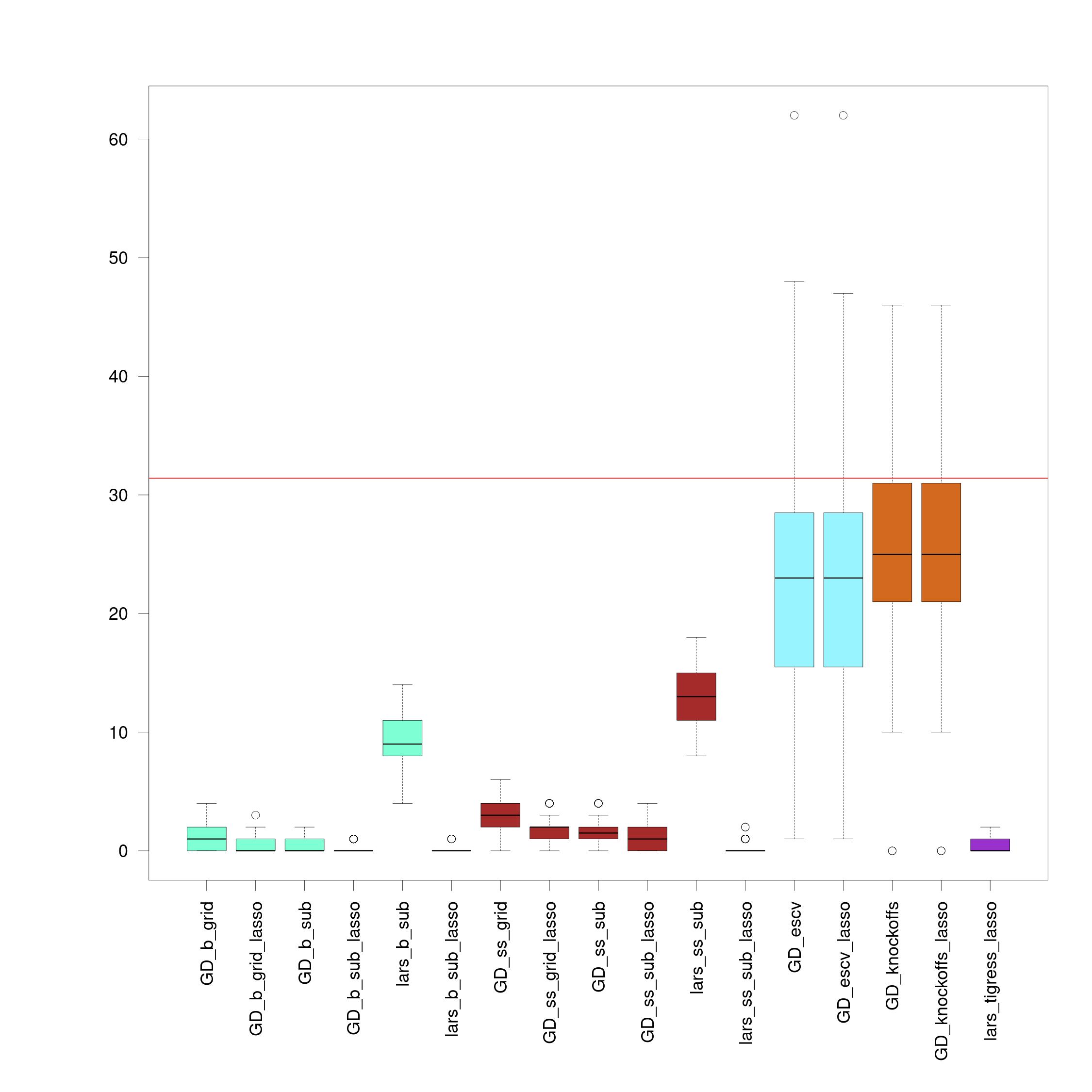}
    \caption*{\textit{scale-free-max}}
\end{subfigure}
\begin{subfigure}{0.49\linewidth}
    \centering
    \includegraphics[width=1\linewidth]{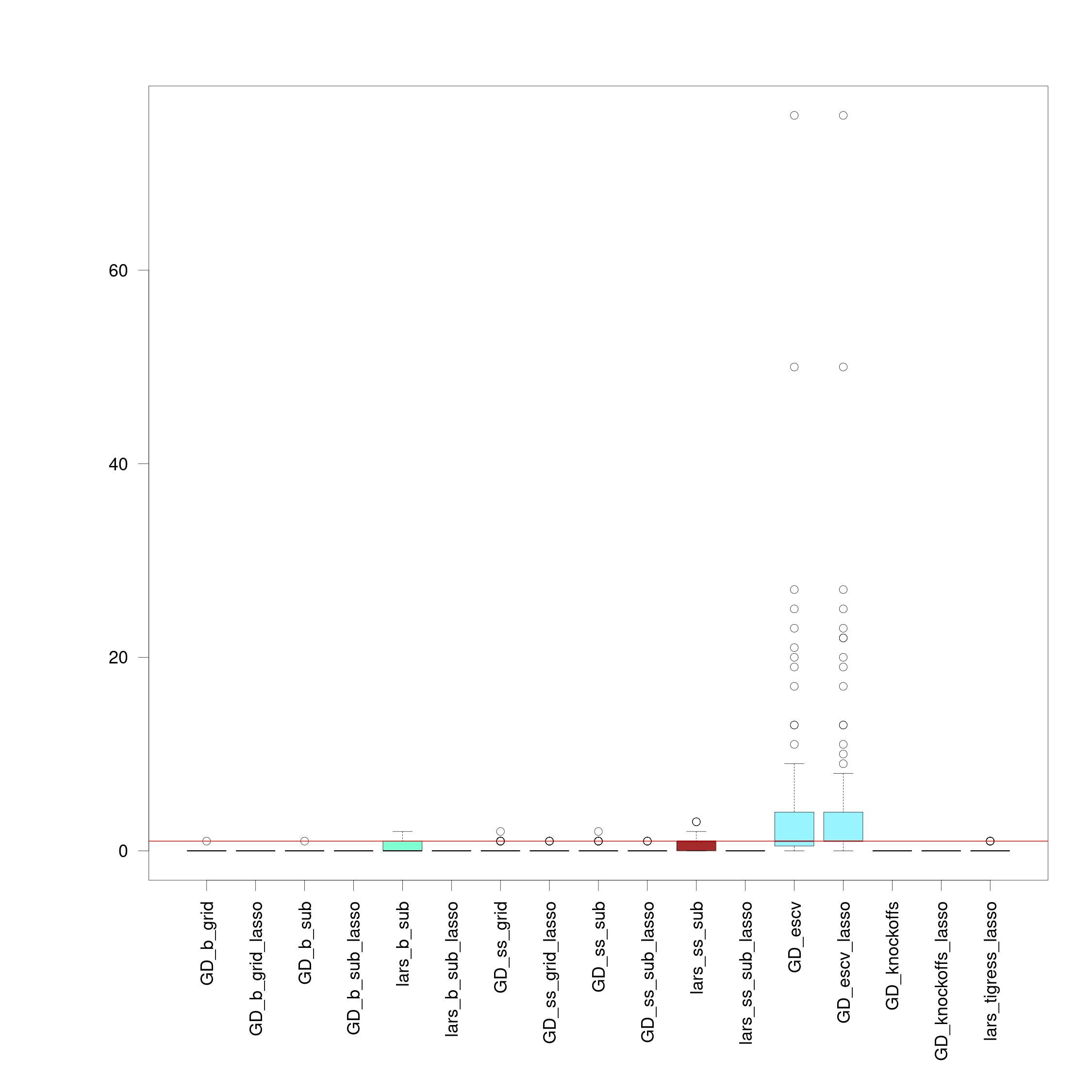}
    \caption*{\textit{scale-free-min}}
\end{subfigure}
\caption{Boxplots of the size of the selected variables subset obtained by the variable identification procedures from dataset of size $n=150$ in the four different settings. The red line indicates the true number of  active variables. }
\label{support2}
\end{figure}

\begin{figure}[h!]
\begin{subfigure}{0.49\linewidth}
    \centering
    \includegraphics[width=1\linewidth]{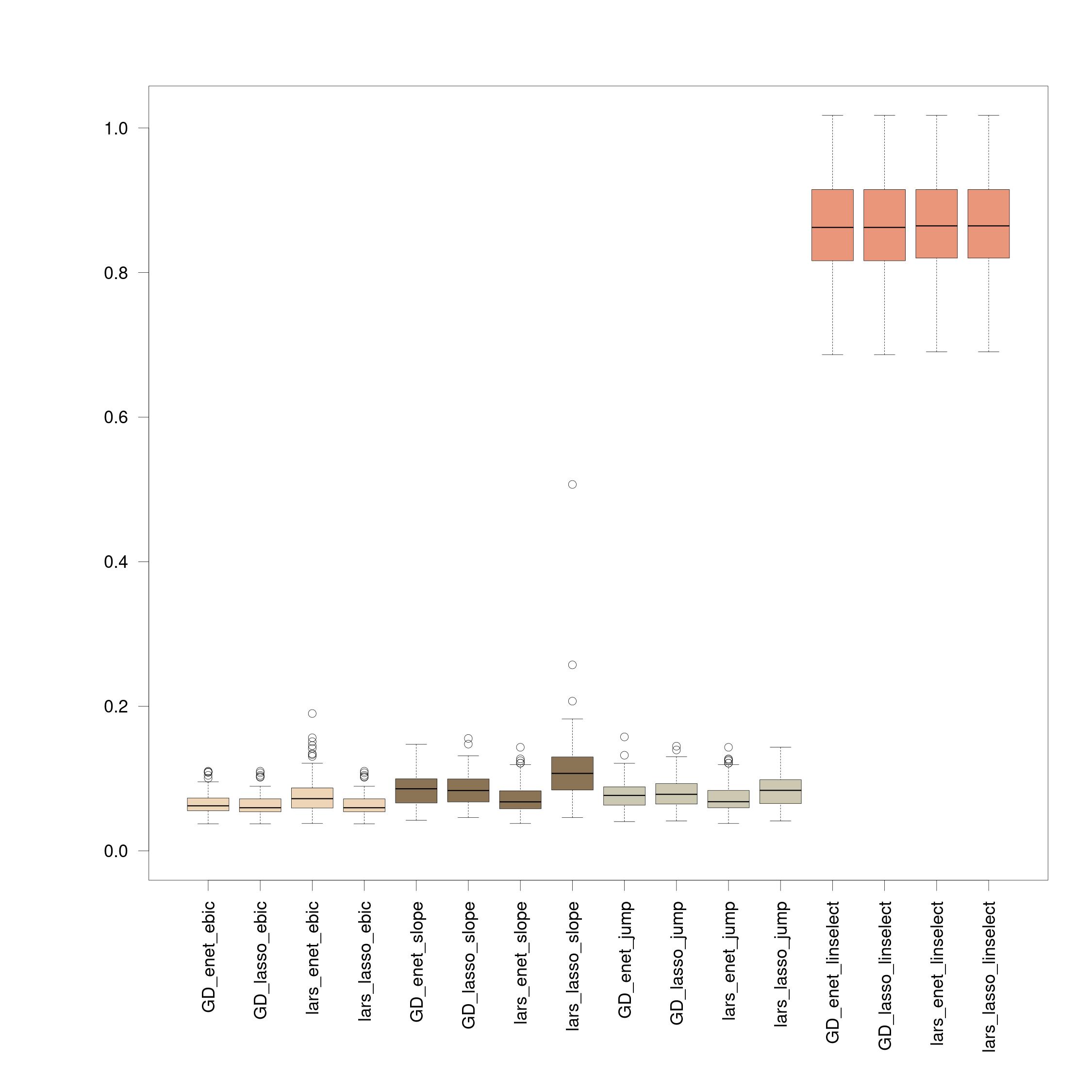}
    \caption*{\textit{independent}}
\end{subfigure}
\begin{subfigure}{0.49\linewidth}
    \centering
    \includegraphics[width=1\linewidth]{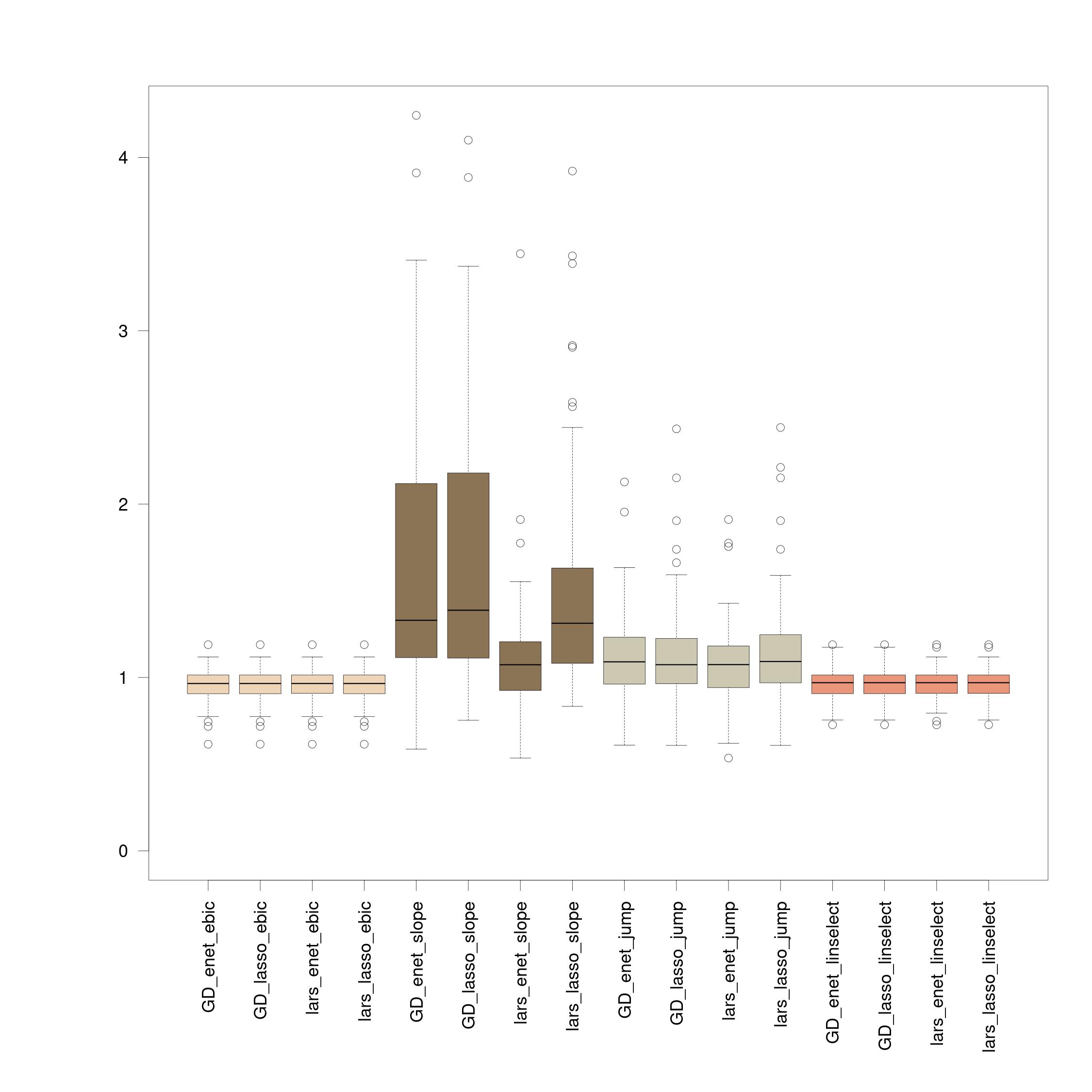}
    \caption*{\textit{cluster}}
\end{subfigure}
\begin{subfigure}{0.49\linewidth}
    \centering
    \includegraphics[width=1\linewidth]{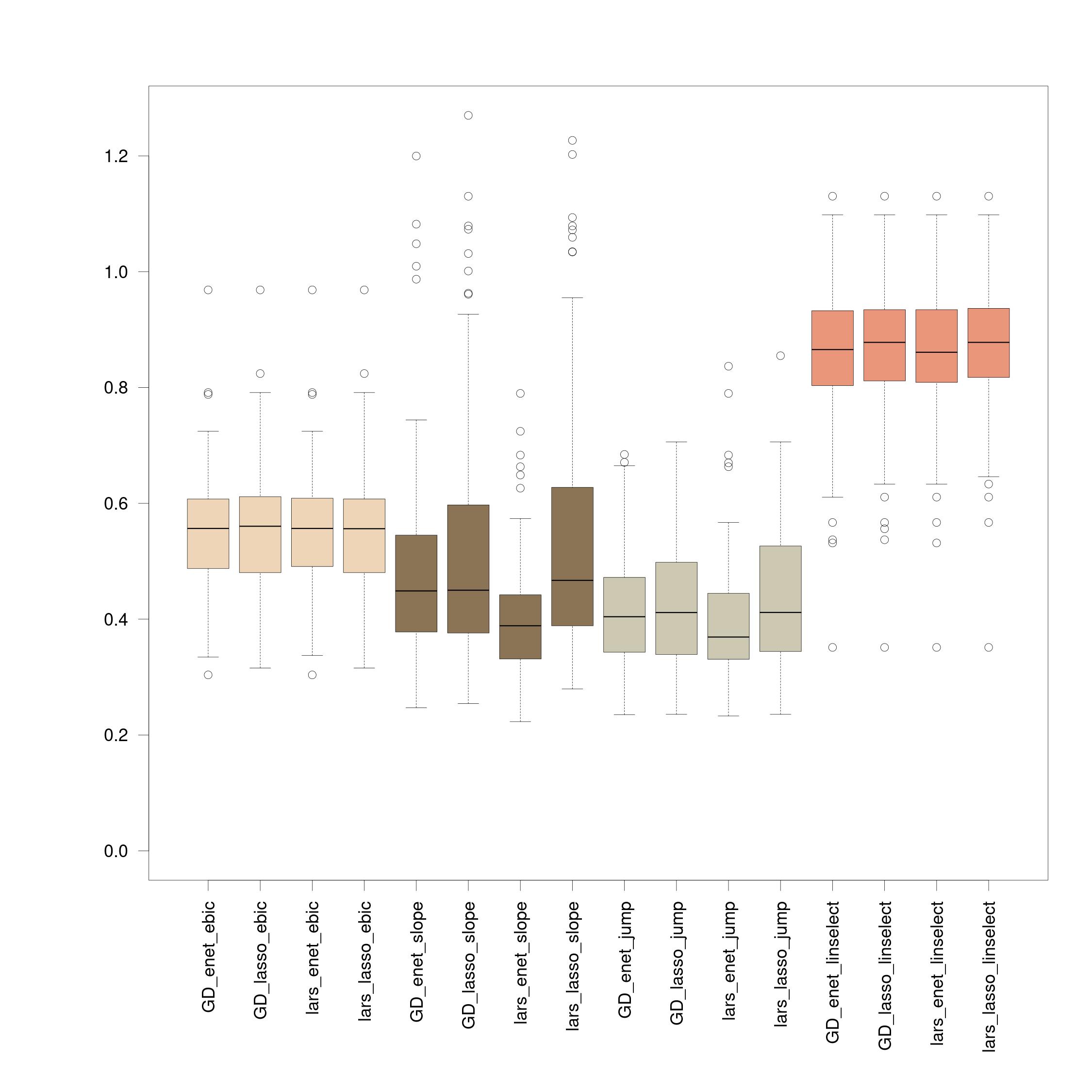}
    \caption*{\textit{scale-free-max}}
\end{subfigure}
\begin{subfigure}{0.49\linewidth}
    \centering
    \includegraphics[width=1\linewidth]{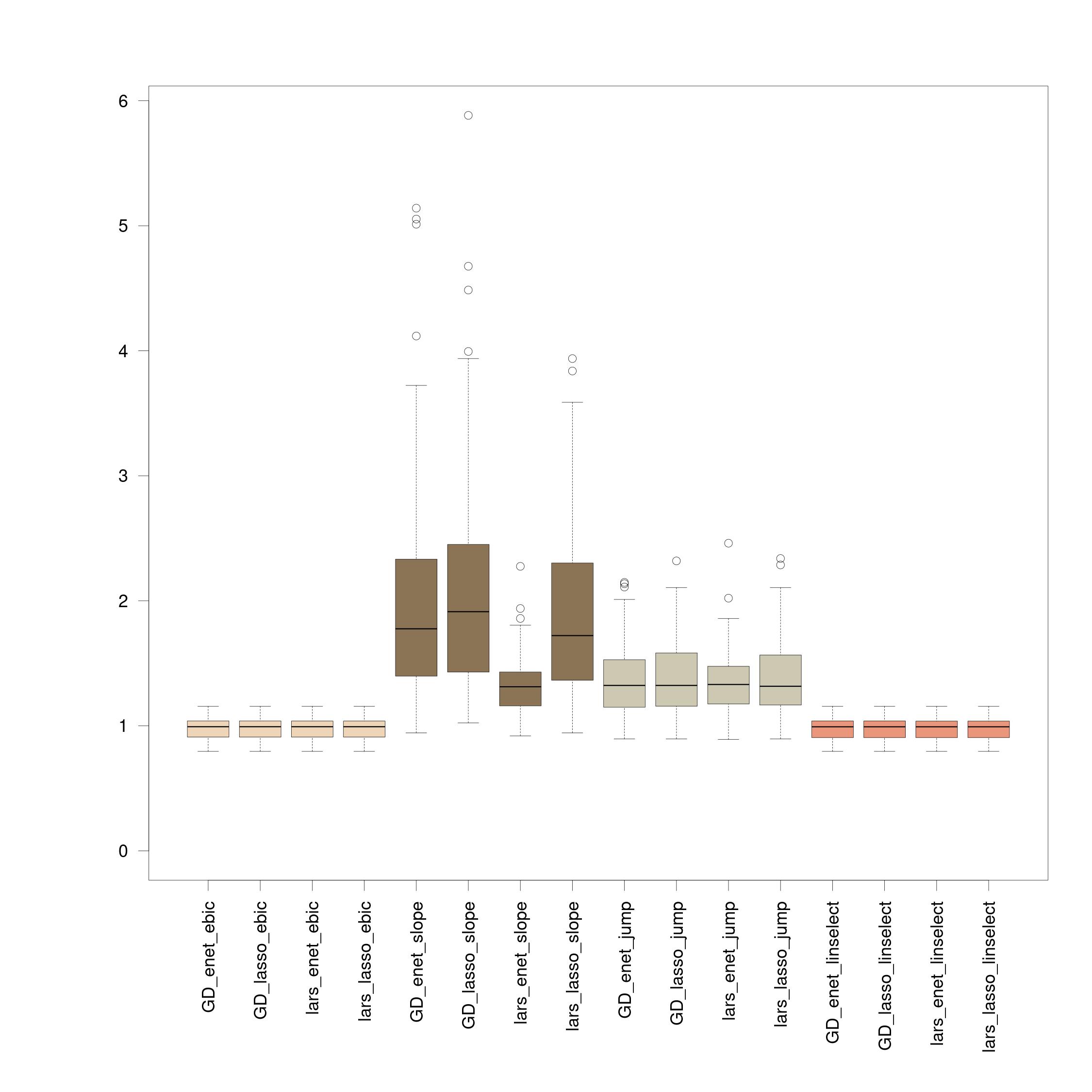}
    \caption*{\textit{scale-free-min}}
\end{subfigure}
\caption{Boxplots of the MSE values obtained by the model selection procedures from dataset of size $n=150$ in the four different settings.}
\label{MSE1}
\end{figure}

\begin{figure}[h!]
\begin{subfigure}{0.49\linewidth}
    \centering
    \includegraphics[width=1\linewidth]{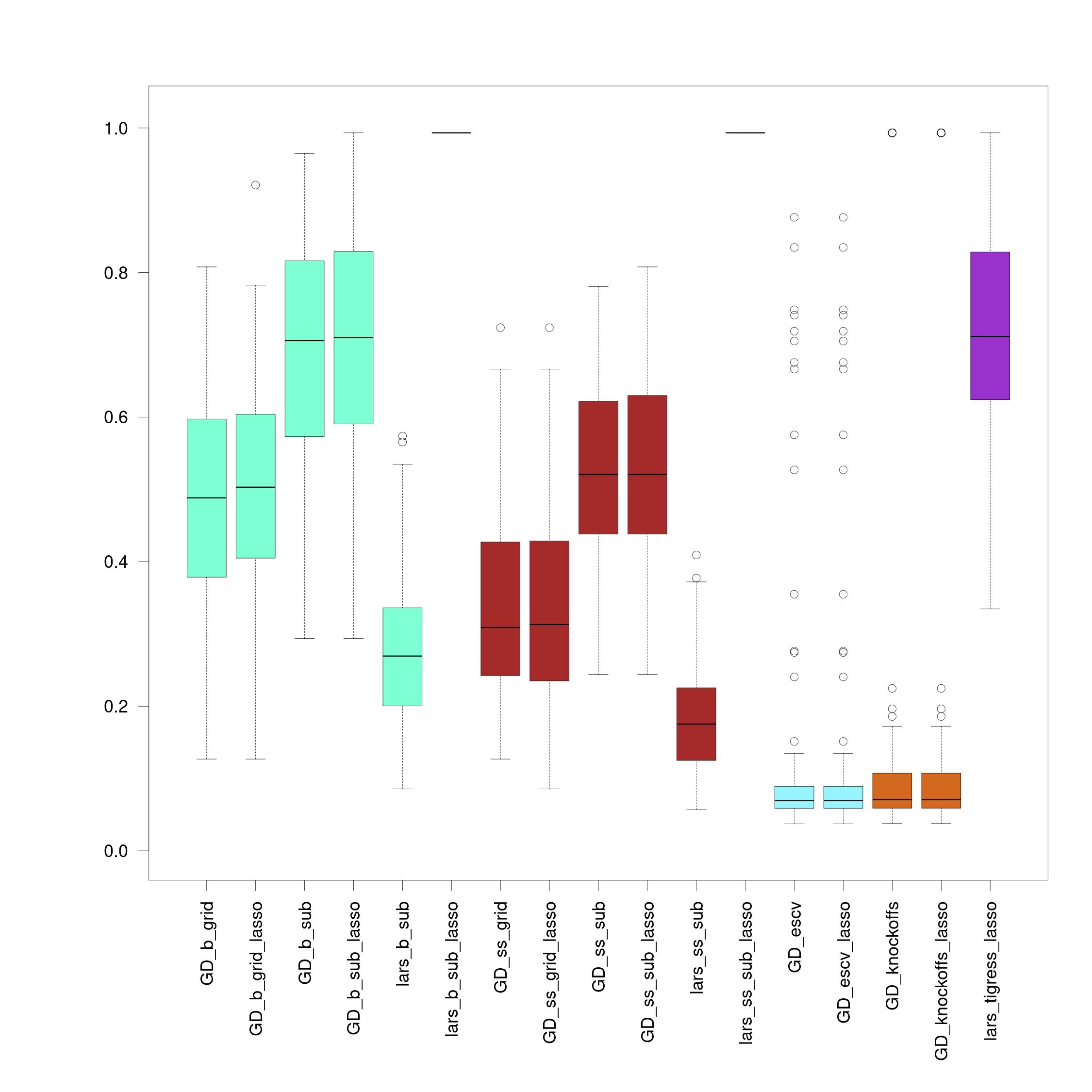}
    \caption*{\textit{independent}}
\end{subfigure}
\begin{subfigure}{0.49\linewidth}
    \centering
    \includegraphics[width=1\linewidth]{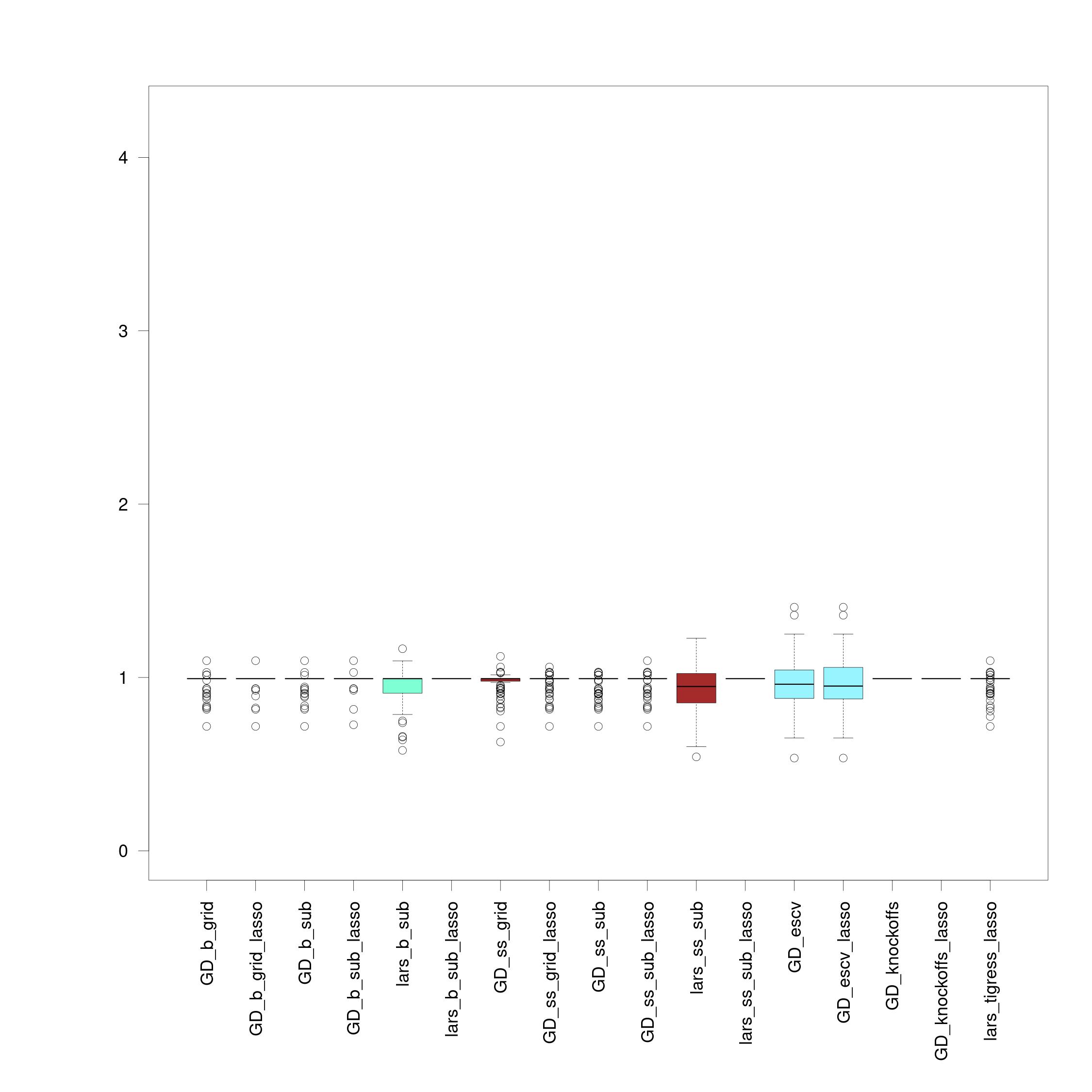}
    \caption*{\textit{cluster}}
\end{subfigure}
\begin{subfigure}{0.49\linewidth}
    \centering
    \includegraphics[width=1\linewidth]{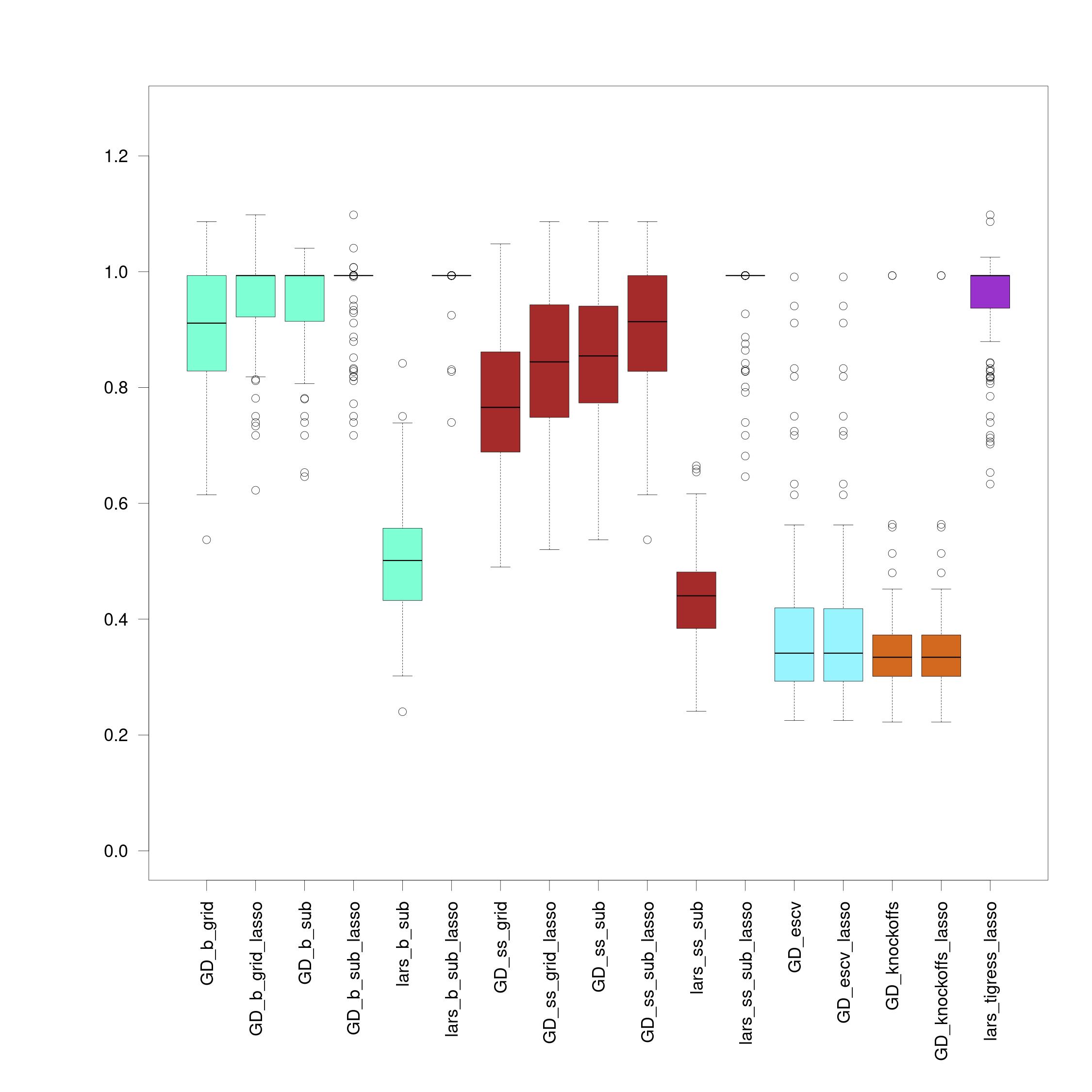}
    \caption*{\textit{scale-free-max}}
\end{subfigure}
\begin{subfigure}{0.49\linewidth}
    \centering
    \includegraphics[width=1\linewidth]{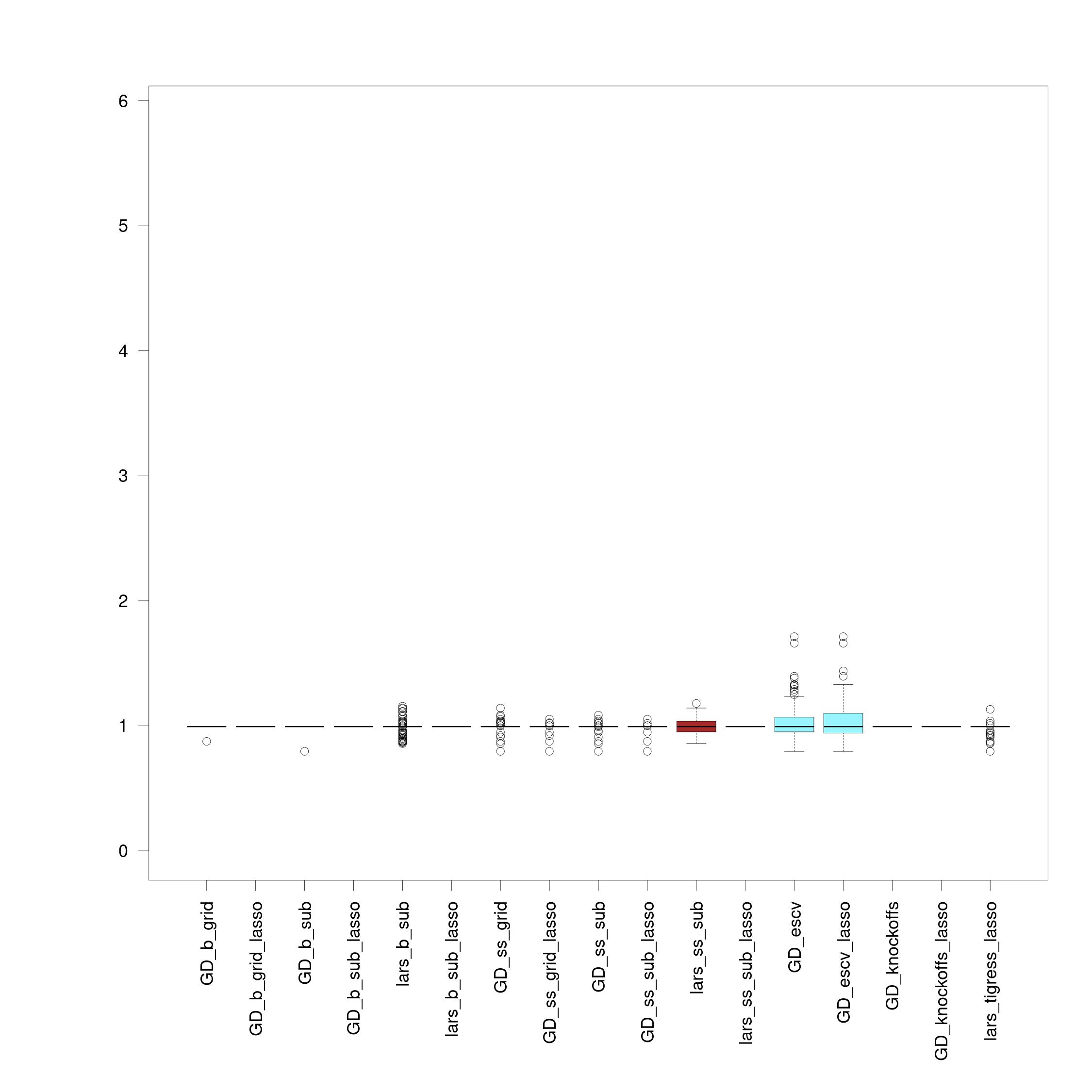}
    \caption*{\textit{scale-free-min}}
\end{subfigure}
\caption{Boxplots of the MSE values obtained by the variable identification procedures from dataset of size $n=150$ in the four different settings.}
\label{MSE2}
\end{figure}

\begin{figure}[h!]
\begin{subfigure}{0.49\linewidth}
    \centering
    \includegraphics[width=1\linewidth]{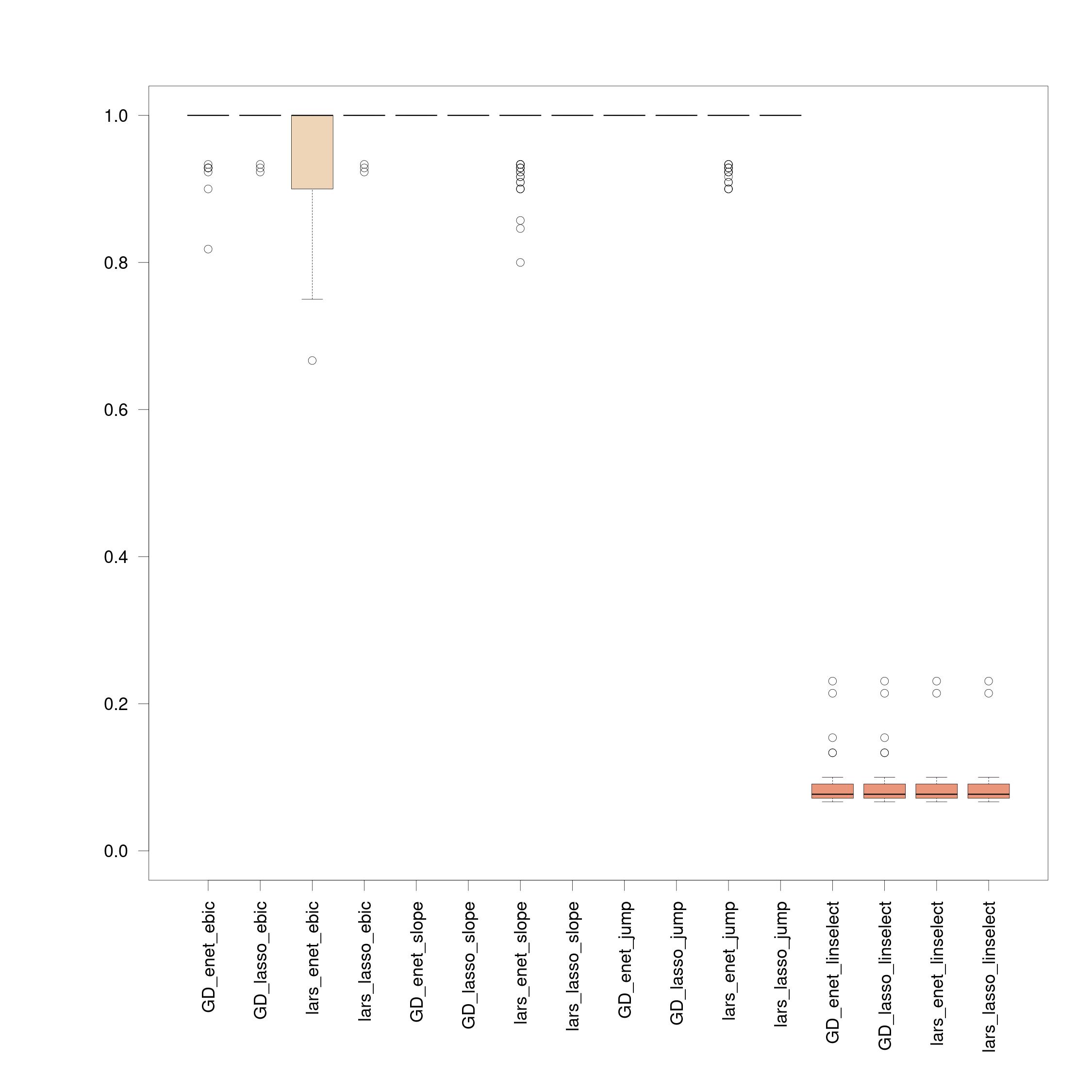}
    \caption*{\textit{independent}}
\end{subfigure}
\begin{subfigure}{0.49\linewidth}
    \centering
    \includegraphics[width=1\linewidth]{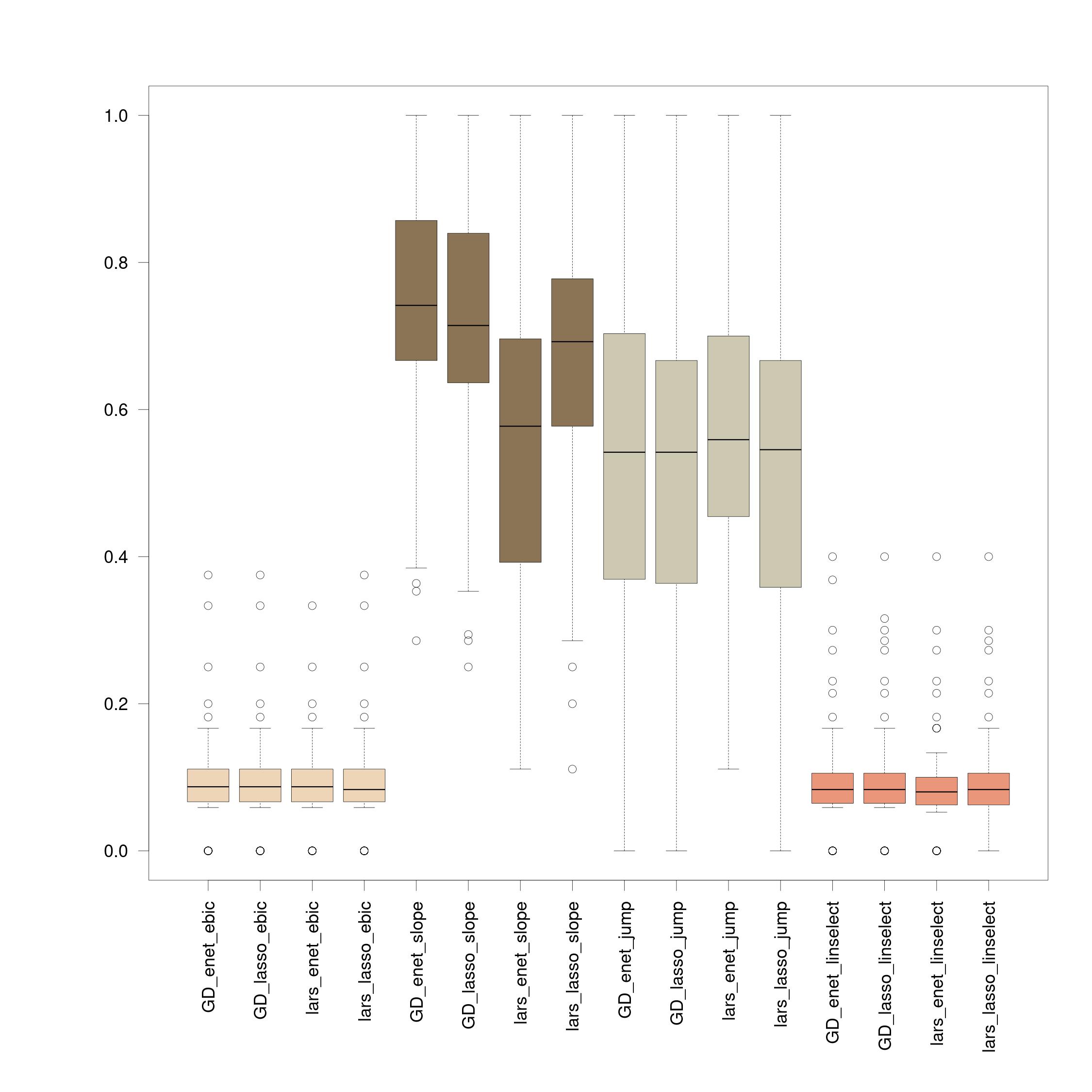}
    \caption*{\textit{cluster}}
\end{subfigure}
\begin{subfigure}{0.49\linewidth}
    \centering
    \includegraphics[width=1\linewidth]{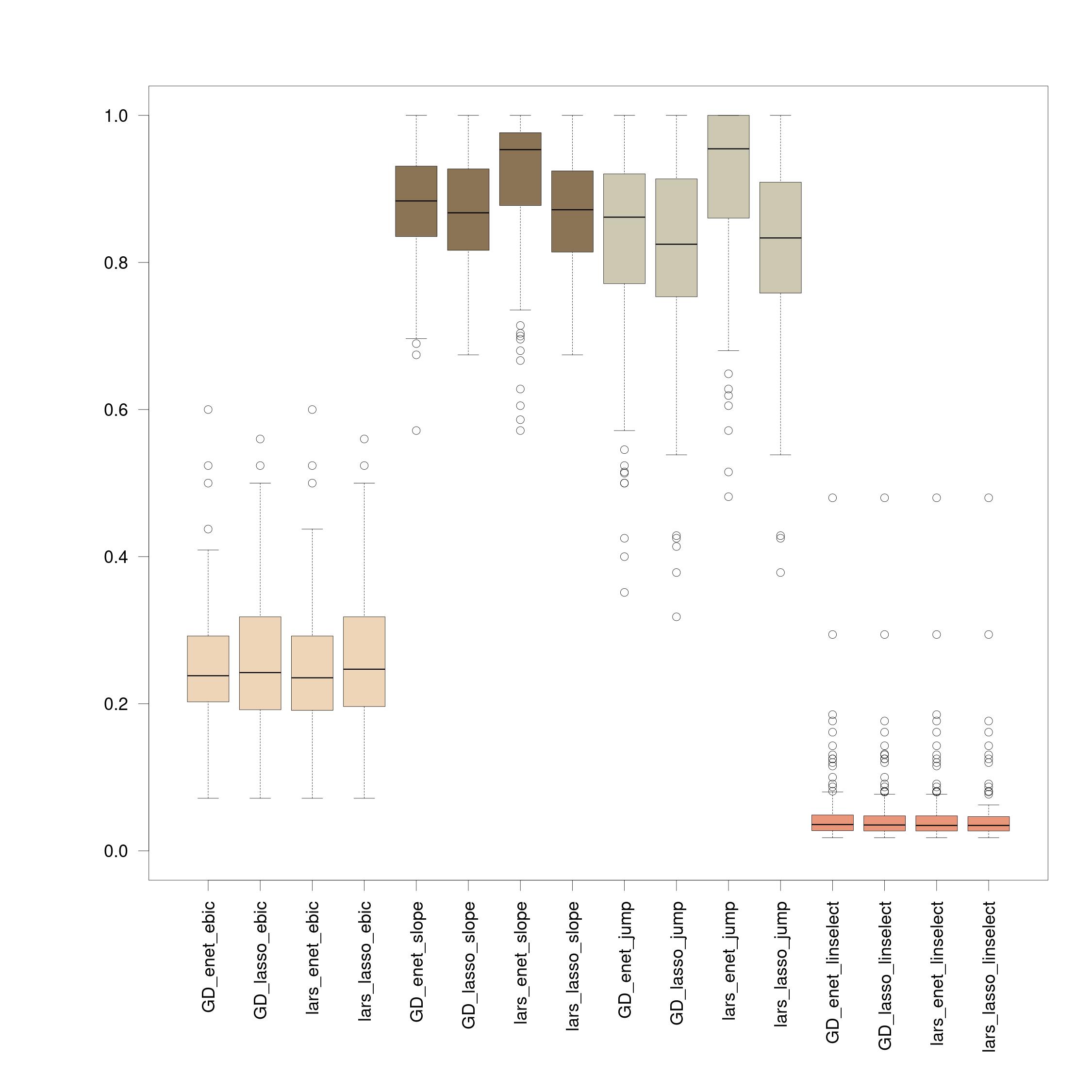}
    \caption*{\textit{scale-free-max}}
\end{subfigure}
\begin{subfigure}{0.49\linewidth}
    \centering
    \includegraphics[width=1\linewidth]{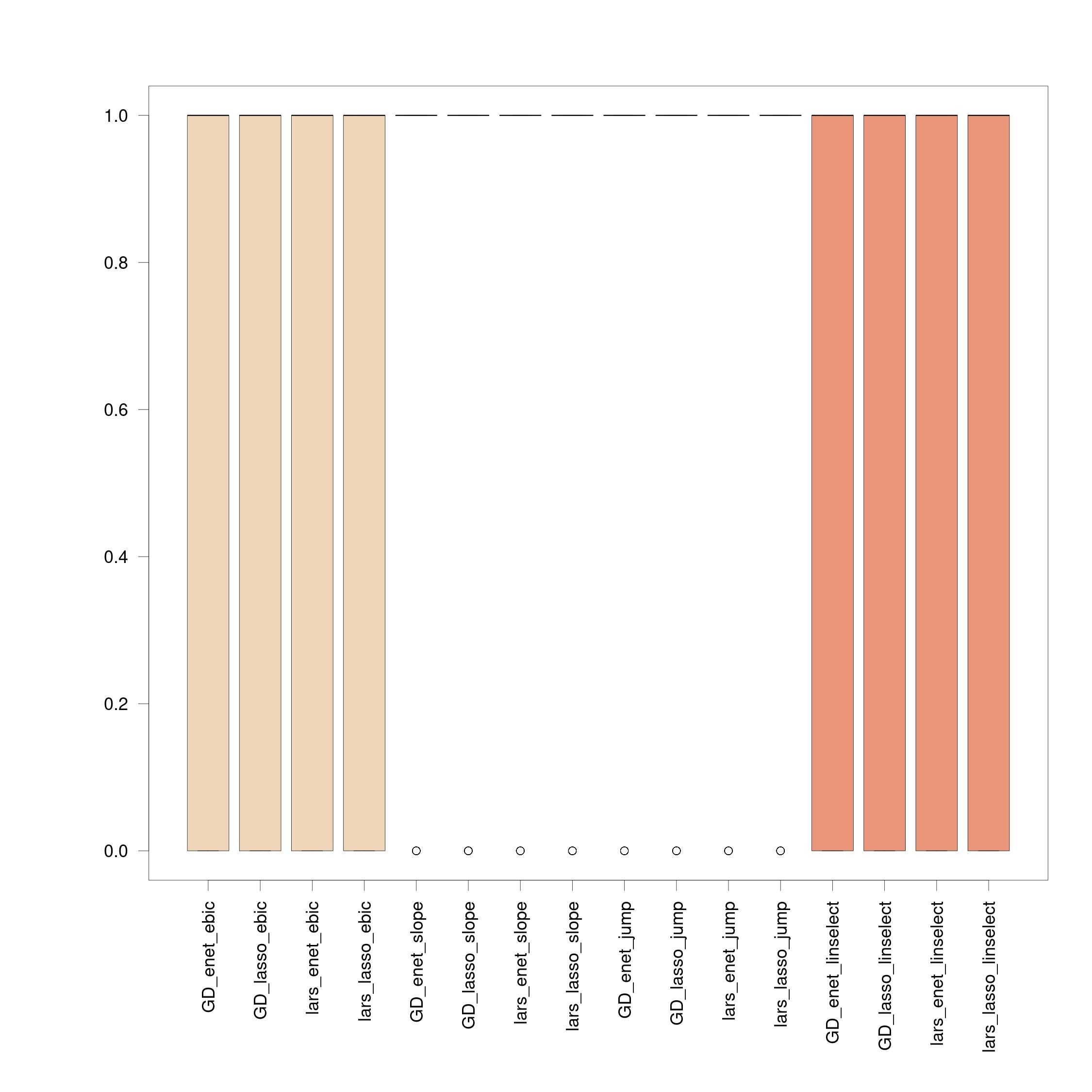}
    \caption*{\textit{scale-free-min}}
\end{subfigure}
\caption{Boxplots of the recall values obtained by the model selection procedures from dataset of size $n=150$ in the four different settings.}
\label{recall1}
\end{figure}

\begin{figure}[h!]
\begin{subfigure}{0.49\linewidth}
    \centering
    \includegraphics[width=1\linewidth]{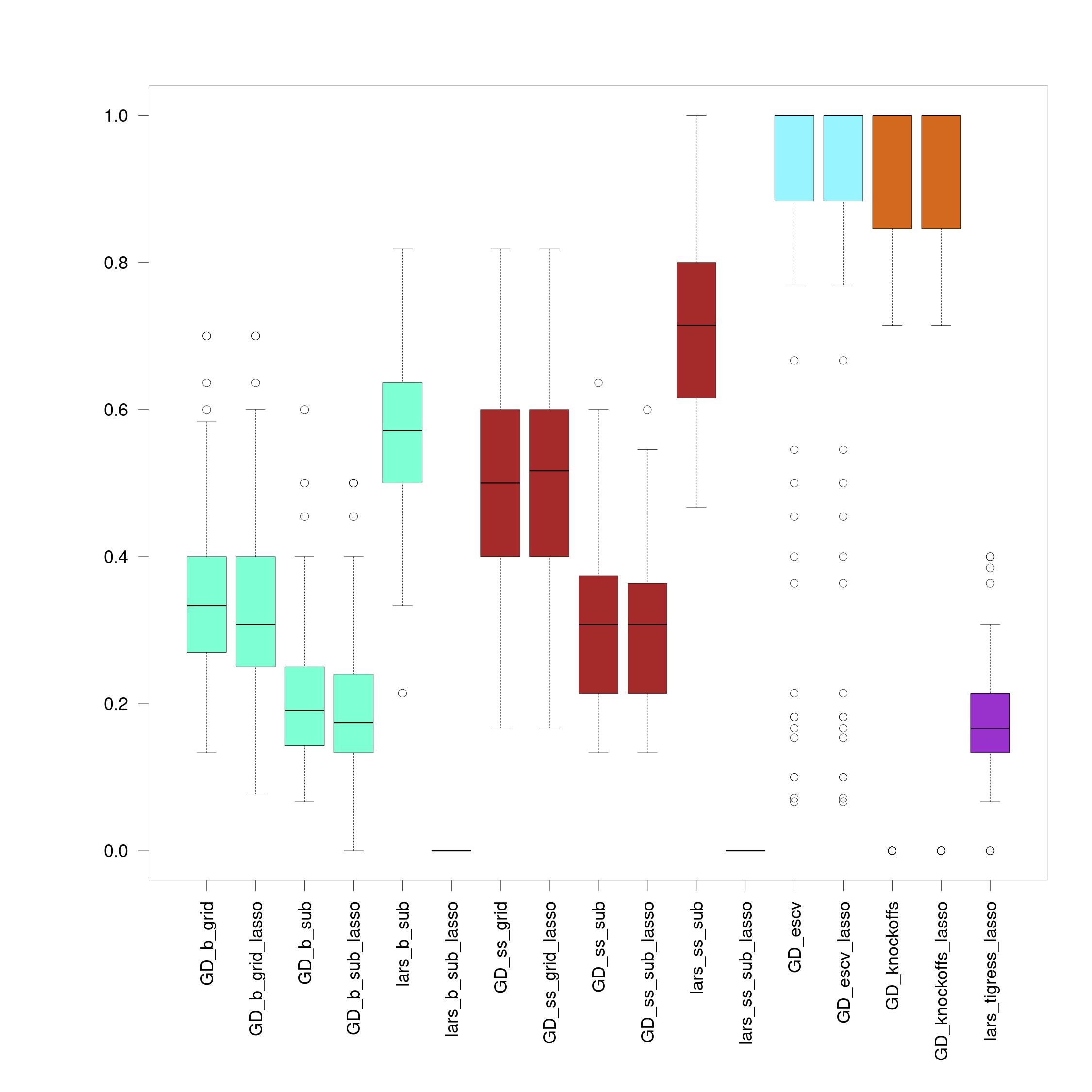}
    \caption*{\textit{independent}}
\end{subfigure}
\begin{subfigure}{0.49\linewidth}
    \centering
    \includegraphics[width=1\linewidth]{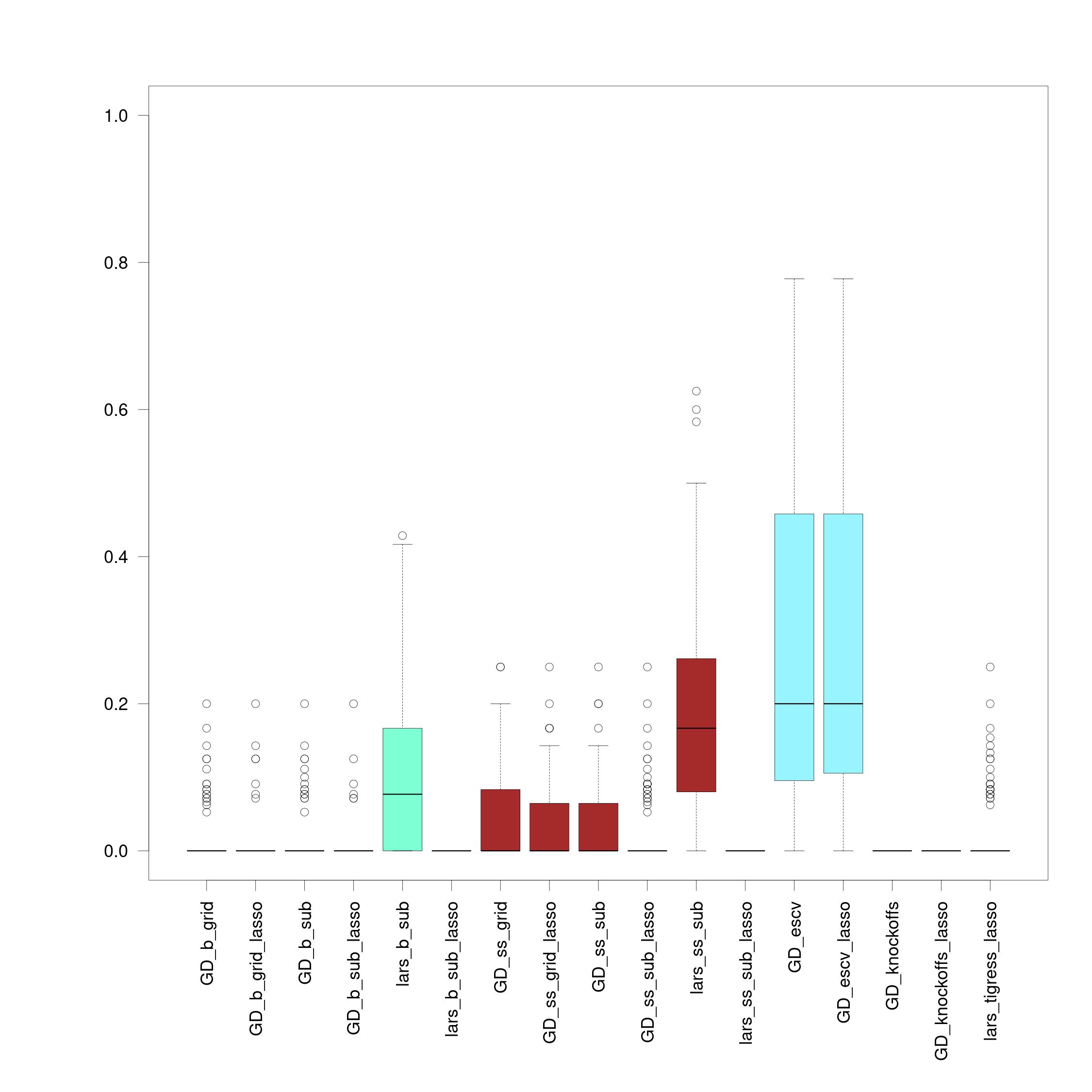}
    \caption*{\textit{cluster}}
\end{subfigure}
\begin{subfigure}{0.49\linewidth}
    \centering
    \includegraphics[width=1\linewidth]{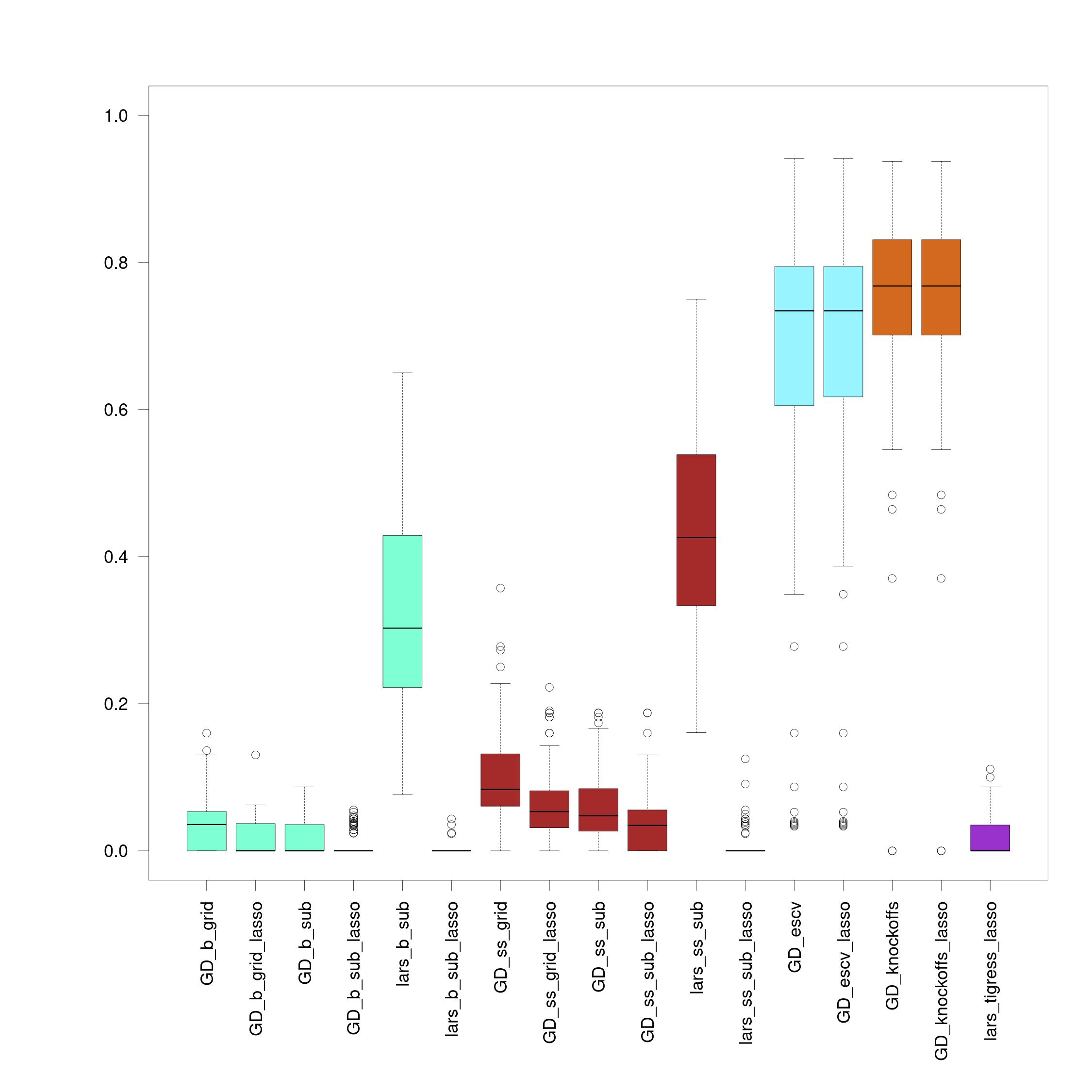}
    \caption*{\textit{scale-free-max}}
\end{subfigure}
\begin{subfigure}{0.49\linewidth}
    \centering
    \includegraphics[width=1\linewidth]{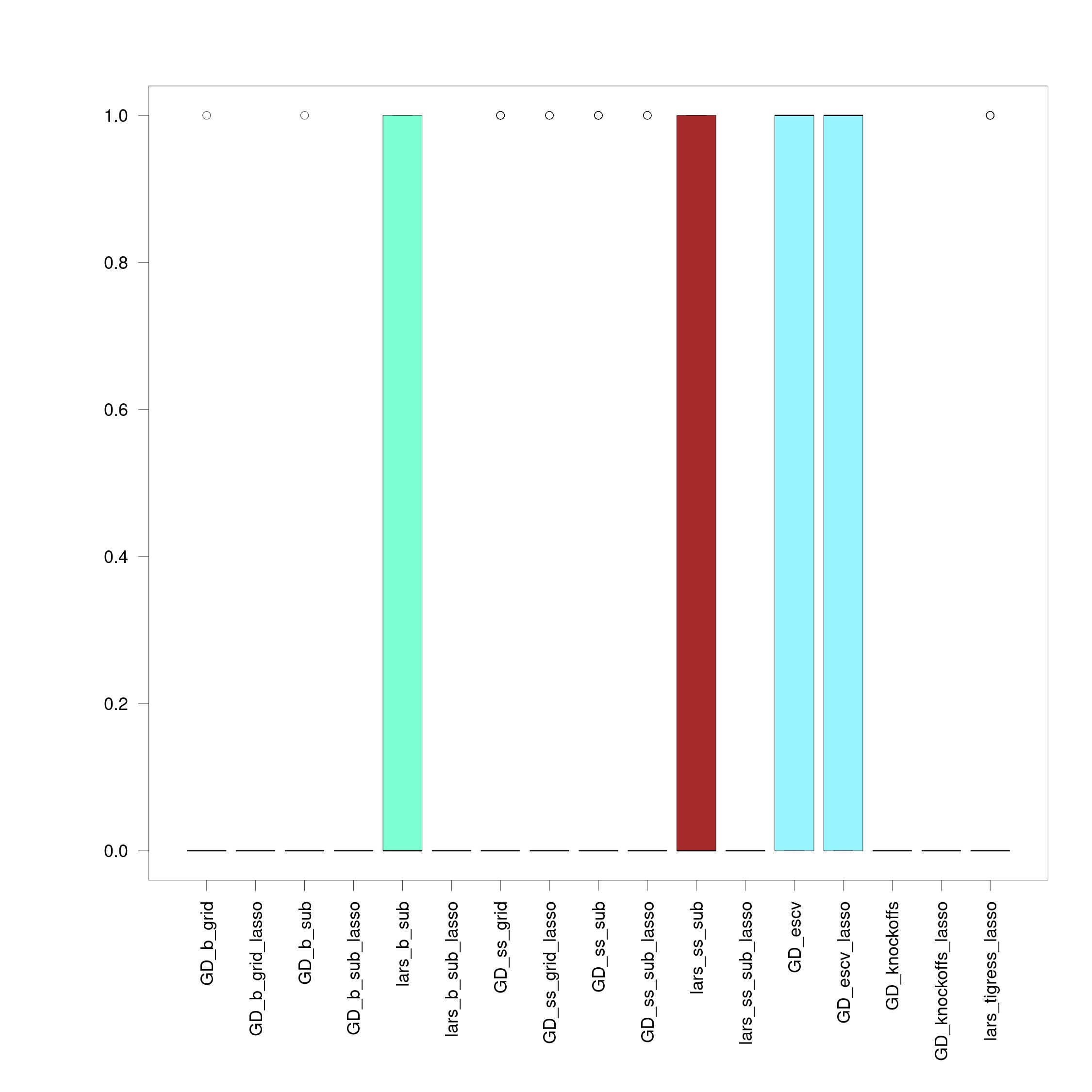}
    \caption*{\textit{scale-free-min}}
\end{subfigure}
\caption{Boxplots of the recall values obtained by the variable identification procedures from dataset of size $n=150$ in the four different settings.}
\label{recall2}
\end{figure}

\begin{figure}[h!]
\begin{subfigure}{0.49\linewidth}
    \centering
    \includegraphics[width=1\linewidth]{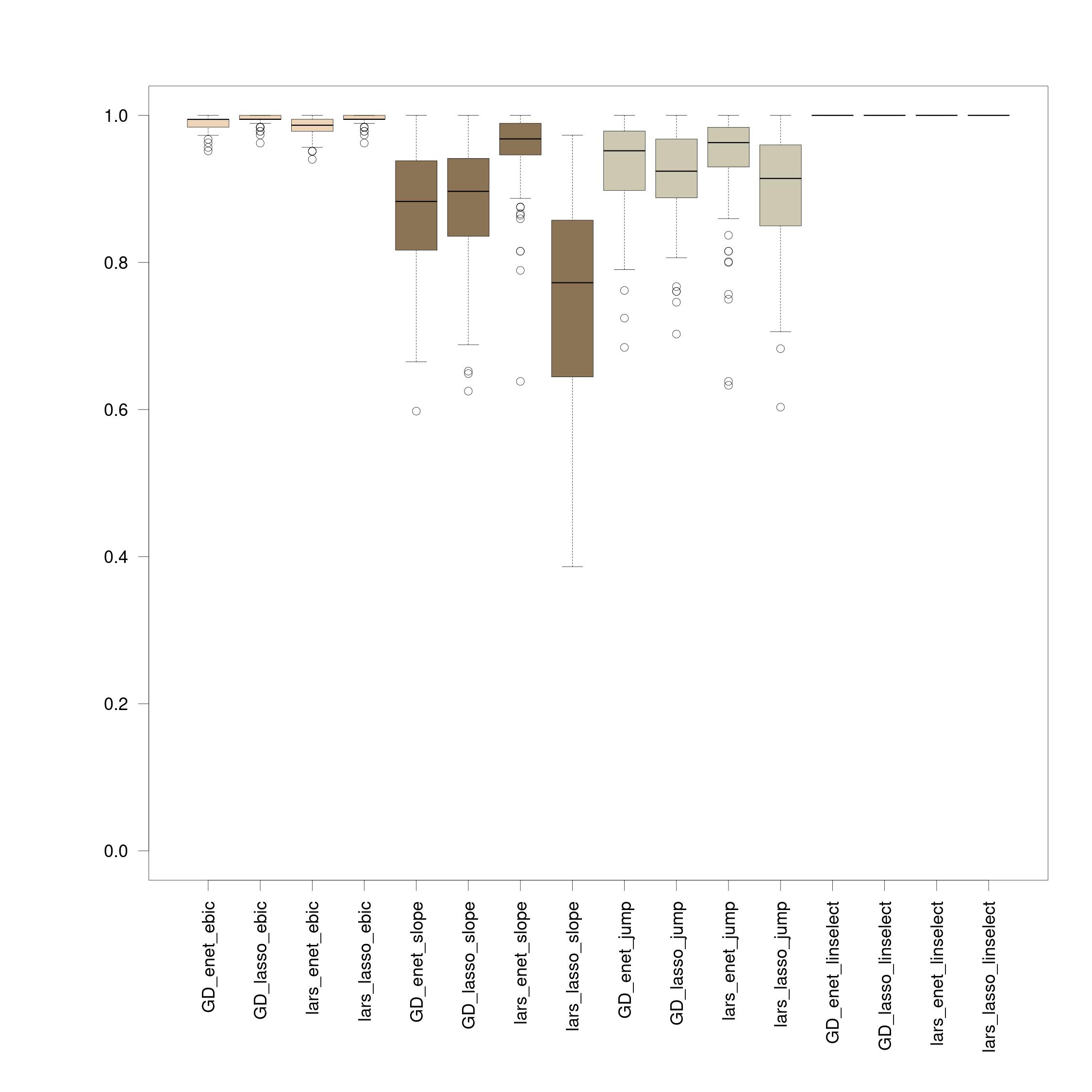}
    \caption*{\textit{independent}}
\end{subfigure}
\begin{subfigure}{0.49\linewidth}
    \centering
    \includegraphics[width=1\linewidth]{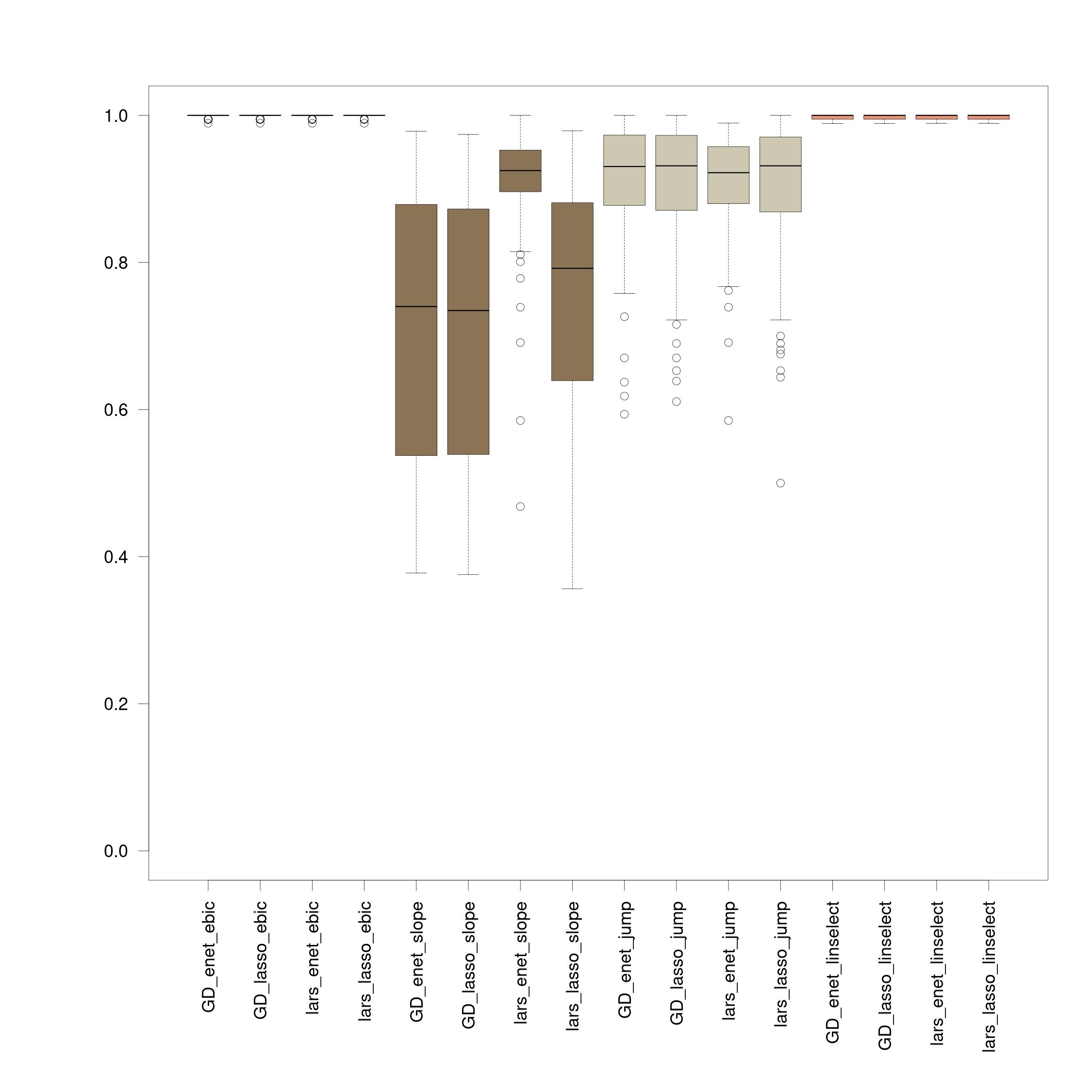}
    \caption*{\textit{cluster}}
\end{subfigure}
\begin{subfigure}{0.49\linewidth}
    \centering
    \includegraphics[width=1\linewidth]{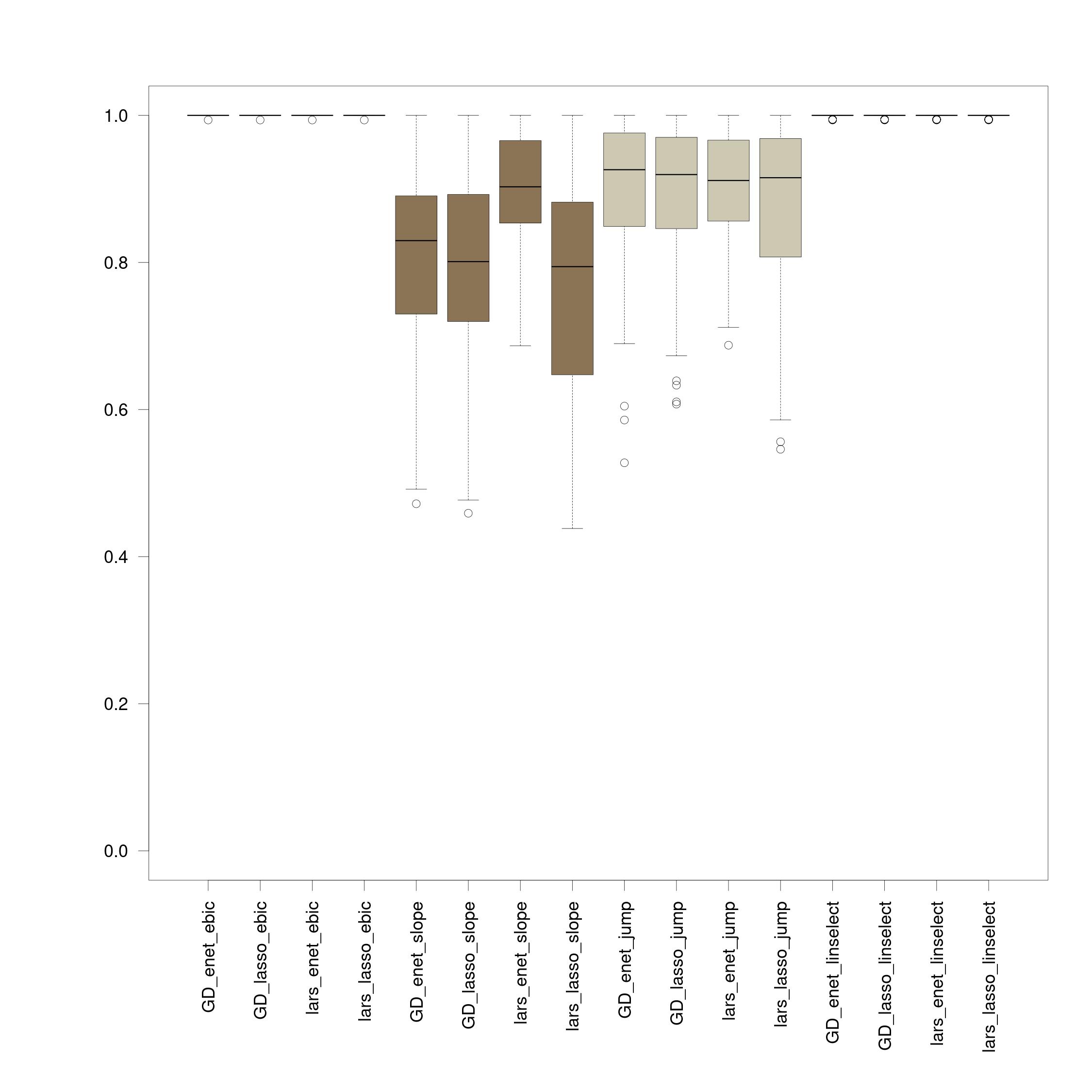}
    \caption*{\textit{scale-free-max}}
\end{subfigure}
\begin{subfigure}{0.49\linewidth}
    \centering
    \includegraphics[width=1\linewidth]{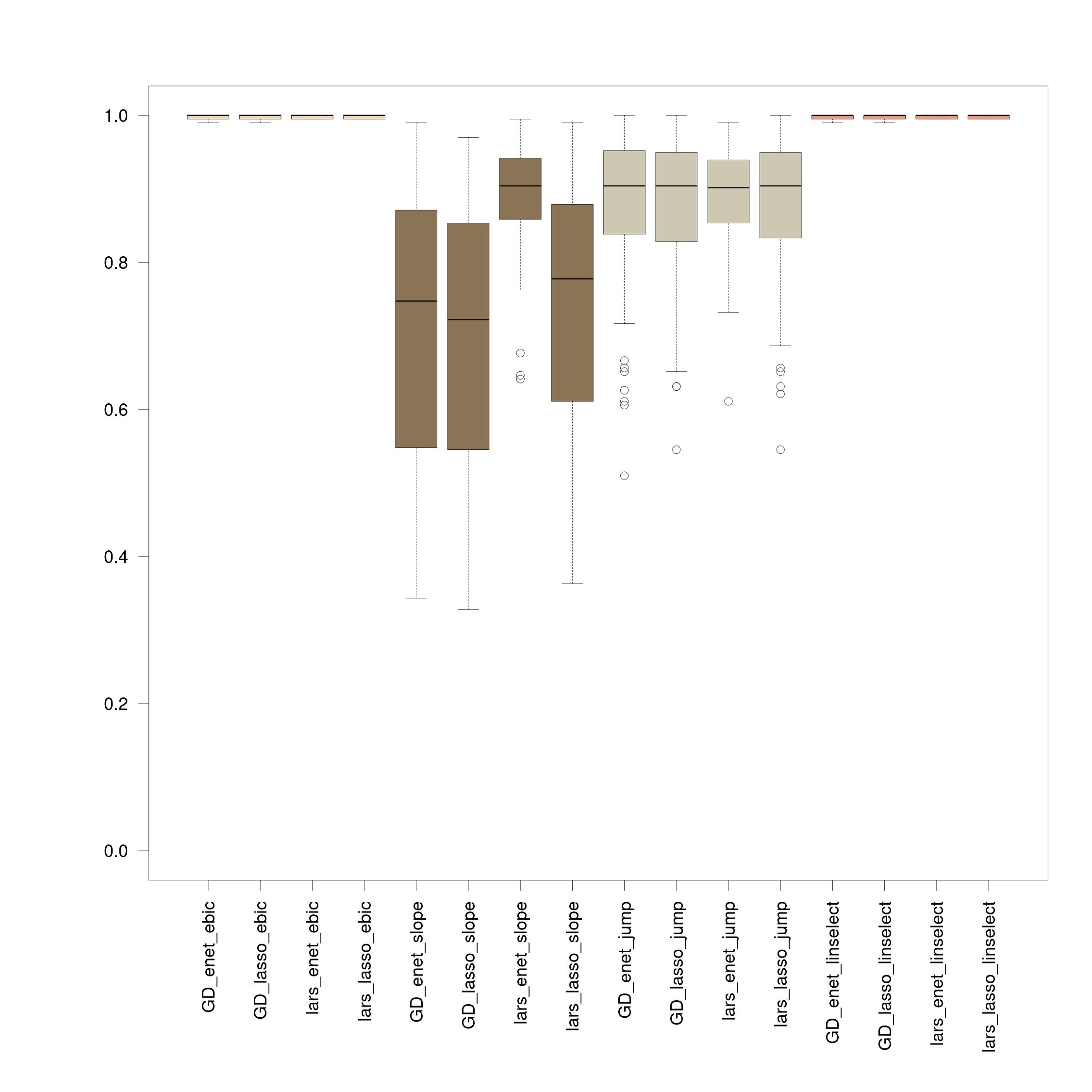}
    \caption*{\textit{scale-free-min}}
\end{subfigure}
\caption{Boxplots of the specificity values obtained by the model selection procedures from dataset of size $n=150$ in the four different settings.}
\label{specificity1}
\end{figure}

\begin{figure}[h!]
\begin{subfigure}{0.49\linewidth}
    \centering
    \includegraphics[width=1\linewidth]{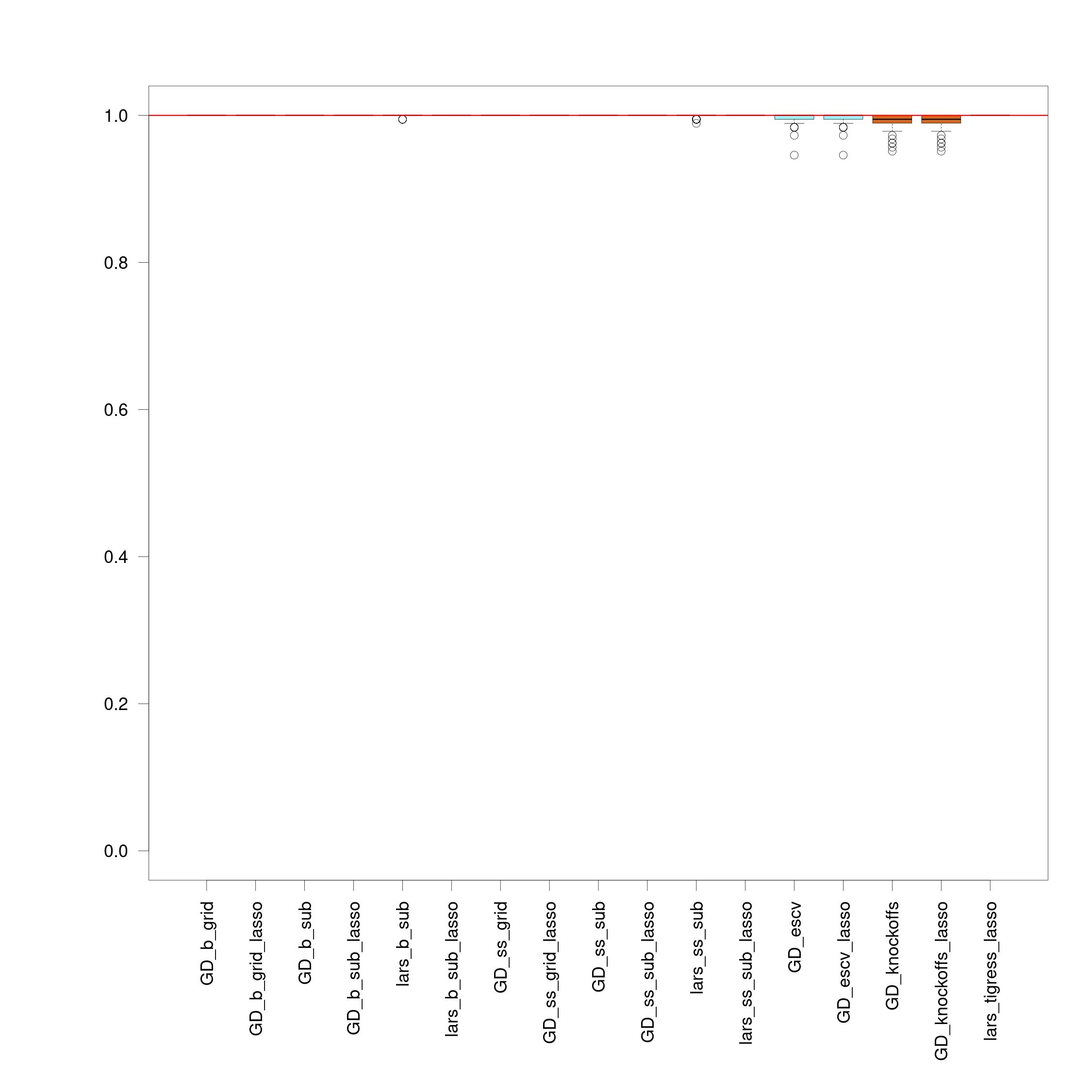}
    \caption*{\textit{independent}}
\end{subfigure}
\begin{subfigure}{0.49\linewidth}
    \centering
    \includegraphics[width=1\linewidth]{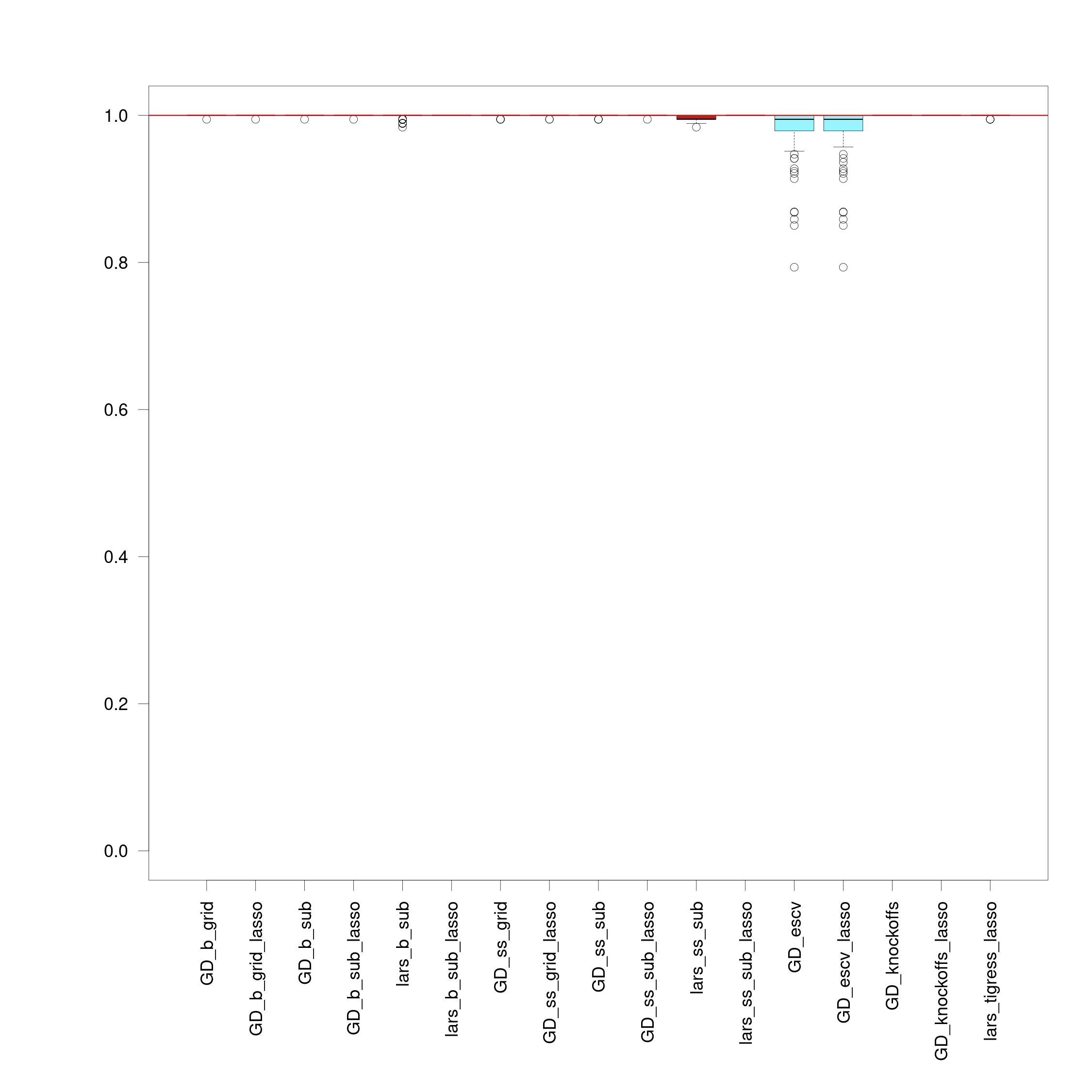}
    \caption*{\textit{cluster}}
\end{subfigure}
\begin{subfigure}{0.49\linewidth}
    \centering
    \includegraphics[width=1\linewidth]{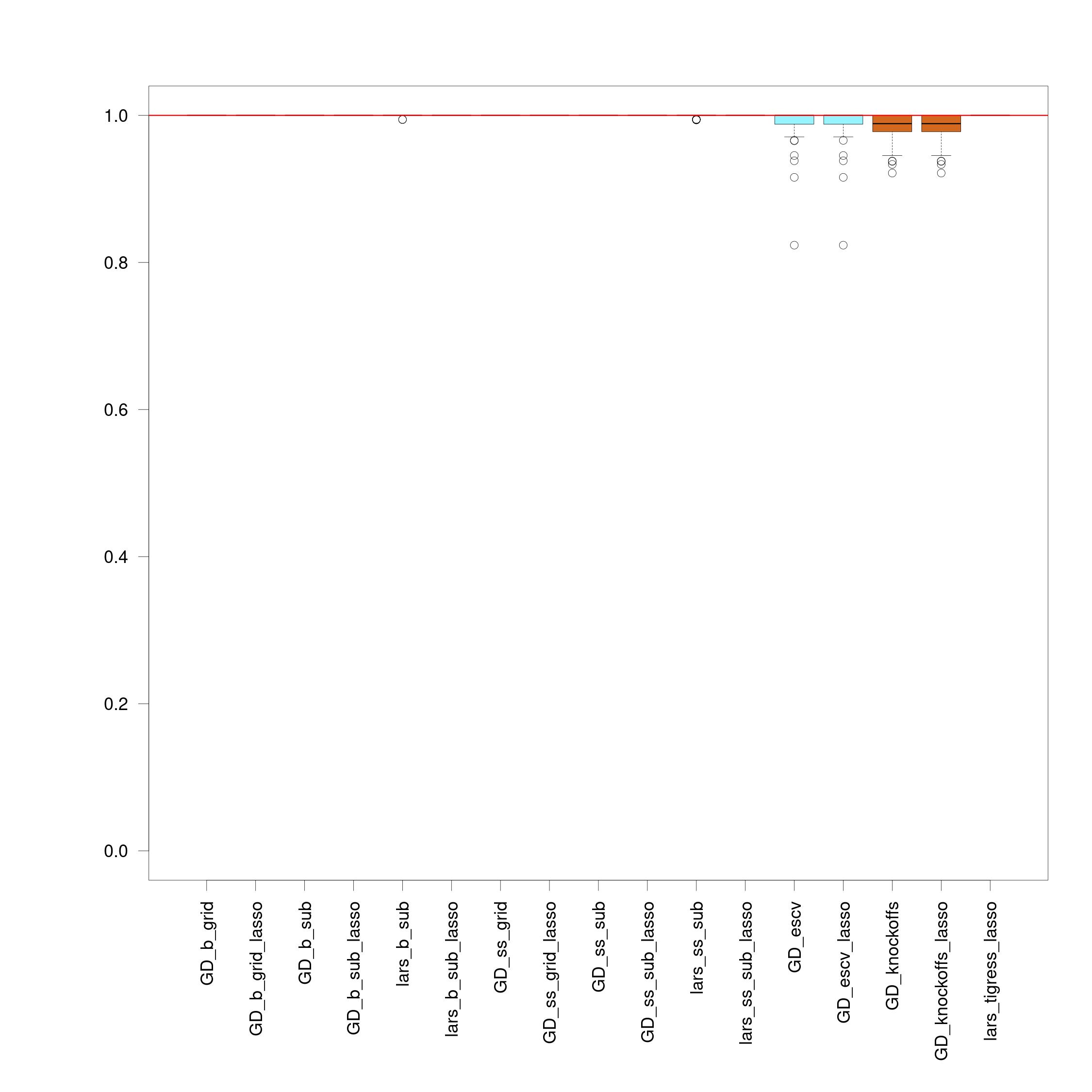}
    \caption*{\textit{scale-free-max}}
\end{subfigure}
\begin{subfigure}{0.49\linewidth}
    \centering
    \includegraphics[width=1\linewidth]{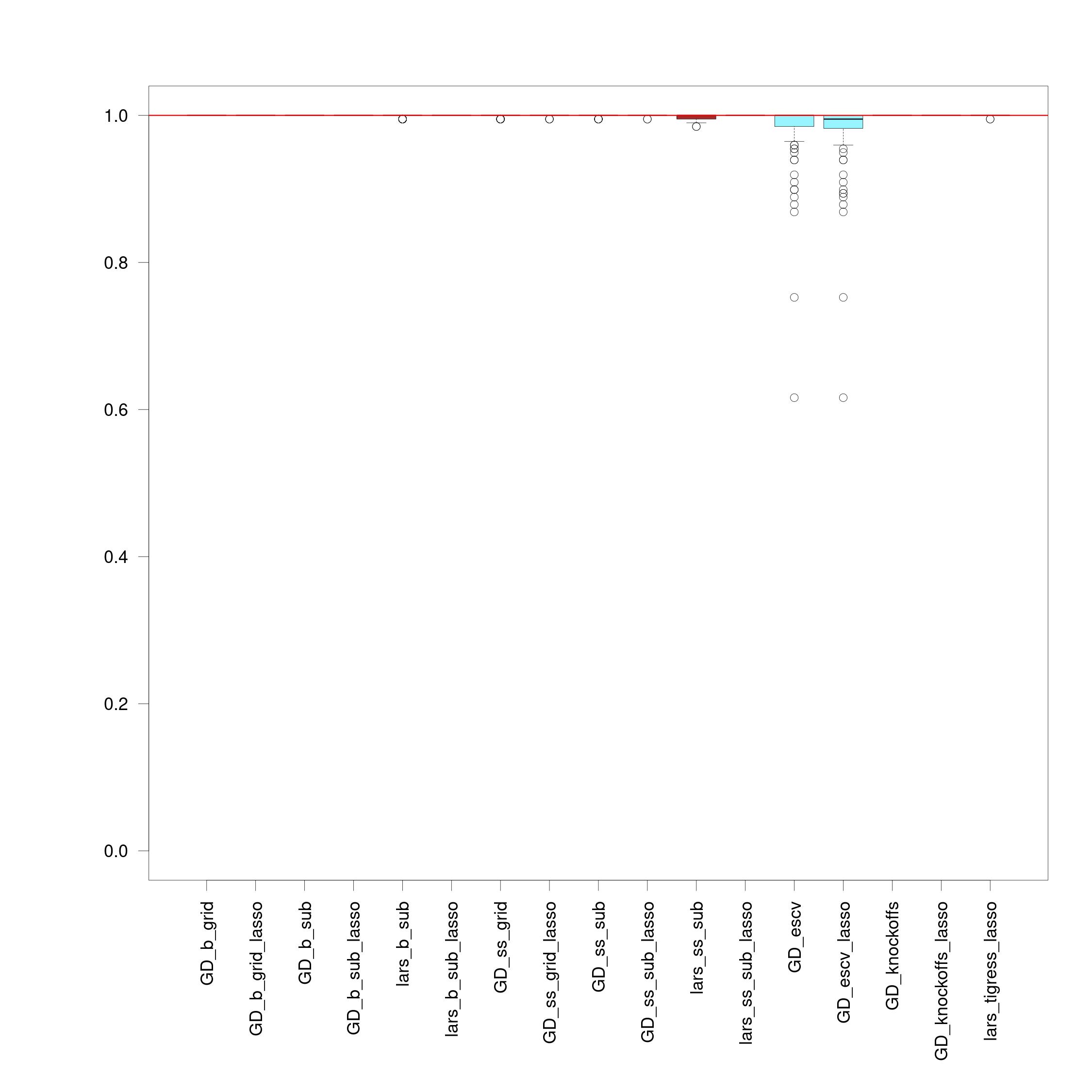}
    \caption*{\textit{scale-free-min}}
\end{subfigure}
\caption{Boxplots of the specificity values obtained by the variable identification procedures from dataset of size $n=150$ in the four different settings.}
\label{specificity2}
\end{figure}

\begin{figure}[h!]
\begin{subfigure}{0.49\linewidth}
    \centering
    \includegraphics[width=1\linewidth]{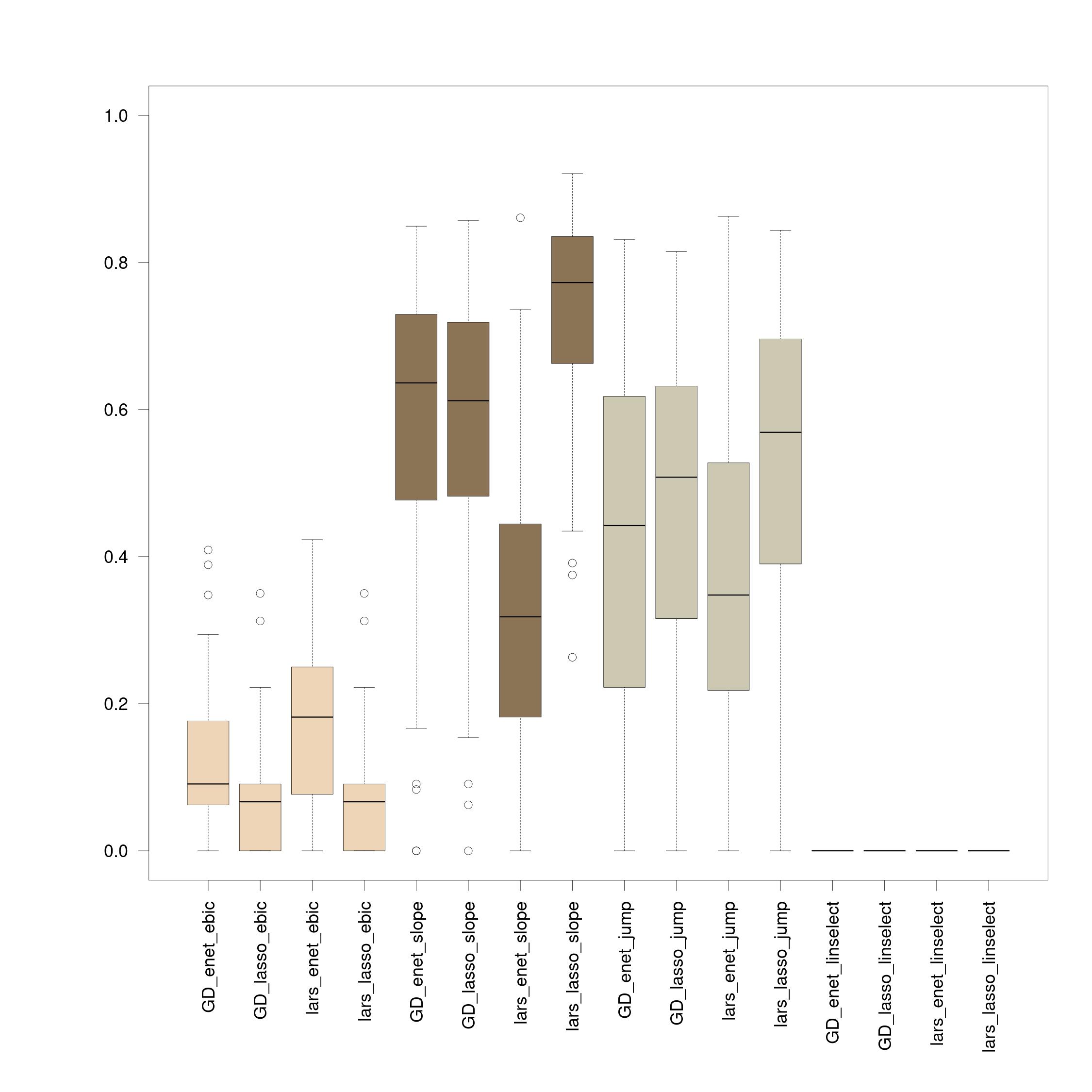}
    \caption*{\textit{independent}}
\end{subfigure}
\begin{subfigure}{0.49\linewidth}
    \centering
    \includegraphics[width=1\linewidth]{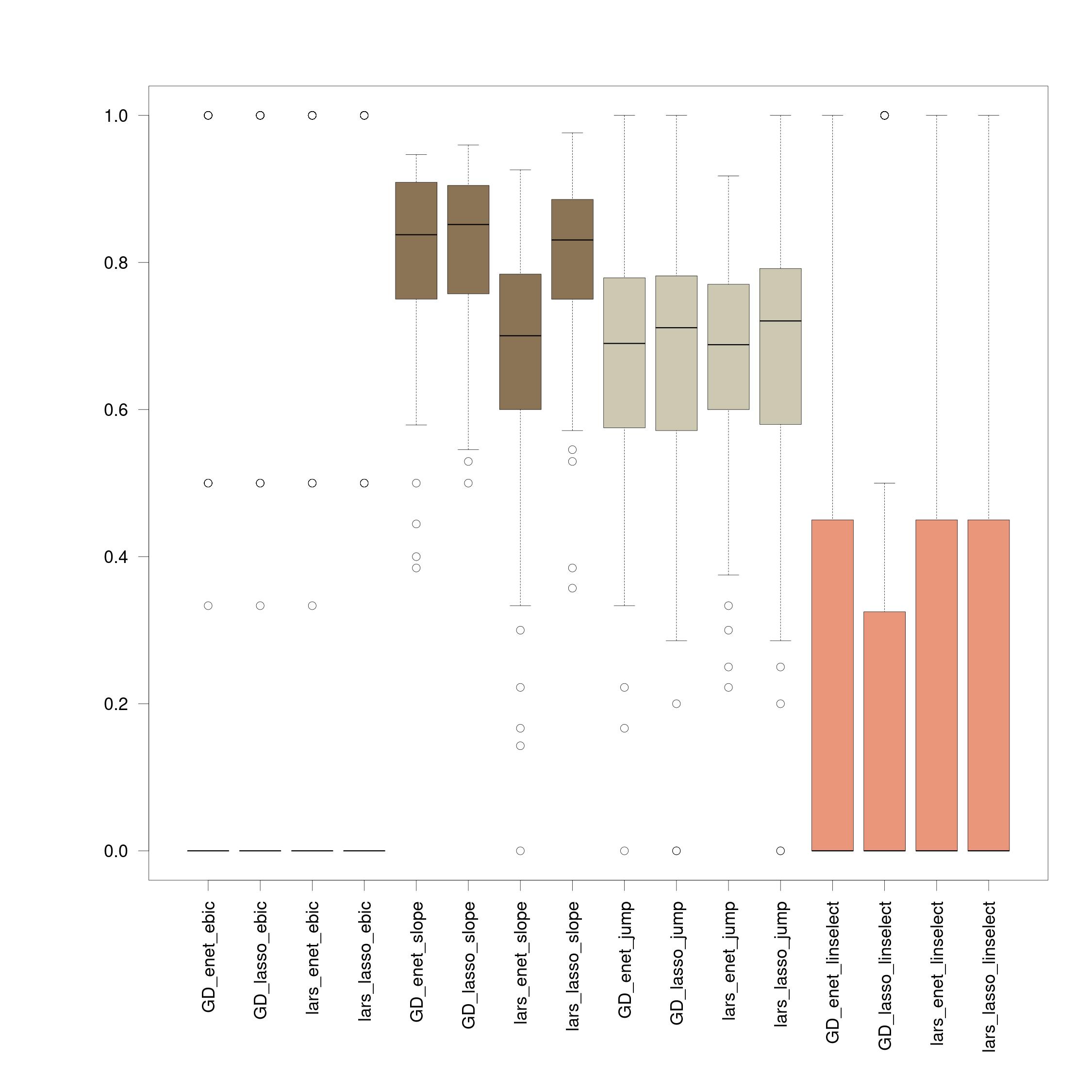}
    \caption*{\textit{cluster}}
\end{subfigure}
\begin{subfigure}{0.49\linewidth}
    \centering
    \includegraphics[width=1\linewidth]{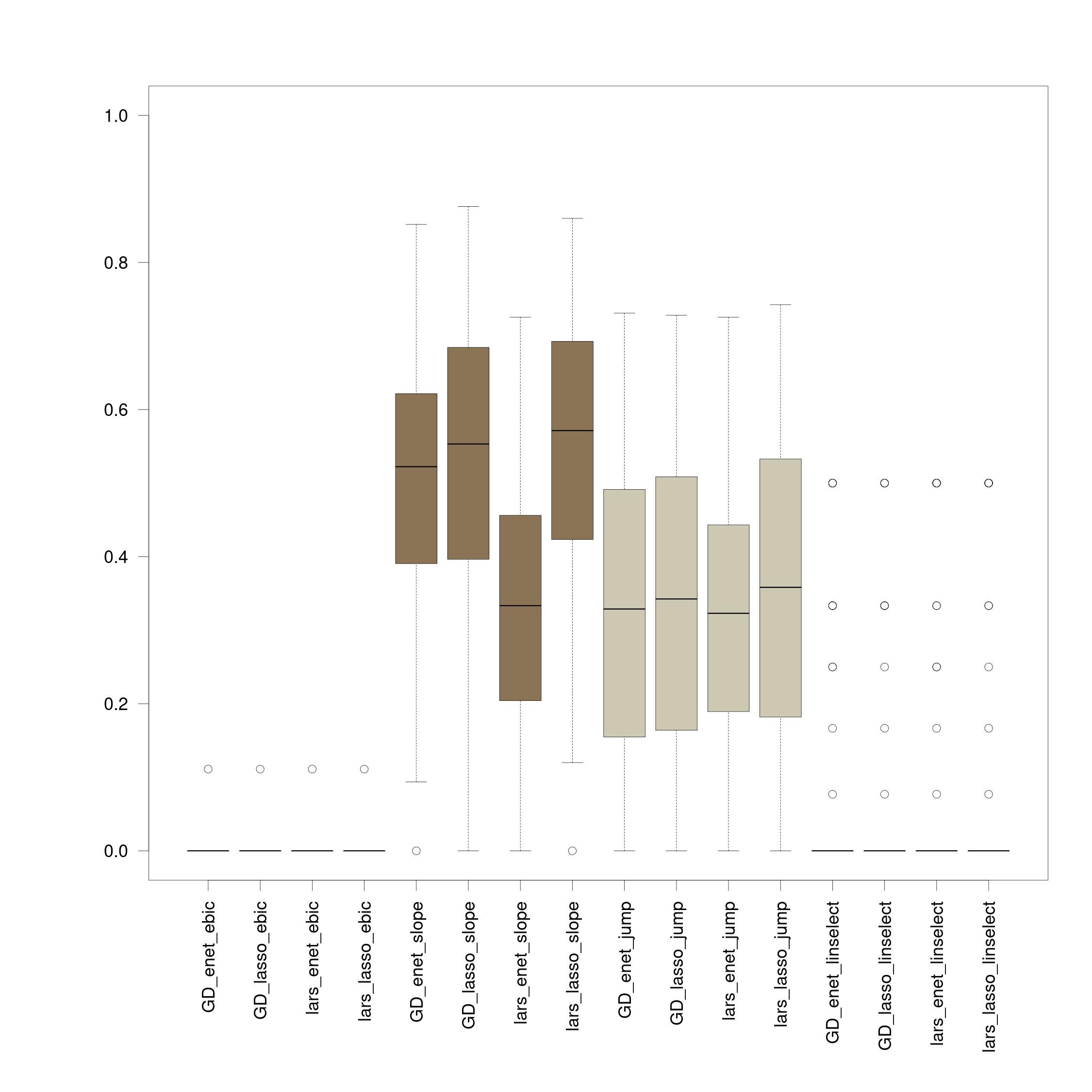}
    \caption*{\textit{scale-free-max}}
\end{subfigure}
\begin{subfigure}{0.49\linewidth}
    \centering
    \includegraphics[width=1\linewidth]{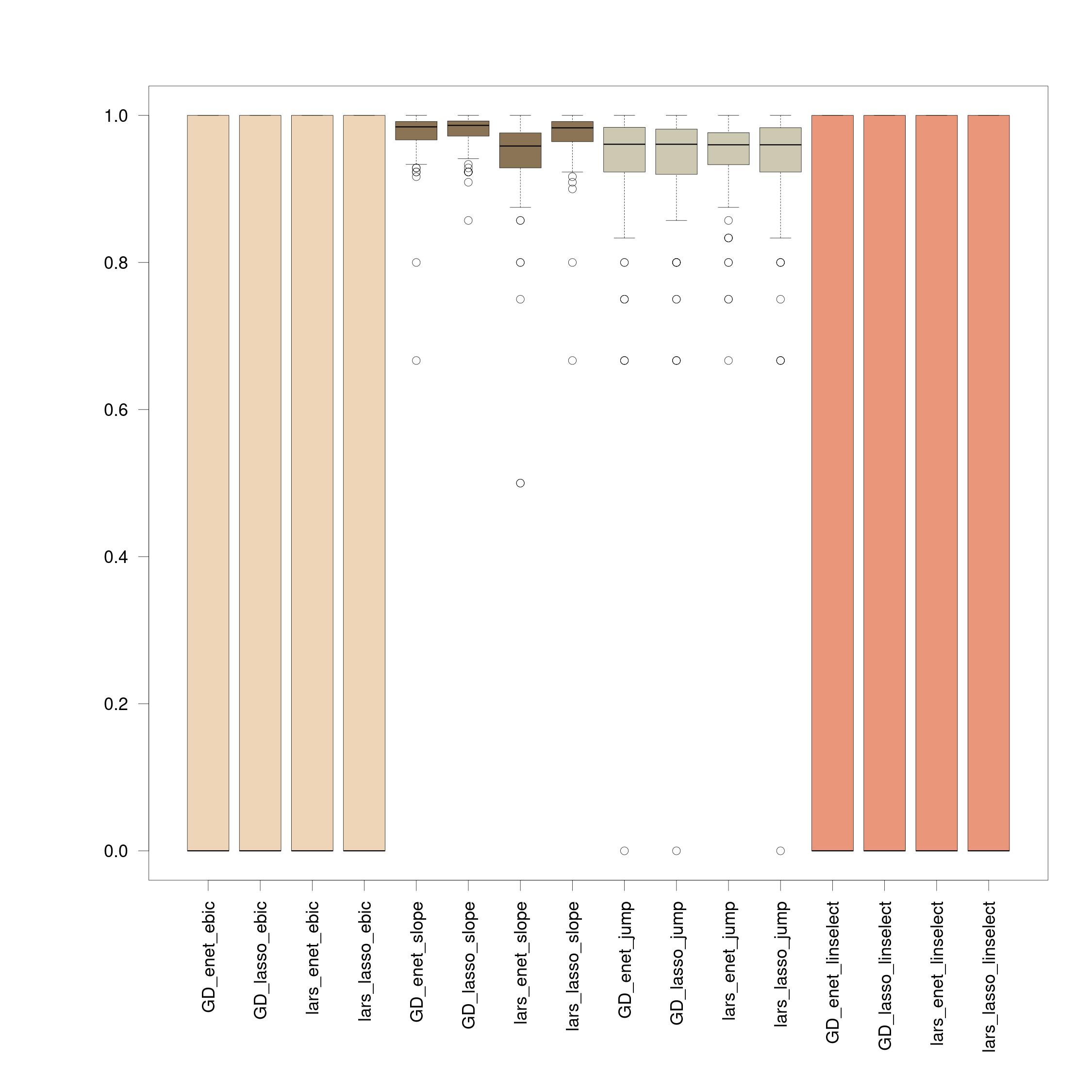}
    \caption*{\textit{scale-free-min}}
\end{subfigure}
\caption{Boxplots of the FDP values obtained by the model selection procedures from dataset of size $n=150$ in the four different settings.}
\label{fdp1}
\end{figure}

\begin{figure}[h!]
\begin{subfigure}{0.49\linewidth}
    \centering
    \includegraphics[width=1\linewidth]{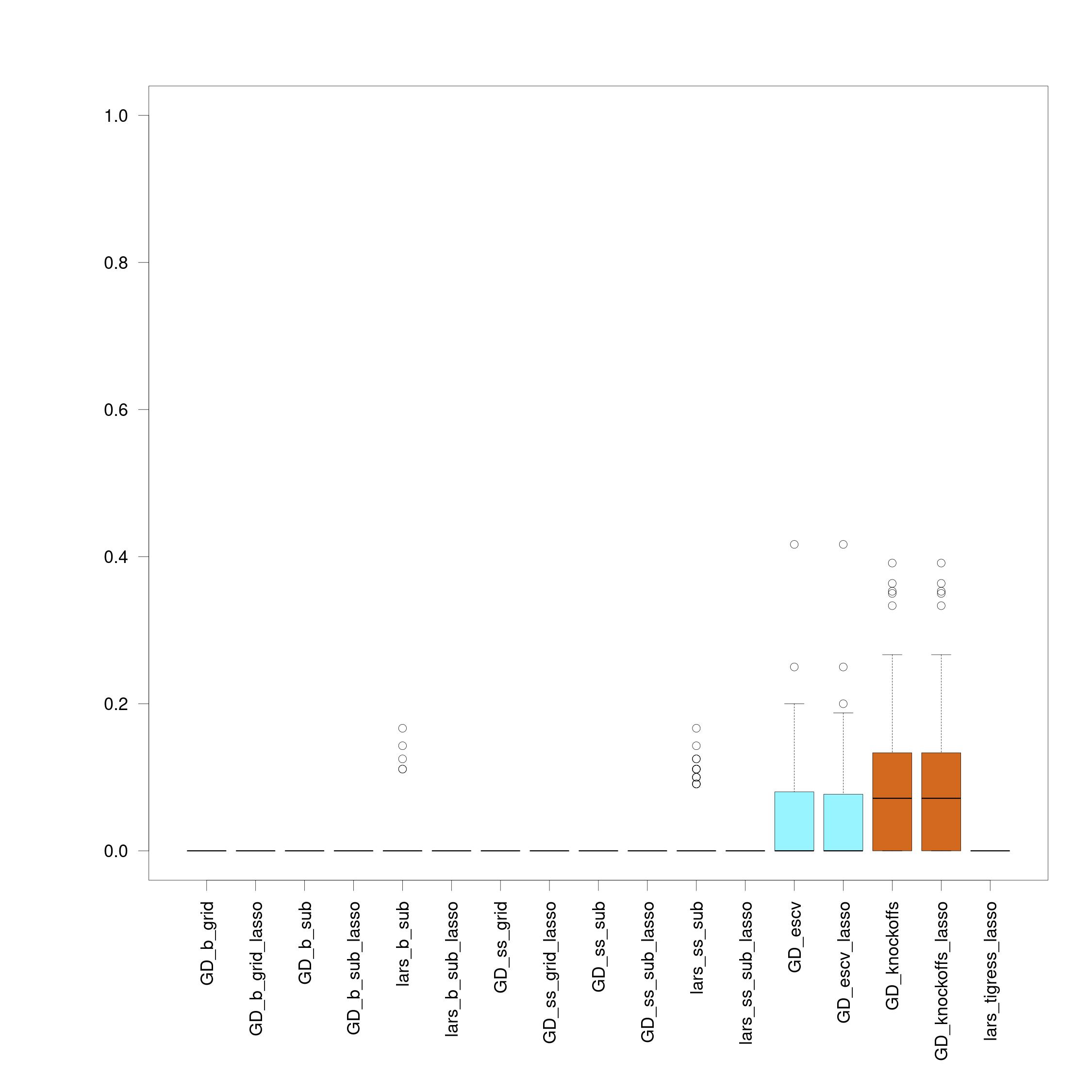}
    \caption*{\textit{independent}}
\end{subfigure}
\begin{subfigure}{0.49\linewidth}
    \centering
    \includegraphics[width=1\linewidth]{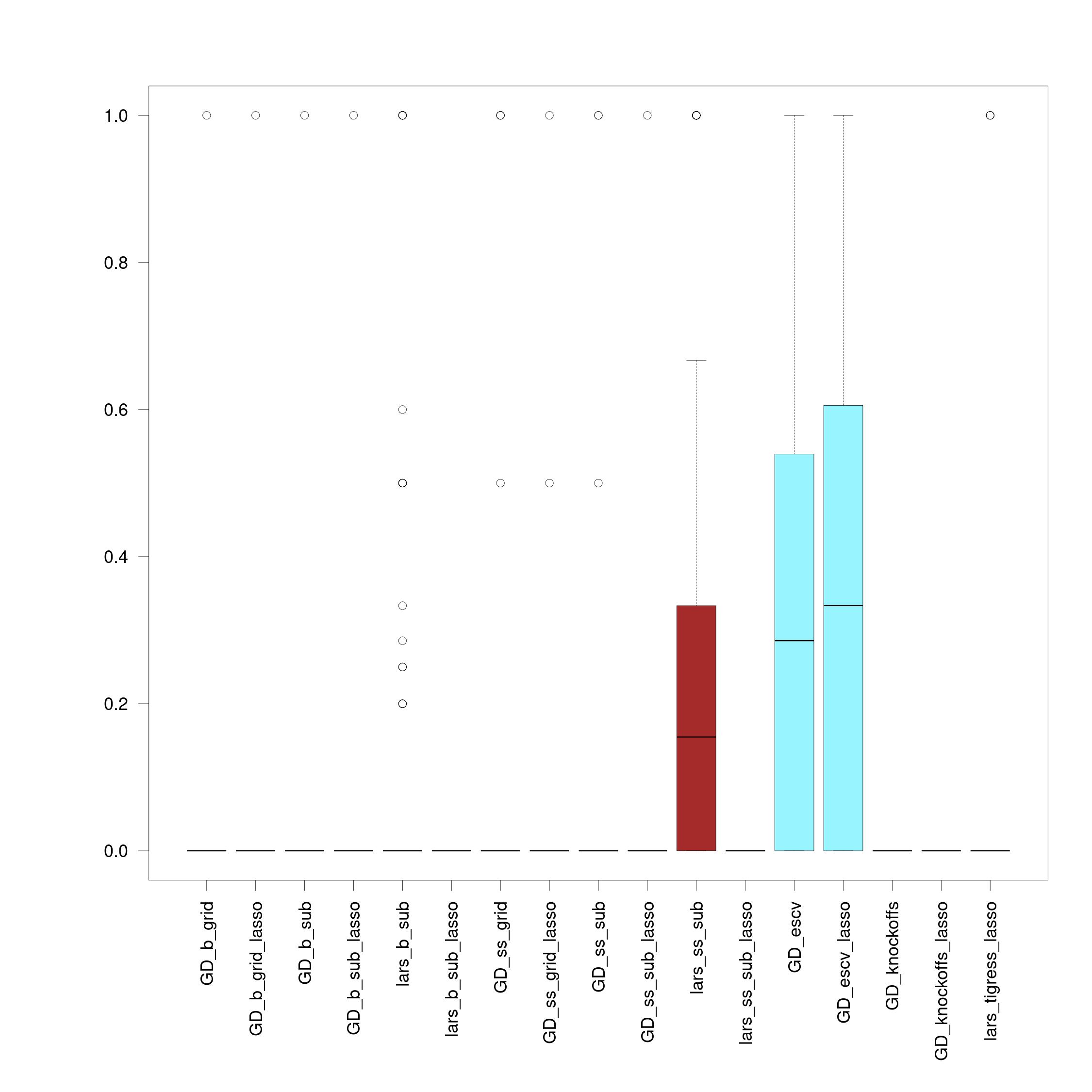}
    \caption*{\textit{cluster}}
\end{subfigure}
\begin{subfigure}{0.49\linewidth}
    \centering
    \includegraphics[width=1\linewidth]{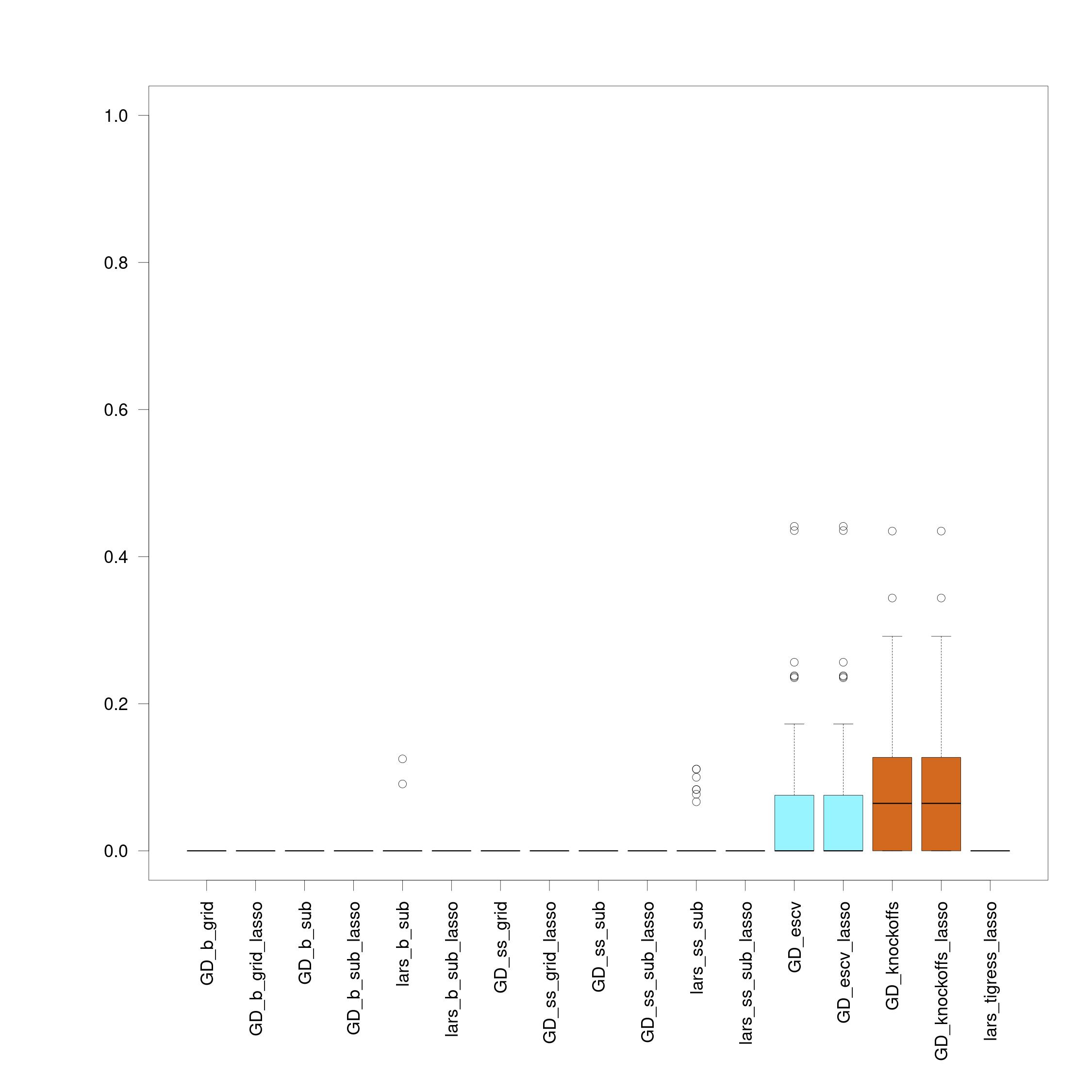}
    \caption*{\textit{scale-free-max}}
\end{subfigure}
\begin{subfigure}{0.49\linewidth}
    \centering
    \includegraphics[width=1\linewidth]{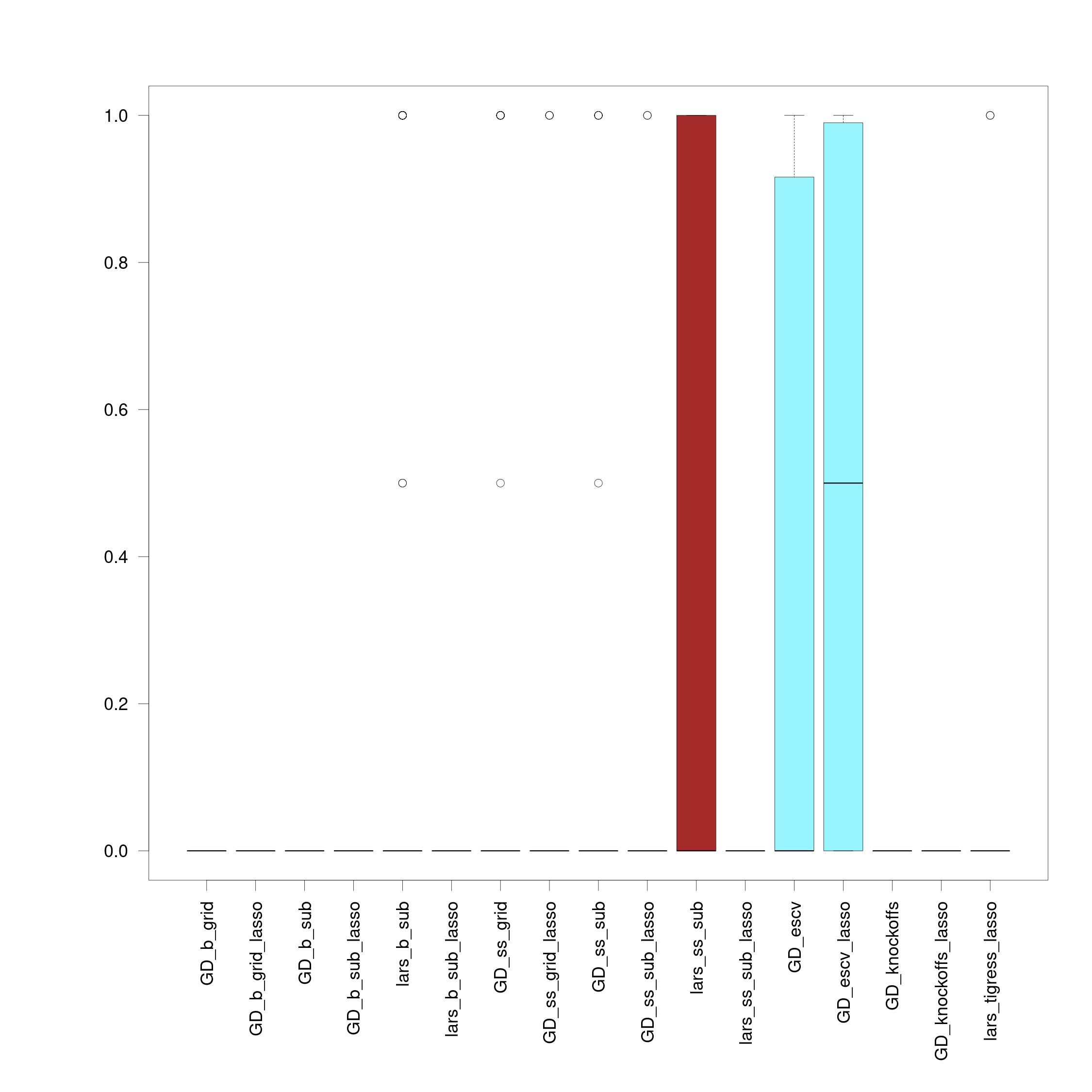}
    \caption*{\textit{scale-free-min}}
\end{subfigure}
\caption{Boxplots of the FDP values obtained by the variable identification procedures from dataset of size $n=150$ in the four different settings.}
\label{fdp2}
\end{figure}

\begin{figure}[h!]    
\centering
\includegraphics[width=1\linewidth]{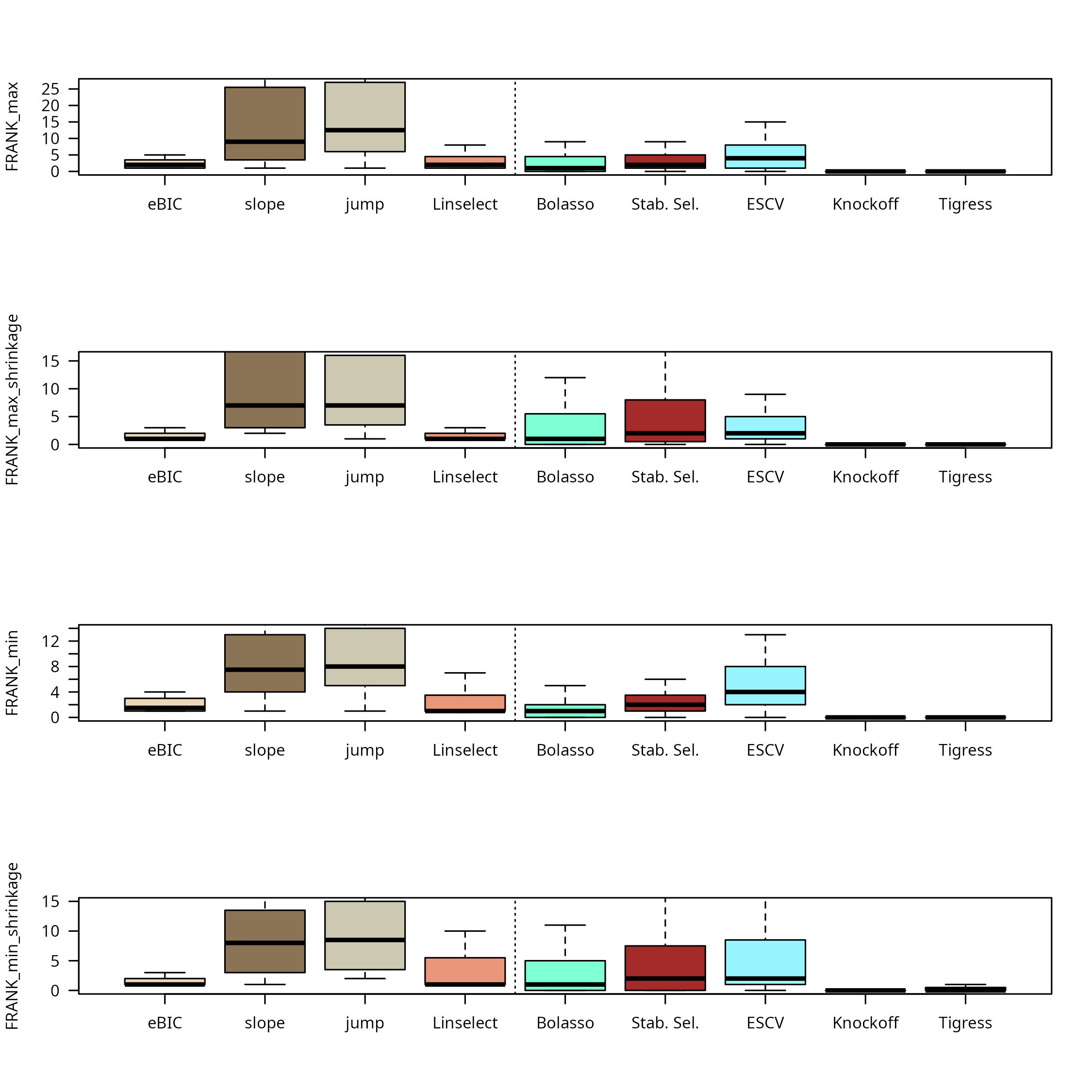}
\caption{Boxplots of the size of the selected variables subset from dataset of size $n=150$ in the FRANK setting. Results are presented with LARS with \textit{E-Net} regularization for the model selection methods, Bolasso and Stability Selection. For these latter, the sampling strategy is $sub$. For ESCV and the knockoff method which are based on the cyclic coordinate descent algorithm, the size is showed with \textit{E-Net}. Tigress is implemented with LARS and Lasso.
The red line indicates the number of active variables.}
\label{support_FRANK}
\end{figure}

\begin{figure}[!htb]
    \centering
    \includegraphics[width=1\linewidth,scale=0.75]{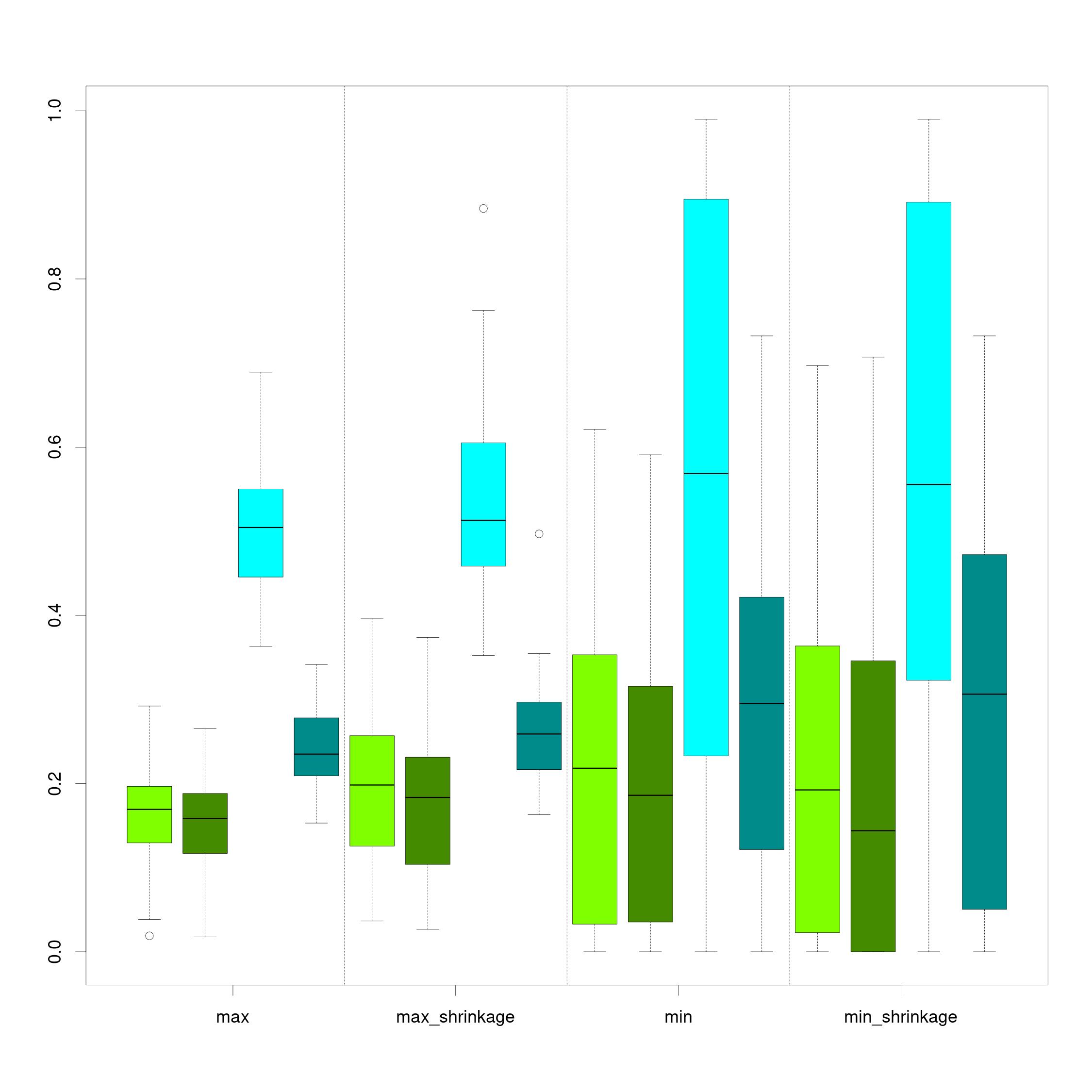}
\caption{Boxplots of the pROC-AUC values calculated on 40 samples of size $n=150$. Max is the FRANK-max setting. Max\_shrinkage is the FRANK-max setting where the nonparanormal transformation is applied on data.
Min is the FRANK-min setting. Min\_shrinkage is the FRANK-min setting where the nonparanormal transformation is applied on data.
The  cyclic coordinate descent algorithm combined with \textit{E-Net} is colored light green, the  cyclic coordinate descent algorithm combined with $\ell_1$ regularization is colored dark green, LARS combined with \textit{E-Net} is colored cyan and LARS combined with $\ell_1$ regularization function is colored dark cyan.}
\label{franck_auc}
\end{figure}

\begin{figure}[!htb]
    \centering
    \includegraphics[width=1\linewidth,scale=0.5]{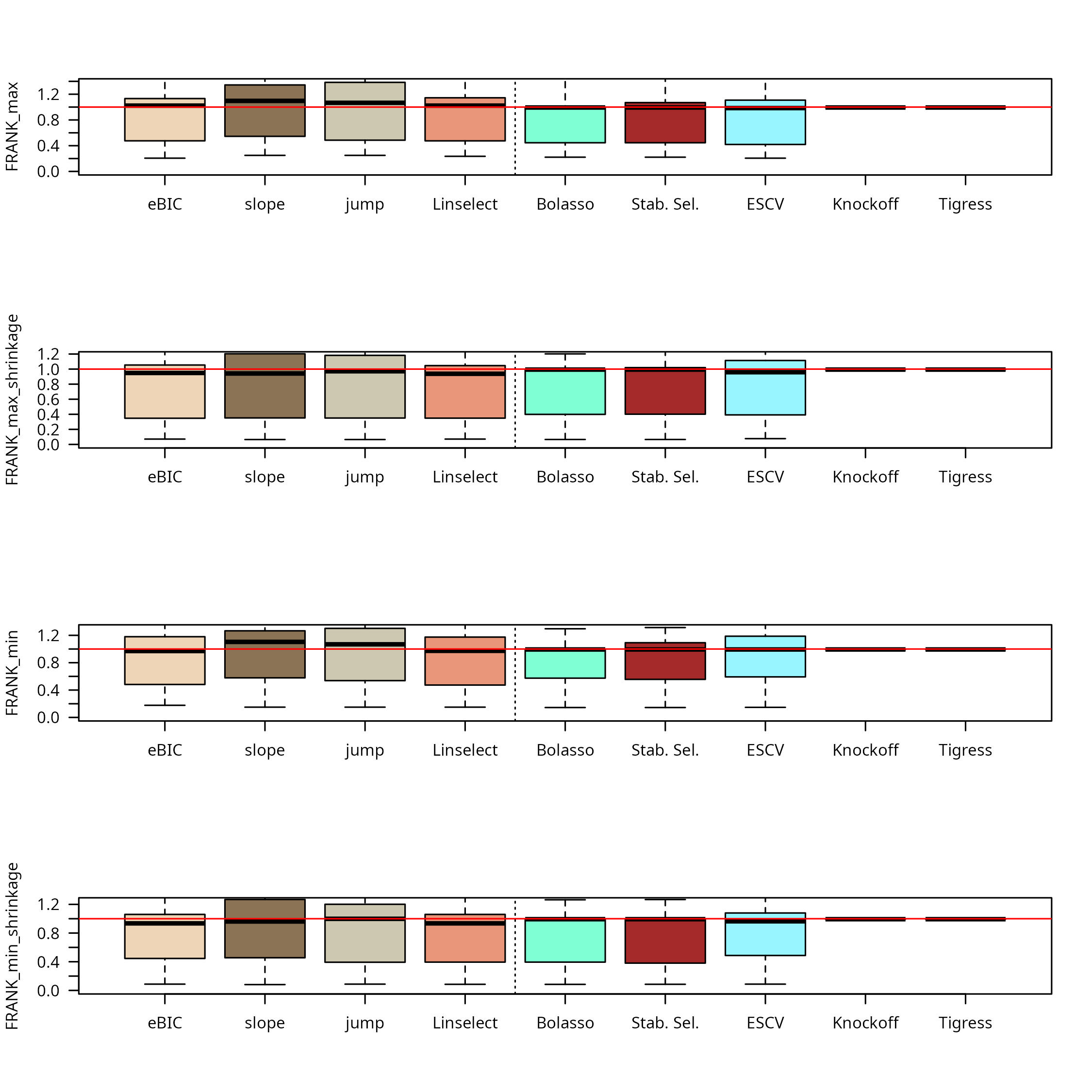}
\caption{Boxplots of the MSE calculated on 40 samples of size $n=150$. Max is the FRANK-max setting. Max\_shrinkage is the FRANK-max setting where the nonparanormal transformation is applied on data. 
Min is the FRANK-min setting. Min\_shrinkage is the FRANK-min setting where the nonparanormal transformation is applied on data.
The red line 1, the value below which the methods have a prediction ability. 
Results are presented with LARS with \textit{E-Net} regularization for the model selection methods, Bolasso and Stability Selection. For these latter, the sampling strategy is $sub$. For ESCV and the knockoff method which
are based on the cyclic coordinate descent algorithm, the MSE is showed with \textit{E-Net}. Tigress is implemented with LARS and Lasso. }
\label{franck_mse_150}
\end{figure}

\begin{figure}[!htb]
    \begin{subfigure}{0.49\textwidth}
        \centering
        \includegraphics[width=1\linewidth,scale=0.5]{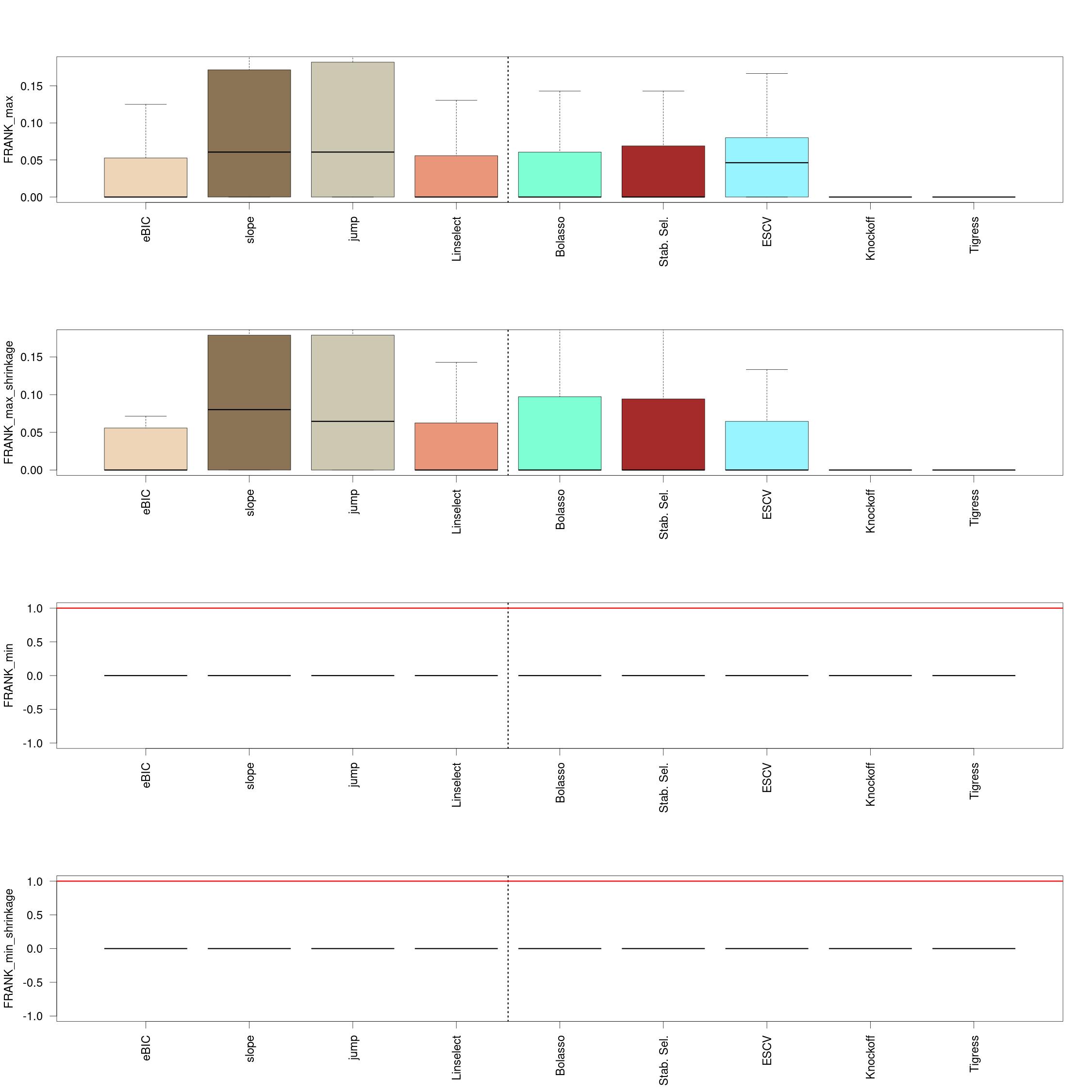}
    \end{subfigure}
    \hfill
    \begin{subfigure}{0.49\textwidth}
        \centering
        \includegraphics[width=1\linewidth,scale=0.5]{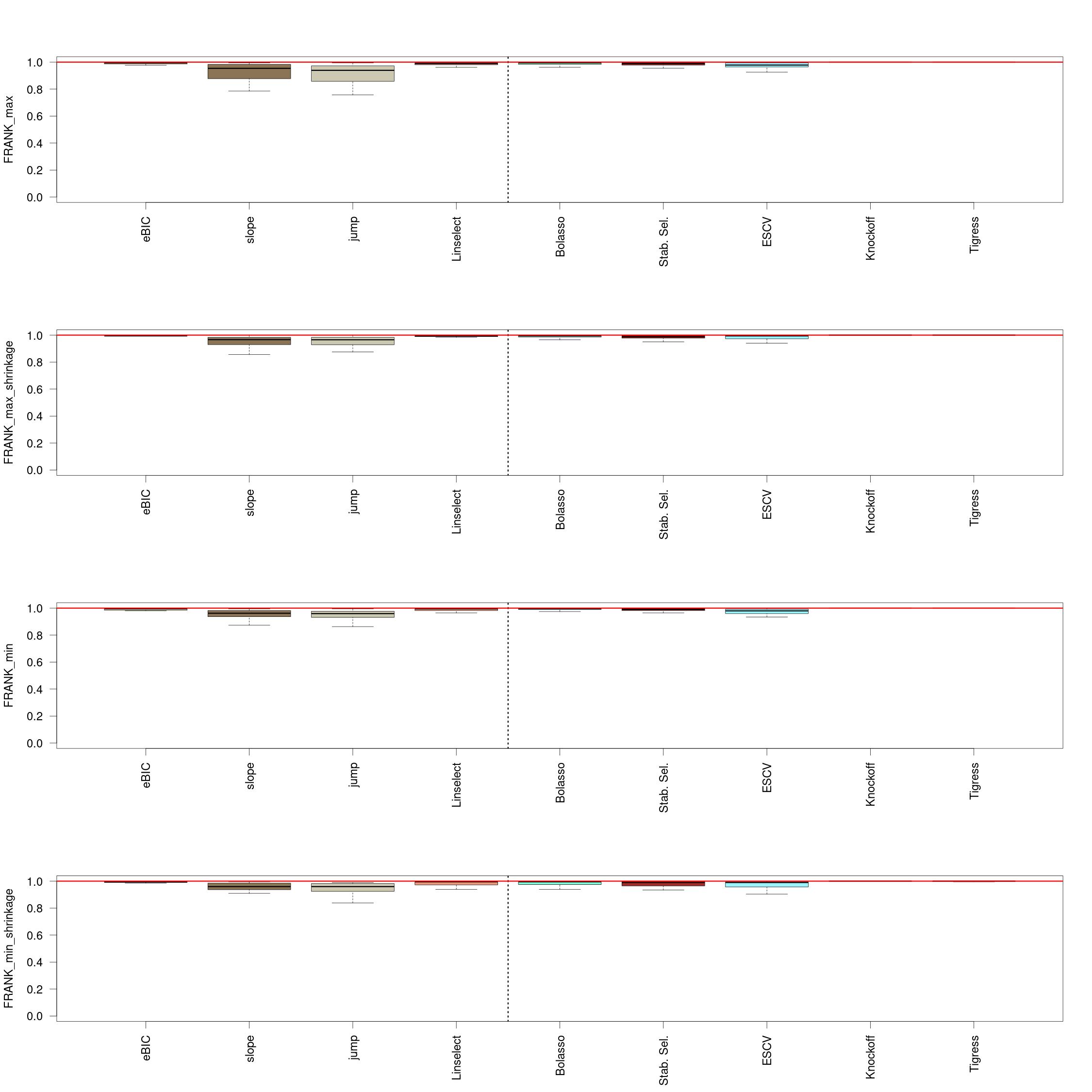}
    \end{subfigure}
\caption{Column A : Boxplots of the recall calculated on 40 samples of size $n=150$ in the four settings. The red line at 1 indicates the ability to recover all the active variables. Column B : Boxplots of the specificity  calculated on the same 40 samples of size $n=150$ in the four settings. The red line at 1 indicates the ability to not select all the non-active variables.
Max is the FRANK-max setting. Max\_shrinkage is the FRANK-max setting where the nonparanormal transformation is applied on data. 
Min is the FRANK-min setting. Min\_shrinkage is the FRANK-min setting where the nonparanormal transformation is applied on data.
Results are presented with LARS with \textit{E-Net}
regularization for the model selection methods, Bolasso and Stability Selection. For these latter, the sampling strategy is $sub$. For ESCV and the knockoff method which
are based on the cyclic coordinate descent algorithm, the recall and the specificity are showed with \textit{E-Net}. Tigress is implemented with LARS and Lasso.}
\label{franck_recall-150}
\end{figure}

\begin{figure}[!htb]
    \centering
    \includegraphics[width=1\linewidth,scale=0.5]{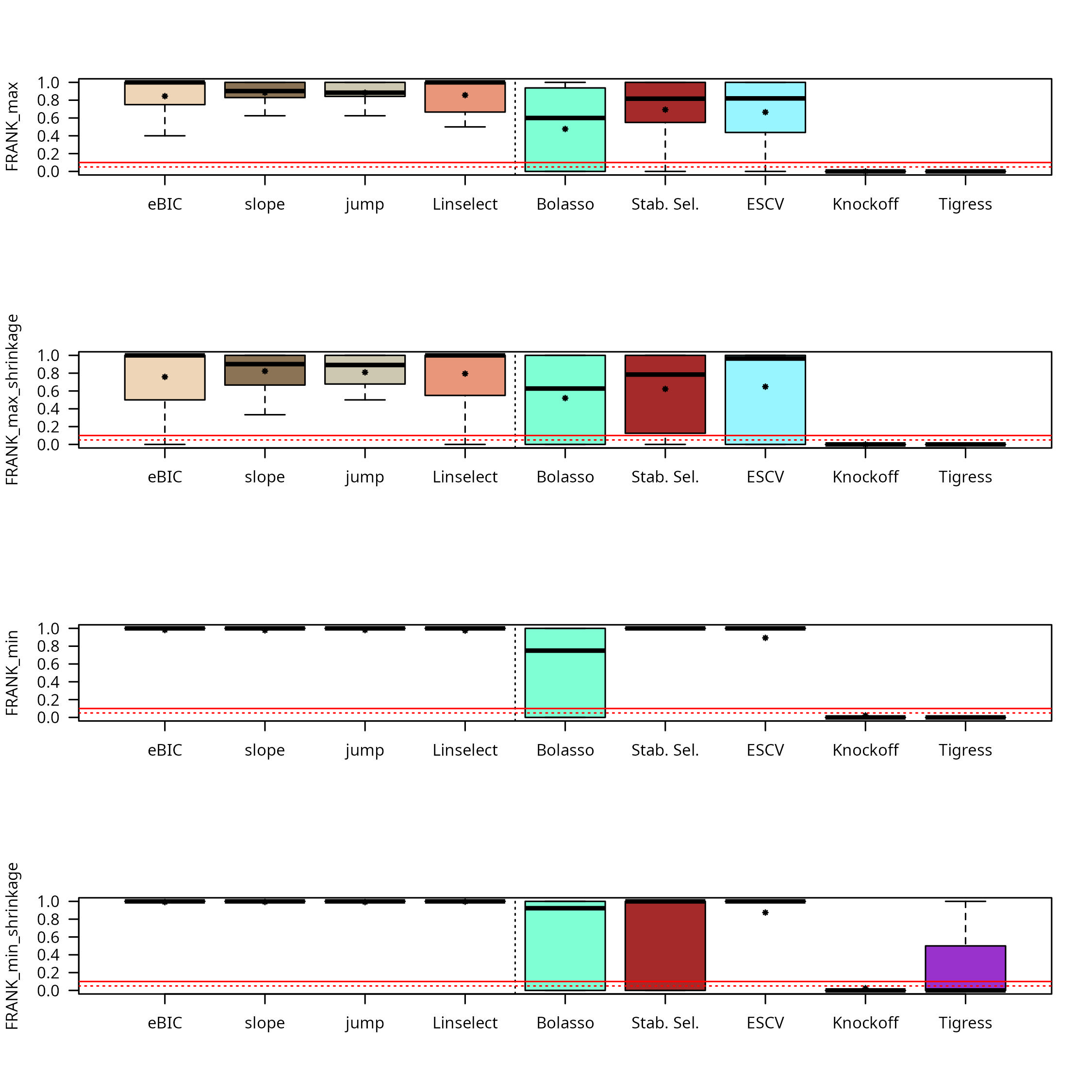}
\caption{Boxplots of the FDP calculated on 40 samples of size $n=150$. Max is the FRANK-max setting. Max\_shrinkage is the FRANK-max setting where the nonparanormal transformation is applied on data. 
Min is the FRANK-min setting. Min\_shrinkage is the FRANK-min setting where the nonparanormal transformation is applied on data.
The star indicates the expected threshold of the FDR (expectation of the FDP). The red line indicates $0.1$, the threshold for the knockoffs method. The dashed red line indicates $0.05$. 
Results are presented with LARS with \textit{E-Net} regularization for the model selection methods, Bolasso and Stability Selection. For these latter, the sampling strategy is $sub$. For ESCV and the knockoffs method which are based on the cyclic coordinate descent algorithm, the FDP is showed with \textit{E-Net}. Tigress is implemented with LARS  and Lasso.}
\label{franck_fdp-150}
\end{figure}

\end{document}